\documentclass[opre,nonblindrev]{informs3} 

\OneAndAHalfSpacedXI

\usepackage[plainpages=false,hyperfootnotes=false]{hyperref}
\hypersetup{
  colorlinks   = true, 
  urlcolor     = blue, 
  linkcolor    = red, 
  citecolor   = blue 
}

\usepackage{natbib}
 \bibpunct[, ]{(}{)}{,}{a}{}{,}%
 \def\bibfont{\small}%
\TheoremsNumberedThrough     
\ECRepeatTheorems

\EquationsNumberedThrough    

\MANUSCRIPTNO{}

\makeatletter
\long\def\@makecaption#1#2{
  \vskip\abovecaptionskip
  \sbox\@tempboxa{#1: #2}
  \ifdim \wd\@tempboxa >\hsize
    #1: #2\par
  \else
    \global \@minipagefalse
    \hb@xt@\hsize{\hfil\box\@tempboxa\hfil}
  \fi
  \vskip\belowcaptionskip}
\makeatother
\usepackage{graphicx}
\usepackage{amssymb,amsmath}
\usepackage{enumitem}
\usepackage{tikz}
\usepackage{siunitx}
\usepackage{tikzsymbols}
\usepackage{parskip}
\usepackage{floatrow}
\usepackage{dsfont}
\usepackage{setspace}
\usepackage{booktabs}
\usepackage{todonotes}
\usepackage{multirow}
\usepackage[linesnumbered,boxruled,lined,noend]{algorithm2e}
\SetAlFnt{\footnotesize}
\SetKw{Break}{break}
\SetKwComment{tcp}{$\triangleright$\ }{}

\definecolor{mgreen}{rgb}{0,0.455,0.247}
\definecolor{mblue}{rgb}{0.098,0.18,0.357}
\definecolor{mred}{rgb}{0.902,0.4157,0.0196}
\definecolor{mgrey}{rgb}{0.90196,0.90,0.90}
\definecolor{mblack}{rgb}{0,0,0}

\newcommand{\ba}{{\boldsymbol{\beta}}}
\newcommand{\A}{{\mathcal{A}_c}}
\newcommand{\At}{{\mathcal{A}_{c,t}}}
\newcommand{\Ao}{{\mathcal{A}_0}}

\newcommand{\B}{{\mathcal{B}}}

\newcommand{\Gat}{{G{\^a}teaux} }
\newcommand{\D}{{\mathcal D}}
\newcommand{\F}{{\mathcal F}}

\renewcommand{\L}{{\mathcal L}}
\newcommand{\M}{{\mathcal M}}
\newcommand{\N}{{\mathcal N}}

\newcommand{\T}{{\mathcal T}}
\newcommand{\bT}{\overline{{\mathcal T}}}
\newcommand{\Z}{{\mathcal Z}}

\newcommand{\rNN}{{\mathfrak{r}}}
\newcommand{\mfA}{{\mathfrak{A}}}
\newcommand{\mfB}{{\mathfrak{B}}}

\newcommand{\mfS}{{\mathfrak{S}}}

\newcommand{\R}{{\mathds{R}}}
\newcommand{\Id}{{\mathds{1}}}

\newcommand{\E}{{\mathbb{E}}}
\renewcommand{\P}{{\mathbb{P}}}

\newcommand{\bnu}{\boldsymbol{{\nu}}}

\newcommand{\bw}{\boldsymbol{{w}}}

\newcommand{\bTheta}{\boldsymbol{{\Theta}}}

\newcommand{\btheta}{{\boldsymbol\theta}}
\newcommand{\bvtheta}{{\boldsymbol\vartheta}}
\newcommand{\bvphi}{{\boldsymbol\varphi}}

\newcommand{\bpsi}{{\boldsymbol\psi}}

\newcommand{\tT}{t \in \T}

\newcommand{\1}{{\boldsymbol{1}}}

\newcommand{\bX}{{\boldsymbol X}}
\newcommand{\bx}{{\boldsymbol x}}
\newcommand{\bY}{{\boldsymbol Y}}
\newcommand{\bZ}{{\boldsymbol Z}}
\newcommand{\bmZ}{{\boldsymbol{\mathcal{Z}}}}

\newcommand{\RC}{RC}
\newcommand{\RM}{{\mathfrak{R}}}

\newcommand{\mfT}{{\mathfrak{T}}}
\newcommand{\mfF}{{\mathfrak{F}}}
\newcommand{\mff}{{\mathfrak{f}}}
\newcommand{\mfg}{{\mathfrak{g}}}

\newcommand{\dtheta}{{\delta\theta}}
\newcommand{\dbtheta}{{\delta\boldsymbol\theta}}

\newcommand{\VaR}{\textrm{VaR}}
\newcommand{\ES}{\textrm{ES}}

\newcommand{\dZ}{{\zeta}}

\newcommand{\bdX}{{\Delta \boldsymbol{X}}}
\newcommand{\dX}{{\Delta {X}}}

\renewcommand{\be}{{\boldsymbol e}}

\newcommand{\ep}{{\varepsilon}}

\newcommand\numberthis{\addtocounter{equation}{1}\tag{\theequation}}

\tikzstyle{theta}=[shape=circle,draw=mgreen,fill=mgreen!10]
\tikzstyle{state}=[shape=circle,draw=mred,fill=mred!10]
\tikzstyle{hidden}=[shape=circle,draw=mblue,fill=mblue!10]
\tikzstyle{gru}=[shape=rectangle,draw=black!50,fill=lime!20]
\tikzstyle{FFN}=[shape=rectangle,draw=black!50,fill=lime!20]
\tikzstyle{obs}=[shape=circle,draw=mblue,fill=mblue!10]
\tikzstyle{lightedge}=[<-,dotted]
\tikzstyle{mainstate}=[state,thick]
\tikzstyle{mainedge}=[<-,thick]	

\begin{document}


\RUNAUTHOR{Pesenti, Jaimungal, Saporito, Targino}

\RUNTITLE{Risk Budgeting Allocation for Dynamic Risk Measures}

\TITLE{Risk Budgeting Allocation for Dynamic Risk Measures}

\ARTICLEAUTHORS{%
\AUTHOR{Silvana M. Pesenti}
\AFF{Department of Statistical Sciences, University of Toronto, Canada, \\
\EMAIL{silvana.pesenti@utoronto.ca}}
\AUTHOR{Sebastian Jaimungal }
\AFF{Department of Statistical Sciences, University of Toronto, Canada, \& \\
Oxford-Man Institute, University of Oxford 
\\
\EMAIL{sebastian.jaimungal@utoronto.ca}} 
\AUTHOR{Yuri F. Saporito}
\AFF{School of Applied Mathematics (EMAp), Getulio Vargas Foundation (FGV), Brazil, \\
\EMAIL{yuri.saporito@fgv.br}}
\AUTHOR{Rodrigo S. Targino}
\AFF{School of Applied Mathematics (EMAp), Getulio Vargas Foundation (FGV), Brazil, \\
\EMAIL{rodrigo.targino@fgv.br}}
\vspace{1em}
February 29, 2024\footnote{First version: May 18, 2023.}
}

\ABSTRACT{%
  We define and develop an approach for risk budgeting allocation -- a risk diversification portfolio strategy -- where risk is measured using a dynamic time-consistent risk measure. For this, we introduce a notion of dynamic risk contributions that generalise the classical Euler contributions and which allow us to obtain dynamic risk contributions in a recursive manner. We prove that, for the class of coherent dynamic distortion risk measures, the risk allocation problem may be recast as a sequence of strictly convex optimisation problems. Moreover, we show that self-financing dynamic risk budgeting strategies with initial wealth of $1$ are scaled versions of the solution of the sequence of convex optimisation problems. Furthermore, we develop an actor-critic approach, leveraging the elicitability of dynamic risk measures,  to solve for risk budgeting strategies using deep learning.  
  }%


\maketitle

\section{Introduction}\label{sec:intro}
The ``risk parity'' portfolio has been pioneered by Bridgewater Associates, when in 1996 it launched the \textit{All Weather} asset allocation strategy -- a portfolio strategy withstanding all weathers -- although the term risk parity was only coined in 2005 in the white paper by \cite{Qian2005PanagoraAM}. Risk parity originated from the desire of a diversified portfolio and the realisation that an equally weighted portfolio is diversified in asset allocation but not in the extent in which each asset contributes to the overall portfolio risk \citep{Qian2011JI}. Emphasised by the 2008 financial crisis, the call for ``maximally'' diversifying a portfolio's risk was born, see e.g. \cite{Choueifaty2008JPM}. Risk parity enjoys widespread popularity in industry as numerous portfolio (performance) comparison studies illustrate, see, e.g., \cite{Chaves2011JI}, \cite{Lee2011JPM}, and \cite{Asness2012FAJ}. An early mathematical formalisation of risk parity strategies can be found in \cite{Maillard2010JPM} and \cite{roncalli2013book}.

Risk parity strategies and more broader risk budgeting strategies are portfolio allocations where the contribution of each asset to the overall portfolio risk is prespecified, e.g. for risk parity each assets contributes equally to the portfolio risk. Thus, central to risk budgeting is the way the risk of a portfolio is quantified. While most of the extant literature measures risk using the portfolio variance and further restrict to assets that follow multivariate Gaussian distributions, recent works relax these assumptions. \cite{Bruder2016SSRN} and \cite{Jurczenko2019II} study the Expected Shortfall (ES; also called Conditional Value-at-Risk) risk measure under the assumption that assets are multivariate Gaussian distributed, resulting in explicit formulae for risk contributions. Further works on risk budgeting include \cite{Ji2018AOR} who utilise the downside risk measure, 
\cite{Bellini2021EJOR} who consider expectile risk measures, \cite{anis2022EJOR} who incorporate asset selection, and \cite{Freitas2022SSRN} who propose algorithms based on the cutting planes methodology to calculate risk budgeting strategies for coherent risk measures. \cite{Haugh2015JCF} combine risk budgeting of (overlapping) groups of asset with simultaneously maximising return and minimising risk.
Variations of risk budgeting portfolio strategies are considered in \cite{Bai2016QF} who propose alternative optimisation problems to solve for risk parity portfolios. \cite{Meucci2015Risk} and \cite{Roncalli2016QF} construct risk factor budgeting portfolios, that are portfolios where each (uncorrelated) factor, rather than asset, contributes equally to the portfolio variance. \cite{Lassance2022OR} continues this line of work by including independent component analysis.

None of these works, however, addresses the dynamic nature of investments, i.e., that portfolio strategies are typically holistically considered over a time horizon larger than one period; we  henceforth refer to the one period setting as the ``static'' setting. In this paper, we develop a dynamic setting in which an investor trades over a finite time horizon using a self-financing risk budgeting strategy. Specifically, the investor aims to create a portfolio strategy, such that at every decision point each asset contributes a prespecified percentage to the future risk of the portfolio. This means that the investor's problem is a multi-period decision problem. Whenever decisions occur over multiple periods, the investor's ``optimal'' choices should be coherent over time. This can be  achieved, for example, by optimising a time-consistent criterion.\footnote{There are a number of alternative approaches to time-inconsistencies, see, e.g., \cite{bjork2021time}, we however, opt to use time-consistent criteria.} Indeed, when decisions are stemming from a time-inconsistent objective (e.g. a static risk measure), the ``optimal'' decision at a future point in time and state may not be optimal when one arrives at that future point in time in that very state \citep{bielecki2018unified}. This is in contrast to time-consistent objectives (e.g., dynamic time-consistent risk measures), which result in decisions that are coherent across time and state. Thus, we consider dynamic time-consistent risk measures to evaluate the risk of a portfolio strategy.

Dynamic time-consistent risk measures have been extensively studied to evaluate the risk of a sequence of random costs, such as the profit and loss (P\&L) of a portfolio strategy; indicatively see \cite{Cheridito2006EJP,ruszczynski2010risk,bielecki2022risk,coache2022conditionally,Bielecki2023WP}. For dynamic risk budgeting, however, we further require the allocations of the dynamic risk to each asset and each time point, a topic whose literature is sparse. An early work for allocations of coherent dynamic risk measures is \cite{Cherny2009MF} and for  
BSDE-based dynamic time-consistent risk measures we refer to \cite{kromer2014TAF,kromer2017TAF}, and \cite{matrogiacomo2021ArXiv}. Related but conceptually different is the work of \cite{Schilling2020MS} who axiomatically study how to decompose a risk dynamically. While working in a dynamic setting, their risk is the portfolio loss itself and not a dynamic risk measure applied to it. 

In this work, we consider the class of dynamic time-consistent risk measures that arise from conditional one-step distortion risk measures. A case in point is the ES whose security level may depend on the investor's wealth  or asset price. For this class, we define their dynamic risk contributions via \Gat derivatives and derive explicit formulae. While most of our results hold for conditional distortion risk measures, we focus on the subset of conditional coherent distortion risk measures, as defining risk allocations for non-coherent risk measures provide an ``incentive for infinite fragmentation of portfolios'' \citep{Tsanakas2009IME}. In the static setting, \Gat derivatives enjoy a long history as risk contributions, also in connection to cooperative game theory. We provide a detailed literature review in Section \ref{sec:risk-contribution}. With this definition of dynamic risk contributions at hand, we define a dynamic risk budgeting portfolio as a strategy whose risk contributions at each point in time are a predefined percentage of the future risk of the strategy. We prove, under mild conditions, that a self-financing dynamic risk budgeting strategy with initial wealth of $1$ is a scaled version of the solution of a sequence of strictly convex optimisation problems. Furthermore, we develop an actor-critic approach to solve the sequence of optimisation problems using deep learning techniques and provide illustrative examples. 

This manuscript is organised as follows. Section \ref{sec:dynamic-rm} introduces dynamic time-consistent risk measures and in Section \ref{sec:rm-self-finacing}, we apply a dynamic time-consistent risk measure to a self-financing strategy and derive a recursive representation. In Section \ref{sec:risk-contribution} we define dynamic risk contributions via the \Gat derivative and derive explicit formulae for the class of dynamic distortion risk measures. Section \ref{sec:dynamic-risk-budgeting} is devoted to dynamic risk budgeting portfolio strategies and we show in Theorem \ref{thm:opt} that a self-financing dynamic risk budgeting strategy with initial wealth of $1$ can be obtained as a scaled version of the solution to a collection of strictly convex optimisation problems. Section \ref{sec:nuermical-implemenatation} discusses how to solve this family of optimisation problems using neural networks  leveraging elicitability of conditional risk measures (Subsection \ref{subsec:elicitable}). Illustrations of risk budgeting strategies are provided in Section \ref{sec:num-examples}. Delegated to the appendix are auxiliary definitions and results (\ref{app:aux-res}), additional technical proofs (\ref{app:sec:proof}), elaborations on conditional elicitability (\ref{sec:appendix-elicitability}), and details on the numerical implementation (\ref{app:numerical}).

\section{Dynamic Risk Assessment}\label{sec:dynamic-rm}

We work on a filtered and completed probability space $(\Omega, \F, (\F_t)_{t\in\bT}, \P)$, where $\bT:=\{0,1,\dots,T+1\}$, and $T \in \mathds{N}$ is a known and finite time horizon. The information available to the investor is encapsulated in the filtration $(\F_t)_{t \in \bT}$, and we assume that $\F_0=(\emptyset,\Omega)$ is the trivial $\sigma$-algebra, and simply write $\E[\cdot]:= \E[\cdot~|\F_0]$. We further denote the spaces of square-integrable random variables (rvs) and sequences by $\Z := \{Z\in\F: \E[Z^2]<+\infty\}$, $\Z_t:=\{Z_t\in\Z: Z_t\in\F_t\}$, and $\Z_{t:T+1}:=\{(Z_t,Z_{t+1},\ldots,Z_{T+1})\in\Z_t\times \Z_{t+1}\times\cdots\times\Z_{T+1}\}$, for all $t\in\bT$. Similarly, we define the spaces of $n$-dimensional random vectors and sequences by $\bmZ := \{\bZ = (Z_1, \ldots, Z_n) : Z_i \in \Z\,, \; \forall i = 1, \ldots, n\}$, $\bmZ_t:=\{\bZ_t\in\bmZ: \bZ_t\in\F_t\}$, and $\bmZ_{t:T+1}:=\{(\bZ_t,\bZ_{t+1},\ldots,\bZ_{T+1})\in\bmZ_t\times \bmZ_{t+1}\times\cdots\times\bmZ_{T+1}\}$, for all $t\in\bT$. {We further define $\L^\infty:= \L^\infty(\Omega, \F,\P)$ and $\L_t^\infty:= \L^\infty(\Omega, \F_t, \P)$ for all $t \in\bT$. Unless otherwise stated, all (in)equalities of random vectors are to be understood component-wise and in a $\P$-almost sure (a.s.) sense.

\subsection{Dynamic Risk Measures}

The agent assesses the risk associated with a trading strategy by a dynamic time-consistent risk measure, which are families of conditional risk measures that satisfy the property of strong time-consistency; see Definition \ref{def:time-consistent} below. 
We adopt the setting of \cite{Cheridito2006EJP}  and  \cite{ruszczynski2010risk} for dynamic risk measures and refer the interested reader to those works and reference therein.

\begin{definition}[Dynamic Risk Measures]\label{def:dynamic-rm}
\fontfamily{lmss}\selectfont
A dynamic risk measure on $\bT$ is a family $\{\rho_{t,T+1}\}_{t \in \bT}$, where for each $t\in\bT$, the conditional risk measure is a mapping $\rho_{t,T+1} \colon \Z_{t:T+1} \to \Z_t$. 
We say that a dynamic risk measure possesses one of the following properties, if for all $t \in \bT$:
\begin{enumerate}[label=\roman*)]
    \item \textbf{Normalisation:}
    $\rho_{t,T+1} (0,\ldots,0) = 0$. 

    \item \textbf{Monotonicity:} $\rho_{t,T+1}(Z_{t:T+1}) \le \rho_{t,T+1}(Y_{t:T+1})$, for all $Z_{t:T+1}, Y_{t:T+1} \in \Z_{t:T+1}$ with $Z_{t:T+1} \le Y_{t:T+1}$.
    
    \item \textbf{Translation invariance:} 
    $\rho_{t,T+1}(Z_{t:T+1}) = Z_{t} + \rho_{t,T+1}(0, Z_{t+1}, \ldots, Z_{T+1})$, for all $Z_{t:T+1}\in\Z_{t:T+1}$.

    \item \textbf{Convexity:} $\rho_{t,T+1}(\lambda \, Z_{t:T+1} + (1-\lambda)\,  Y_{t:T+1}) \le \lambda\, \rho_{t,T+1}( Z_{t:T+1}) + (1-\lambda)\, \rho_{t,T+1}(  Y_{t:T+1})$, for all  $\lambda \in \F_t$ with $0\le \lambda\le 1$ and $Y_{t:T+1}, Z_{t:T+1}\in\Z_{t:T+1}$.

    \item \textbf{Positive homogeneity:} $\rho_{t,T+1}(\lambda\,  Z_{t:T+1}) = \lambda\, \rho_{t,T+1}(Z_{t:T+1})$, for all $\lambda \in \L_t^\infty$ with $\lambda > 0$ and $Z_{t:T+1}\in\Z_{t:T+1}$.

    \item \textbf{Coherency:} $\rho_{t,T+1}$ is monotone, translation invariant, convex, and positive homogeneous.
\end{enumerate}
\end{definition}
The mapping $\rho_{t,T+1}$ thus assesses the risk of the sequence $Z_{t:T+1}\in\Z_{t:T+1}$ viewed from time $t$, by mapping it to an $\F_t$-measureable rv. The investor may view this as the $\F_t$-measurable quantity they are willing to exchange in place of the sequence of future risks.

Next, we recall the notion of strong time-consistency, which, for simplicity, we refer to as time-consistency. A dynamic risk measure is time-consistent if it compares risks coherently over time. Specifically, if the time-$s$ risk of one stochastic process is larger than another, then the former should also be riskier at an earlier time $t<s$, if the processes are a.s. equal at all times $u$ satisfying $t\le u<s$. Thus, 
time-consistency is a property that results in optimal sequential decisions that are coherent when optimised at different points in time, and that further leads to a dynamic programming principle for optimising dynamic risk measures. 
\begin{definition}[Strong time-consistency -- \cite{Cheridito2006EJP}]\label{def:time-consistent}
\fontfamily{lmss}\selectfont
A dynamic risk measure  $\{\rho_{t,T+1}\}_{t \in \bT}$ is (strong) time-consistent if for all $Z_{t:T+1}, Y_{t:T+1} \in \Z_{t:T+1}$ that satisfy for some $s\in\{t, \ldots, T+1\}$ 
\begin{equation*}
    Z_{t:s} = Y_{t:s}
    \quad \text{and}\quad
    \rho_{s,T+1}(Z_{s:T+1}) \le \rho_{s,T+1}(Y_{s:T+1}) 
\end{equation*}
it holds that 
\begin{equation*}
    \rho_{t,T+1}(Z_{t:T+1})
    \le
    \rho_{t,T+1}(Y_{t:T+1})\,,
\end{equation*}
and where $Z_{t:s}:= (Z_t, \ldots, Z_s, 0,\ldots, 0)$ is understood as the projection of $Z_{t:T+1}$ onto $\Z_{t} \times \cdots \times \Z_s$.
\end{definition}

While not apparent at first, the theorem below shows that time-consistency creates a connection between dynamic risk measures and so-called one-step risk measures. In particular, the theorem states that a dynamic time-consistent risk measure induces a family of one-step risk measures and conversely, any family of one-step risk measures defines a dynamic time-consistent risk measure. The following theorem is due to \cite{Cheridito2006EJP} and \cite{ruszczynski2010risk}.

\begin{theorem}[Recursive Relation]
    \label{thm:recursive-relation}
    Let $\{\rho_{t,T+1}\}_{t \in \bT}$ be a dynamic risk measure which is monotone, normalised, and translation invariant. 
    Then $\{\rho_{t,T+1}\}_{t \in \bT}$ is time-consistent if and only if the following recursive representation holds:
    \begin{equation}
    	\rho_{t,T+1} (Z_{t}, \ldots, Z_{T+1}) = Z_{t} +
    	\rho_{t} \Bigg( Z_{t+1} +
    	\rho_{t+1} \bigg( Z_{t+2} +
    	\cdots +
    	\rho_{T-1} \Big( Z_{T} +
    	\rho_{T} \big( Z_{T+1} \big) \Big) \cdots \bigg) \Bigg)\,,\label{eq:dynamic-risk}
    \end{equation}
    where the
     one-step risk measures $\{\rho_{t}\}_{\tT}$ are mappings $\rho_t \colon \Z_{t+1} \to \Z_t$, defined by $\rho_t(Z_{t+1}):=\rho_{t,T+1}(0, Z_{t+1}, 0, \ldots, 0)$, for $Z_{t+1} \in \Z_{t+1}$. Moreover the one-step risk measures are monotone, normalised, and translation invariant.
\end{theorem}

By Theorem \ref{thm:recursive-relation}, any family of mappings $\rho_t \colon \Z_{t+1} \to \Z_t$ that are  monotone, normalised, and translation invariant, for all $\tT$,$\T:=\{0,1,\dots,T\}$, gives rise to a dynamic time-consistent risk measure and vice-versa. Thus, without loss of generalisation, we make a slight abuse of terminology and call $\{\rho_t\}_{t\in \T}$ a dynamic time-consistent risk measure (DRM) with representation \eqref{eq:dynamic-risk}.

For defining risk budgeting strategies we further require the DRM to be positive homogeneous and convex; thus, coherent. Throughout the remainder of the exposition we focus on coherent distortion DRMs, which are a generalisation of the class of distortion risk measures to the dynamic setting. Coherent distortion risk measures span the subclass of law-invariant coherent risk measures that are comonotonic additive \citep{Kusuoka2001AME}. Dynamic distortion risk measures with deterministic distortion function have been considered in \cite{Bielecki2023WP} who focus on coherent dynamic acceptability indices generated by families of distortion functions. Here, we allow the distortion function to be both time and state dependent, which differs from earlier works. For this we first define for each $\tT$ the regular cumulative distribution function (cdf) of $Z \in \Z_{t
+1}$ conditional on $\F_t$ as $F_{Z|\F_t}(z) := \P(Z \le z~|~\F_t)$, and the regular (left-) quantile function of $Z$ conditional on $\F_t$ as $F_{Z|F_t}^{-1}(u):= \inf\{y\in \R ~|~ F_{Z|\F_t}(u) \ge y\}$. Moreover, define $U_{Z|\F_t}:=F_{Z|\F_t}(Z)$,
which is $\F_{t+1}$-measurable and, conditional on $\F_t$, uniform distributed on $(0,1)$.\footnote{If $F_{Z|\F_t}(\cdot)$ is continuous, then $U_{Z|\F_t}$ is, conditionally on $\F_t$, a uniform rv. If $F_{Z|\F_t}(\cdot)$ is discontinuous, define $\tilde{F}_{Z|\F_t}(z, \lambda):= \P\left(Z< z ~|~\F_t\right) + \lambda \,\P\left(Z = z ~|~\F_t\right)$. Next, let $V$ be a uniform $\F_{t+1}$-measurable rv that is, conditional on $\F_t$, independent of $Z$. Then $\tilde{U}_{Z|\F_t}:= \tilde{F}_{Z|\F_t}(Z, V)$ is, conditional on $\F_t$, a uniform rv, see \cite[Def. 1.2.]{Ruschendorf2013book}. For simplicity, we use the notation $U_{Z|\F_t}:= \tilde{U}_{Z|\F_t}$. 
} We refer the reader to Appendix \ref{app:aux-res}, Definition \ref{def:distortion-Choquet} for further discussion of one-step distortion risk measures and their representation via Choquet integrals.

\begin{definition}[Coherent One-step Distortion Risk Measures]
\label{def:distortion}
\fontfamily{lmss}\selectfont
For each $\tT$, let $\gamma_t: [0,1] \times \Omega \to \R_+$  be a (state dependent) distortion weight function. This means that  $\int_0^1 \gamma_t(u,\omega)\,du=1$ and $\gamma_t (\cdot, \omega)$ is non-decreasing for  all $\omega\in\Omega$, and that the rv $\gamma_t(u,\cdot) : \Omega \to \R_+$ is $\F_{t}$-measurable for every $u\in[0,1]$ and for all $\tT$. Then, the coherent distortion dynamic risk measure with weight functions $\{\gamma_t\}_{\tT}$ is the family $\{\rho_t\}_{\tT}$, where for each $\tT$ and $Z\in \Z_{t+1} $, the one-step risk measure $\rho_t$ is defined as
\begin{equation}\label{eq:distortion-rm}
\rho_t(Z) := \int_0^1 F_{Z|\F_t}^{-1}(u) \;\gamma_t(u)\, du
=
\E\Big[\;Z\,\gamma_t\left(F_{Z|\F_t}(Z)\right) \;\Big|\, \F_t\;\Big]
=
 \E\Big[ \;Z\,\gamma_t\left(U_{Z|\F_t}\right)\; \Big|\, \F_t\Big]\,,
\end{equation}
where, as usual, we suppress the dependence of $\gamma_t$ on its second argument. 
\end{definition}

Allowing for a time- and state-dependent distortion weight function includes, e.g., Expected Shortfall at level $\alpha_t\in\F_t$, $\alpha_t \in [0,1)$,  in which case $\gamma_t(u) = \frac{1}{1-\alpha_t}\Id_{u\ge \alpha_t}$. The level $\alpha_t$ may, e.g., decrease as wealth decreases to express the fact that the investor becomes more risk averse if their wealth drops.

One-step distortion risk measures that are coherent give raise to coherent DRM via representation \eqref{thm:recursive-relation}. Throughout, we  work with coherent dynamic DRM.

\begin{assumption} 
\label{assumption:gamma}
We assume that $\{\rho_t\}_{t\in \T}$ is a coherent distortion DRM with representation \eqref{eq:distortion-rm} that satisfy  $\int_0^1 \gamma_t(u)^2\, du <C$ a.s., for some $C<+\infty$, for all $\tT$. 
\end{assumption}

The integrability condition on $\{\gamma_t\}_{\tT}$ guarantees  by Lemma \ref{lemma:rho-finite}, that any coherent one-step distortion risk measure is a mapping $\rho_t \colon \Z_{t+1} \to \Z_t$, $\tT/\{0\}$ and $\rho_0 \colon \Z_1 \to \R$.

\subsection{Risk-to-go of a Strategy}
\label{sec:rm-self-finacing}
We denote by $\bX=(\bX_t)_{t\in\bT} \, \in \bmZ_{0:T+1}$ the (strictly) positive a.s. $n$-dimensional price process of the universe of assets and consider an investor who invokes a long-only self-financing trading strategy and invests in all assets. We also denote by $\btheta=(\btheta_t)_{t\in\T}\, \in \bmZ_{0:T}$ a (not necessarily self-financing) strategy, where $\btheta_t = (\theta_{t,1}, \ldots, \theta_{t,n})$ is an $n$-dimensional, positive a.s. random vector representing the amount of shares invested in each asset at time $t$. In the sequel, we often use the ``slice notation'' $\btheta_{t_1:t_2}:=(\btheta_{t_1},\btheta_{t_1+1},\dots,\btheta_{t_2})$ for $0\le t_1<t_2\le T$.

A strategy $\btheta$ induces a self-financing strategy $\bvtheta = (\bvtheta_t)_{\tT}$ -- referred to as the \emph{induced self-financing strategy} -- as follows
\begin{equation*}
    \bvtheta_0 := \btheta_0
    \quad \text{and} \quad 
    \bvtheta_t  := \frac{\bvtheta_{t-1}^\intercal \bX_t}{\btheta_t^\intercal \bX_t }\, \btheta_t
    \,,
    \quad \forall \;\tT / \{0\}.
\end{equation*}
Recall that the investor invokes a long-only strategy and invests in all assets, thus $\theta_{t,i} > 0$, a.s., for all $i\in \N:=\{1, \ldots, n\}$ and $t\in\T$, and hence $\bvtheta_{0:T}$ is well-defined. The strategy $\bvtheta$ is self-financing, i.e. it satisfies $(\bvtheta_t - \bvtheta_{t-1})^\intercal\, \bX_t = 0$, for all $\tT / \{0\}$. To simplify notation, we define the weight process $\bw^\btheta=(\bw_t^\btheta)_{\tT}$:
\begin{align*}
    w^\btheta_t: = \frac{\btheta^\intercal_{t} \bX_{t+1}}{\btheta^\intercal_{t+1} \bX_{t+1}}\,,
    \quad \forall \; t\in\T \,,
\end{align*}
and notice that, 
\begin{align*}
\bvtheta_t = \left(\prod_{s = 0}^{t-1} w^\btheta_s \right)\, \btheta_t\,,  \quad \forall \;\tT / \{0\}.
\end{align*}
For each $\tT$, $w^{\btheta}_t$ is $\F_{t+1}$-measurable and if the original strategy $\btheta_{0:T}$ is already self-financing, then $w_t^{\btheta}=1$.

We denote the (negative) price increment by $\bdX_{t}:=-(\bX_{t+1}-\bX_{t})$. As the investor aims to invest in a risk budgeting portfolio, they only consider strategies that are long-only and satisfy the following integrability conditions. 
\begin{definition}
    For any $c\ge0$ define the set of admissible strategies by 
    \begin{align*}
        \A:=
        \bigg\{\btheta_{0:T}\in \bmZ_{0:T} ~&|~ \btheta_{0:T}> c \;\;\;a.s.\;, \quad 
        w_{t}^\btheta \in \L_{t+1}^\infty\,, \; \forall \tT \,,
        \quad \text{and} \quad
        \sum_{\tT}\;  \E\left[\big(\btheta_t^\intercal \bdX_t\big)^2 \right] < + \infty\,
        \bigg\}\,.
    \end{align*}
\end{definition}
For any $c\ge0$, by Lemma \ref{lemma:admissble-combination},
$\A$ is a  a convex set and any induced self-financing strategy $\bvtheta$ belongs to $\Ao$. We further use the notation $\At$, $\tT$, to denote the sliced set of admissible strategies at time $t$, that is $\At:=\{ \btheta_t~|~\btheta_{0:T}\in\A\}$ for each $\tT$. We use the terminology that the strategy $\btheta$ is admissible, if there exists a $c\ge0$ such that $\btheta \in \A$.

The investor assesses the risk of an admissible strategy $\btheta_{0:T}\in\A$ at time $t = 0$ with a coherent distortion DRM $\{\rho_t\}_{t \in \T}$ by
\begin{align*}
   \RM[\btheta_{0:T}]
   &= 
    \rho_0
    \bigg(
        \btheta^\intercal_0\, \Delta\bX_0 + 
            \rho_1
            \bigg(
            w^\btheta_0 \;\btheta^\intercal_1\, \Delta\bX_1 
            + 
            \rho_2\bigg(w^\btheta_0 w^\btheta_1 
            \;\btheta^\intercal_2\, \Delta\bX_2 
            + 
               \cdots
    \\
    &
     \hspace*{6.25em}   \left.\left.\left.
        +\,\rho_{T-1}\left(\prod_{s = 0}^{T-2}w^\btheta_s
               \;
               \btheta^\intercal_{T-1} \Delta\bX_{T-1} 
               +\rho_{T}\left( \prod_{s' = 0}^{T-1}w^\btheta_{s'}
    \;
    \btheta^\intercal_{T} \Delta\bX_{T} \right)\right)
                    \cdots
            \right)\right)
    \right)
    \\
   &= 
    \rho_0
    \bigg(
        \bvtheta^\intercal_0\, \Delta\bX_0 + 
            \rho_1
            \bigg(
            \bvtheta^\intercal_1\, \Delta\bX_1 
            + 
            \cdots +
        \rho_{T-1}\left(\bvtheta^\intercal_{T-1} \Delta\bX_{T-1} 
               +\rho_{T}\left( \bvtheta^\intercal_{T} \Delta\bX_{T} \right)\right)
                    \cdots
            \bigg)\bigg)
    \,.
\end{align*}
Hence, $\RM[\btheta_{0:T}]$ is the dynamic risk of the induced self-financing strategy, but parameterised by $\btheta_{0:T}$.
We can view the risk recursively by defining the risk-to-go process $(\RM_{t}[\btheta_{t:T}])_{t \in \bT}$ via
\begin{subequations}\label{eq:risk-to-go-theta-t}
\begin{align}
\RM_{T+1}
     &:=0
     \quad \text{and}
    \\
     \RM_{t}[\btheta_{t:T}] 
    &:=
    \rho_t\left(
    \btheta_t^\intercal \bdX_t + 
    w^\btheta_t\;\RM_{t+1}[\btheta_{t+1:T}]
    \right)\,,
    \qquad \forall \; \tT\,.
\end{align}
\end{subequations}
At time $t = 0$, it holds that $\RM_{0}[\btheta_{0:T}] = \RM[\btheta_{0:T}] $. By Lemma \ref{lemma:risk-to-go-finite}, any admissible strategy $\btheta \in \A$, $c\ge 0$, (and thus also any induced self-financing strategy) has finite risk  $\RM[\btheta_{0:T}] < + \infty$, and satisfies $\RM_t[\btheta_{t:T}] \in\Z_t$ for all $\tT/\{0\}$.

The risk-to-go evaluated along a self-financing strategy $\bpsi_{0:T}\in\bmZ_{0:T}$ satisfies a slightly simpler recursion
\begin{equation*}
     \RM_{t}[\bpsi_{t:T}]  :=\RM_{t}[\btheta_{t:T}]\big|_{\btheta_{t:T} = \bpsi_{t:T}}
    =
    \rho_t\big(
    \bpsi_t^\intercal \bdX_t + 
    \RM_{t+1}[\bpsi_{t+1:T}]
    \big)\,, 
    \qquad \forall \;  \tT,
\end{equation*}
as $w_t^\bpsi=1$, for all $\tT/\{0\}$, for self-financing strategies.

The next proposition connects the risk-to-go process of $\btheta_{0:T}$ with the risk-to-go process of its induced self-financing strategy $\bvtheta_{0:T}$.
\begin{proposition}\label{prop:RM-vtheta}
 Let $\btheta_{0:T}$ be an admissible strategy and denote by $\bvtheta_{0:T}$ its induced self-financing strategy. Then the following holds
 \begin{align}
 \notag
     \RM_{0}[\bvtheta_{0:T}] 
     &= 
     \RM_{0}[\btheta_{0:T}] 
     \quad \text{and}
    \\
    \label{eq:RM-vtheta}
     \RM_{t}[\bvtheta_{t:T}] 
     &=
     \left(\prod_{s = 0}^{t-1} w_s^\btheta\right)\;
     \RM_{t}[\btheta_{t:T}]\,,
     \quad \forall \;\tT / \{0\}\,.
 \end{align}
\end{proposition}
\proof{Proof:}
The equation for $t = 0$ follows by definition. To show the equalities for $\tT / \{0\}$, we proceed by induction starting backwards in time. At time $T$, we use positive homogeneity of the conditional risk measures, since by admissibility of $\btheta_{0:T}$, we have $0 < w_t^\btheta \in \L_{t+1}^\infty$, for all $\tT$. Thus,
\begin{equation*}
    \RM_{T}[\bvtheta_T]
     = \rho_T(\bvtheta_T^\intercal \bdX_T)
     = 
     \left(\prod_{s = 0}^{T-1} w_s^\btheta\right)
     \rho_T(\btheta_T^\intercal \bdX_T)
     =
     \left(\prod_{s = 0}^{T-1} w_s^\btheta\right)
     \RM_{T}[\btheta_T]
     \,.
\end{equation*}
Next, assume Equation \eqref{eq:RM-vtheta} holds for $t+1$ and note that $w_s^\btheta$ is $\F_t$-measurable for all $0\le s< t$. Then at time $t$, we have
\begin{align*}
    \RM_{t}[\bvtheta_{t:T}] 
    &= 
     \rho_t\left(
    \bvtheta_t^\intercal \bdX_t + 
    \,\RM_{t+1}[\bvtheta_{t+1:T}]
    \right)
    \\
    &= 
     \rho_t\left(
    \left(\prod_{s = 0}^{t-1} w_s^\btheta\right)\btheta_t^\intercal \bdX_t + 
    \left(\prod_{s = 0}^{t} w_s^\btheta\right)
    \,\RM_{t+1}[\btheta_{t+1:T}]
    \right)
    \\
    &=
    \left(\prod_{s = 0}^{t-1} w_s^\btheta\right)
    \rho_t\left(
    \btheta_t^\intercal \bdX_t + 
     w_t^\btheta\,
    \,\RM_{t+1}[\btheta_{t+1:T}]
    \right)
    \\
    &=
    \left(\prod_{s = 0}^{t-1} w_s^\btheta\right)
    \RM_{t}[\btheta_{t:T}] \,,
\end{align*}
where the first equality holds from the induction assumption. The second follows from positive homogeneity of the conditional risk measures and that $0 < \prod_{s = 0}^{t-1} w_s^\btheta \in \L_t^\infty$. The last equality follows from Equation \eqref{eq:risk-to-go-theta-t}.
\hfill \Halmos\endproof

Here, as in the static risk budgeting problem, positive homogeneity of the risk measure plays a central role. Therefore, we next discuss the positive homogeneity of the risk-to-go process. For this, it is convenient to split the arguments of $\RM_t$ into two parts, specifically we write $\RM_t[\btheta_{t:T}]=\RM_t[(\btheta_t,\btheta_{t+1:T})]$ to emphasise the difference of the position at $t$, $\btheta_t$, and the remaining ones, $\btheta_{t+1:T}$.

\begin{proposition}[Positive homogeneity of Risk-to-go Process]\label{prop:homo-risk-to-go}
Let $\btheta_{0:T}$ be an admissible strategy. Then, the risk-to-go process is positive homogeneous viewed as a function of $\btheta_t$ and also viewed as a function of $\btheta_{t:T}$, that is for all $\tT$ and for all  $a_t\in \L_t^\infty$, $a_t > 0$,
\begin{align}
\label{eqn:homogeneity-risk-to-go}
    a_t \,\RM_{t}\left[ \btheta_{t:T} \right]
    =
    \RM_{t}\left[( a_t \,\btheta_t, \btheta_{t+1:T}) \right]
    =
     \RM_{t}\left[a_t \,(\btheta_t, \btheta_{t+1:T}) \right]\,.
\end{align}
\end{proposition}
\proof{Proof:}
The first equality, i.e. positive homogeneity of $\RM_t[\btheta_{t:T} ]$ in $\btheta_t$, follows from representation \eqref{eq:risk-to-go-theta-t}, linearity of $w^\btheta_t$ in $\btheta_t$, noting that $\RM_{t+1}[\btheta_{t+1:T} ]$ does not depend on  $\btheta_t$, and from $\rho_t(\cdot)$ being positive homogeneous. 

To see the second equality, we proceed by induction. First, $\RM_T[\btheta_T] = \rho_T(\btheta_T^\intercal \bdX_T)$ is positive homogeneous in $\btheta_T$. Next, as $w_t^{\btheta}$ is invariant under scaling of both $\btheta_{t}$ and $\btheta_{t+1}$, i.e. $w_t^\btheta = \frac{ \btheta_t^\intercal \bX_{t+1}}{\btheta_{t+1}^\intercal \bX_{t+1}} = 
\frac{ a_t\,\btheta_t^\intercal \bX_{t+1}}{a_t\,\btheta_{t+1}^\intercal \bX_{t+1}}=
w_t^{\btheta'}$, $a_t\in \L_t^\infty$, $a_t > 0$, where $\btheta'_{0:T}$ is s.t. $\btheta_t'=a_t\btheta_t$, $\btheta_{t+1}'=a_t\btheta_{t+1}$, with all remaining $\btheta_s'=\btheta_s$ for $s\notin\{t,t+1\}$. Moreover, $\btheta'_{0:T}$ is admissible. 
Now, assume the second equality in \eqref{eqn:homogeneity-risk-to-go} holds for $t+1$, then we have
\begin{align*}
    \RM_{t}\left[a_t (\,\btheta_t, \btheta_{t+1:T} )\right]
    &= 
    \rho_t\left(
    a_t \,\btheta_t^\intercal \bdX_t + 
    w^{\btheta'}_t\, \,\RM_{t+1}[a_t\; (\btheta_{t+1:T})]
    \right)
    \\
    &= 
     \rho_t\left(
    a_t \,\btheta_t^\intercal \bdX_t + 
    w^\btheta_t\,a_t \,\RM_{t+1}[\btheta_{t+1:T}]
    \right)
    =
    a_t \,\RM_{t}\left[\btheta_t; \btheta_{t+1:T} \right]\,,
    \end{align*}
where the first equality follows from \eqref{eq:risk-to-go-theta-t}, the second equality follows from the inductive assumption and that $w_t^{\btheta}=w_t^{\btheta'}$, and the last equality follows by positive homogeneity of $\rho_t(\cdot)$.
\hfill \Halmos\endproof

We next discuss how to allocate the risk-to-go onto its components at each point in time.

\section{Dynamic Risk Contributions} \label{sec:risk-contribution}
The literature on risk contribution -- also called capital (cost) allocation -- in the static setting is extensive. Approaches ranging from performance measurement \citep{Tasche1999Report}, cooperative game theory including Aumann-Shapley allocation (see e.g., \cite{Mirman1982MOR} and \cite{Billera1982MOR} for early works on cost allocation and \cite{Denault2001JR} in a risk management setting) and allocation in the fuzzy core (\cite{Tsanakas2003IME}), as well as \Gat derivatives and Euler allocations \citep{Kalkbrener2005MF}. Using an axiomatic approach, \cite{Kalkbrener2005MF} showed that for any positive homogeneous and sub-additive static risk measure, the only linear and diversifying
capital allocation rule is the \Gat derivative.

Here, we proceed inline with \cite{Kalkbrener2005MF} by defining risk contributions as a sub-differential, specifically through the \Gat derivative. We note that in case of coherent risk measures, the allocation defined via the \Gat derivatives is the same as the Aumann-Shapley allocation \citep{Tsanakas2009IME}.
An advantage of defining risk contributions through \Gat derivatives of a distortion risk measure is that they satisfy \emph{full allocation}, the property that the sum of the risk contributions adds up to the total risk. In the sequel, we show that our dynamic risk contributions also satisfies a dynamic version of the full allocation property. When defining dynamic risk budgeting strategies, the full allocation property is imperative as it allows to allocate the entire risk to its components. 

As we work in a dynamic setting, at each time $\tT$, the investor faces the future risk of the induced self-financing strategy  and aims to allocate the risk-to-go to each asset $i$. Thus for each $\tT$, we define the risk contribution of asset $i$ as the one-sided \Gat derivative of the risk-to-go processes $\RM_t[\btheta_{t:T}]$ in direction $\theta_{t,i}$. This allows to measure the degree to which the risk-to-go is impacted by the investor's position in the $i^{\text{th}}$ asset at time $t$. Moreover, as we show in Corollary \ref{thm:euler-theta}, this approach allows for full allocation. 

We start by recalling the definition of one-sided \Gat derivative of a functional.

\begin{definition}\label{def:rc}
\fontfamily{lmss}\selectfont
For a functional $F_t:\bmZ_{t:T}\to \Z_t$, $t\in \T$, we denote by $\D_{i}^{\dZ} \,F_t$ its one-sided \Gat derivative of $F_t$ to the $i^\text{th}$ component in direction $\dZ \in\Z_t$. That is, for $\tT$, and $\bZ_{t:T}\in\bmZ_{t:T}$
\begin{equation*}
    \D_{i}^{\dZ}\, F_t[\bZ_{t:T}] := 
    \lim\limits_{\ep\downarrow 0} \;\frac{1}{\ep}\Big(F_t[\bZ_{t:T} + \ep\,\1_{t,i} \dZ] - F_t[\bZ_{t:T}]\Big)\,,
\end{equation*}
where $(\1_{t,i})_{\tT}$ is the stochastic process taking value $1$ in component $i$ at time $t$, and $0$ otherwise.
\end{definition}

We consider the one-sided \Gat derivative as we work with long-only portfolios. Thus, when taking directional derivatives with respect to $\theta_{t,i}$, its perturbed values should also  be positive, which is guaranteed with non-negative $\ep$ in Definition \ref{def:rc}.

\begin{definition}
\fontfamily{lmss}\selectfont
For each $\tT$, we define the risk contribution of the risk-to-go to the $i^\text{th}$ investment as
\begin{equation*}
    \RC_{t,i}[\btheta_{t:T}]
    := \D_{i}^{\theta_{t,i}}\,\RM_t[\btheta_{t:T}]\,.
\end{equation*}    
\end{definition}
Note that the risk contributions $\RC_{t,i}[\btheta_{t:T}]$ are $\F_t$-measurable rvs.

The next result shows that also the risk contribution are positive homogeneous.

\begin{proposition}[Positive homogeneity of Risk Contributions]\label{prop:homo--RC}
The risk contributions of an admissible strategy $\btheta_{0:T}$ to the $i^{\text{th}}$ investment at time $\tT$ are positive homogeneous in the following way. For all $\tT$ and for all $a_t\in \L^\infty_t$, $a_t >0$, we have that
 \begin{equation}
 \label{eqn:homogeneity-rc}
    a_t\, \RC_{t,i}\left[\btheta_{t:T}\right]
    =
    \RC_{t,i}\left[\left(a_t \,\btheta_t,\,\btheta_{t+1:T}\right)\right]
    =
    \RC_{t,i}\left[\left(a_t \,\btheta_{t:T}\right)\right]
    \,.
\end{equation}
 \end{proposition}
\proof{Proof:}
This follows as $\RC_{t,i}$ are the \Gat derivatives of a positive homogeneous function. For completeness we provide a short proof. 
For any $\tT$ and $i \in\N$, we obtain from positive homogeneity of $\RM_t[\btheta_{t:T}]$ in $\btheta_t$, see Proposition \ref{prop:homo-risk-to-go},
\begin{align*}
     \RC_{t,i}\left[\left(a_t \,\btheta_t,\,\btheta_{t+1:T}\right)\right]
     &= 
    \lim\limits_{\ep\downarrow 0} \frac{1}{\ep}\big(\RM_t\left[\left(a_t\left(\btheta_t + \ep\,\be_{i}\, \theta_{t,i}\right),\, \btheta_{t+1:T}\right)\right] - \RM_t[(a_t\,\btheta_t, \btheta_{t+1:T})]\big)
     \\
     &= 
    \lim\limits_{\ep\downarrow 0}\; a_t\, \frac{1}{\ep}\big(\RM_t\left[\left(\left(\btheta_t + \ep\,\be_{i}\, \theta_{t,i}\right),\, \btheta_{t+1:T}\right)\right] - \RM_t[(\btheta_t, \btheta_{t+1:T})]\big)
     \\     
     &=
    a_t\,\RC_{t,i}[ \btheta_{t:T}]\,,
\end{align*}
where $\be_i$ is the unit vector having value $1$ at position $i$ (and $0$ otherwise). The second equality in \eqref{eqn:homogeneity-rc} follows via similar arguments using positive homogeneity of $\RM_t[\btheta_{t:T}]$ in $\btheta_{t:T}$, see Proposition \ref{prop:homo-risk-to-go}.

\hfill\Halmos\endproof

By positive homogeneity of the risk-to-go process, we obtain for all $\tT$ an Euler-like theorem, as stated in the next corollary, which guarantees full allocation.

\begin{corollary}[Full Allocation]\label{thm:euler-theta}
Let $\{\rho_t\}_{t\in\T}$ be a coherent distortion DRM. Then it holds for any admissible strategy $\btheta_{0:T}$ that
\begin{equation*}
    \RM_t[\,\btheta_{t:T}]
    =
    \sum_{i\in \N }
    \RC_{t,i}[\btheta_{t:T}] \,, 
    \quad \text{a.s.}\,,
    \quad \forall\; \tT\,.
\end{equation*}
\end{corollary}

\proof{Proof:}
This proof follows similar steps as the classical Euler allocation. By positive homogeneity of the risk-to-go, Proposition \ref{prop:homo-risk-to-go}, we have for all real numbers $a>0$
\begin{equation}\label{pf:Euler-1}
    a \,\RM_{t}\left[ \btheta_{t:T} \right]
    =
    \RM_{t}\left[( a \,\btheta_t, \btheta_{t+1:T}) \right]\,, \qquad \tT\,.
\end{equation}
Next, viewing Equation \eqref{pf:Euler-1} as a function of $a$, we take a one-sided partial derivative with respect to $a$ of both sides of the above equation, that is
\begin{subequations}
\begin{align*}
    \RM_{t}\left[ \btheta_{t:T} \right]
    \label{pf:eq:Euler-lhs}
    &=
    \lim_{\ep \downarrow 0} \frac1\ep \Big(\RM_{t}\big[\big( (a+\ep)  \,\btheta_t, \btheta_{t+1:T}\big) \big] - \RM_{t}\big[( a \,\btheta_t, \btheta_{t+1:T}) \big]\Big)
    \\
    &=
    \sum_{i\in\N} \D_{i}^{\theta_{t,i}}\,\RM_{t}\big[( a \,\btheta_t, \btheta_{t+1:T}) \big]
    \,.
\end{align*}
\end{subequations}
Evaluating the above at $a = 1$ concludes the proof.
\hfill \Halmos\endproof

In \cite{tsanakas2004IME} the author derives, in the static setting, a closed form formula for the risk contributions of distortion risk measures. Our next result extends this to the dynamic setting.

\begin{theorem}[Risk Contributions]\label{thm:\RC-theta}
Let $\{\rho_t\}_{t\in\T}$ be a coherent distortion DRM with weight functions $\{\gamma_t\}_{\tT}$. Then, the risk contribution of an admissible strategy $\btheta_{0:T}$ to the $i^\text{th}$-investment at time $t\in\T$ is given by
\begin{equation*}
    \RC_{t,i}[\btheta_{t:T}]
    =
   \E\left[
   \theta_{t,i} \left(\dX_{t,i} +
   \frac{X_{t+1,i}}{\btheta_{t+1}^\intercal \bX_{t+1}}
   \RM_{t+1}[\btheta_{t+1:T}]\right)
    \gamma_t \Big(U_{t}[\btheta_{t:T}]\Big)
    ~\Big|~\F_t\right]\,,
\end{equation*}
where $U_{t}[\btheta_{t:T}]$ is a uniform rv comonotonic to $\btheta_t^\intercal \bdX_t + w^\btheta_t \,\RM_{t+1}[\btheta_{t+1:T}]$. Furthermore, $\E\big[\,|\RC_{t,i}[\btheta_{t:T}]|\,\big]<+\infty$ for all $\tT$ and $i\in\N$.
\end{theorem}

\proof{Proof}
Using Prop. 1 \cite{Tsanakas2016RA} and Prop. 3.2 in \cite{Pesenti2021RA}, we have that for $\bY, \bY^\prime \in \bmZ_{t+1}$, a differentiable function $h\colon \R^d \to \R$, and a conditional distortion risk measure $\rho_t$, it holds that
\begin{equation}\label{eq:proof-derivative-rm}
    \lim_{\ep \downarrow 0}\frac{\rho_t\left(h(\bY^\prime + \ep\, \be_i \,\bY)\right) -\rho_t(h(\bY^\prime))}{\ep}
    =
    \E\left[\,Y_i \,\tfrac{\partial}{\partial y_i} h(\bY^\prime)\, \gamma_t \left(U_{h(\bY^\prime)|\F_t}\right)| \F_t\,\right]\,,
\end{equation}
where $U_{h(\bY^\prime)|\F_t}$ is a uniform rv that is comonotonic to the rv $h(\bY^\prime)$ conditional on the information $\F_t$, see also Equation \eqref{eq:distortion-rm}. In Appendix \ref{app:aux-res}, Proposition \ref{aux:prop-assumption}, we provide an alternative proof of Equation \eqref{eq:proof-derivative-rm}.

Next,  note that the risk contributions are
\begin{align*}
     \RC_{t,i}[\btheta_{t:T}]
     &= 
    \D_{i}^{\theta_{t,i}}\; \RM_t[\btheta_{t:T}]
     \\
     &= 
    \D_{i}^{\theta_{t,i}}\,    \rho_t\left(
    \btheta_t^\intercal \bdX_t + 
    w^\btheta_t\;\RM_{t+1}[\btheta_{t+1:T}]
    \right)
    \\
    &=
    \D_{i}^{\theta_{t,i}}\,    \rho_t\left(
    \sum_{i\in\N}
    \theta_{t,i}\left\{ 
    \dX_{t,i} + 
    \frac{ X_{t+1,i}}{\btheta^\intercal_{t+1} \bX_{t+1}}\;\RM_{t+1}[\btheta_{t+1:T}]
    \right\}
    \right)\,.
\end{align*}
Applying Equation \eqref{eq:proof-derivative-rm} and noting that $\RM_{t+1}[\btheta_{t+1:T}]$ is a function of $\btheta_{t+1:T}$ only, and not a function of $\theta_{t,1}$, concludes the first part of the statement. The last part follows from Lemma \ref{lemma:RC-in-L1}.
\hfill \Halmos\endproof

The next representation of the risk contributions illustrates that a decision at time $t$, via $\btheta_t$, cascades through time and impacts all later decision points. This is because restricting to self-financing strategies implies that the investor's future wealth, and thus also possible investment decisions, depend on the current choice of $\btheta_t$.  The proof is delegated to Appendix \ref{app:impact-of-decision}.

\begin{proposition}[Impact of a Decision]\label{prop:impact-of-decision}
Let $\{\rho_t\}_{t\in\T}$ be a coherent distortion DRM with weight functions $\{\gamma_t\}_{\tT}$. Then, the risk contribution of an admissible strategy $\btheta_{0:T}$ to the $i^\text{th}$-investment at time $\tT$ may be written as
\begin{align*}
    \RC_{t,i}[\btheta_{t:T}]
    &=
   \E\left[
   \theta_{t,i} \,\dX_{t,i} \;\Gamma_t^\btheta    ~\Big|~\F_t\right]
   \\
   &
   \quad    +
   \E\left[
   \frac{\theta_{t,i} \,X_{t+1,i}}{\btheta_{t+1}^\intercal \bX_{t+1}}\;
    (\btheta_{t+1}^\intercal \bdX_{t+1}) \;
    \Gamma^\btheta_t\; \Gamma^\btheta_{t+1}
    ~\Big|~\F_t\right]
    \\
   &
   \quad    +
   \E\left[
   \frac{\theta_{t,i} \,X_{t+1,i}}{\btheta_{t+1}^\intercal \bX_{t+1}}\;
    w_{t+1}^\btheta\; (\btheta_{t+2}^\intercal \bdX_{t+2}) 
    \;\Gamma^\btheta_t\; \Gamma_{t+1}^\btheta\; \Gamma_{t+2}^\btheta
    ~\Big|~\F_t\right]
    \\
   &
   \quad    + \cdots 
   +
   \E\left[
   \frac{\theta_{t,i} \,X_{t+1,i}}{\btheta_{t+1}^\intercal \bX_{t+1}}\;
    w_{t+1}^\btheta\; \cdots w_{T-1}^\btheta\; (\btheta_{T}^\intercal \bdX_{T} )
    \;\Gamma_t^\btheta\; \cdots \Gamma_{T}^\btheta
    ~\Big|~\F_t\right]
\end{align*}
where $\Gamma_s^\btheta : = \gamma_s \big(U_s[\btheta_{s:T}]\big)$, with $U_s[\btheta_{s:T}]$ as defined in Theorem \ref{thm:\RC-theta}, for all $s \in \T$.
\end{proposition}

From the above proposition, we can interpret the first expectation as the time $t$ impact of $\theta_{t,i}$, the second expectation as the effect the choice $\theta_{t,i}$ has at time $t+1$, and so on. 

Note that the risk contribution of investment-$i$ at time $t$ of a induced self-financing strategy $\bvtheta$ is 
\begin{equation*}
     \RC_{t,i}[\bvtheta_{t:T}]
     := 
      D_{i}^{\theta_{t,i}}\;
     \RM_{t}[\btheta_{t:T}] \Big|_{\btheta = \bvtheta}\,.
\end{equation*}
The next statement relates the risk contributions of a strategy with those of its induced self-financing strategy.

\begin{proposition}\label{prop:\RC-vtheta}
Let $\{\rho_t\}_{t\in\T}$ be a coherent distortion DRM with weight functions $\{\gamma_t\}_{\tT}$. Let $\btheta_{0:T}$ be admissible and $\bvtheta_{0:T}$ its induced self-financing strategy. Then the following holds:
\begin{align*}
    \RC_{0,i}[ \bvtheta_{0:T}]
    &=
    \RC_{0,i}[ \btheta_{0:T}]
    \quad \text{and}
    \\
    \RC_{t,i}[\bvtheta_{t:T}]
    &=
    \left( \prod_{s = 0}^{t-1} w_s^\btheta\right)
      \RC_{t,i}[\btheta_{t:T}] \,,
     \qquad \forall \tT/\{0\}\,.
\end{align*}
\end{proposition}

\proof{Proof: }
Case $t = 0$ follow immediately from Proposition \ref{prop:RM-vtheta}. For $\tT / \{0\}$, define the scalar rv $c_t^\btheta := \prod_{s = 0}^{t-1} w_s^\btheta\; \in \L_t^\infty$ and recall that $\bvtheta_t = c_t^\btheta\,\btheta_t$. 
Then, we have that
\begin{align*}
      \RC_{t,i}[ \bvtheta_{t:T}]
      &= 
      \D_{i}^{\theta_{t,i}}\; \RM_t[\btheta_{t:T}]\Big|_{\btheta = \bvtheta}
      \\
    \text{\tiny(by Theorem \ref{thm:\RC-theta})\;} &= 
   \E\left[
   \vartheta_{t,i}  \left(\dX_{t,i} +
   \frac{X_{t+1,i}}{\bvtheta_{t+1}^\intercal \bX_{t+1}}\RM_{t+1}[\bvtheta_{t+1:T}]\right)
    \gamma_t \big(U_t[\bvtheta_{t:T}]\big)
    ~\Big|~\F_t\right]
      \\
    \text{\tiny(by Proposition \ref{prop:RM-vtheta})\;}
    &= 
    \E\left[
    \vartheta_{t,i}  \left(\dX_{t,i} +
    \frac{X_{t+1,i}}{\bvtheta_{t+1}^\intercal \bX_{t+1}}\;c_{t+1}^\btheta\,\RM_{t+1}[\btheta_{t+1:T}]\right)
    \gamma_t \big(U_t[\bvtheta_{t:T}]\big)
    ~\Big|~\F_t\right]
    \\
    &= \,
    c_t^{\btheta}\;\E\left[
    \theta_{t,i}  \left(\dX_{t,i} +
    \frac{X_{t+1,i}}{\btheta_{t+1}^\intercal \bX_{t+1}}\;\RM_{t+1}[\btheta_{t+1:T}]\right)
    \gamma_t \big(U_t[\bvtheta_{t:T}]\big)
    ~\Big|~\F_t\right] \,.
    \numberthis    
    \label{eqn:\RC-intermediate}
\end{align*}

Next, $U_t[\bvtheta_{t:T}]$ may be simplified by noting
\begin{align*}
\bvtheta^\intercal_t\bdX_t + w^\bvtheta_t\,\RM_{t+1}[\bvtheta_{t+1:T}]  
&=\bvtheta^\intercal_t\bdX_t + \RM_{t+1}[\bvtheta_{t+1:T}]  
\qquad \text{\tiny(as $\vartheta$ is self-financing)\;}
\\
\text{\tiny(by Proposition \ref{prop:RM-vtheta})\;}
&=\bvtheta^\intercal_t\bdX_t + c_{t+1}^\btheta\,\RM_{t+1}[\btheta_{t+1:T}]  
\\
\text{\tiny(as $\bvtheta_t=c_t^\btheta\,\btheta_t$)\;}
&=c_t^\btheta\,\btheta^\intercal_t\bdX_t + c_{t+1}^\btheta\,\RM_{t+1}[\btheta_{t+1:T}]  
\\
\text{\tiny(as $c_{t+1}^\btheta=c_t^\btheta\,w_t^\btheta$)\;}
&=c_t^\btheta\left(\btheta^\intercal_t\bdX_t + w_{t}^\btheta\;\RM_{t+1}[\btheta_{t+1:T}]  \right).
\end{align*}
Furthermore, as $c_t^\btheta >0$ a.s., the bivariate vector
\begin{equation*}
   \Big( c_t \left(\btheta_t^\intercal \bdX_t +
    w_t^\btheta\RM_{t+1}[\btheta_{t+1:T}]\right)
    ,\;
    \btheta_t^\intercal \bdX_t +
    w_t^\btheta\RM_{t+1}[\btheta_{t+1:T}]
    \Big)
    \quad \text{is comonotonic}
\end{equation*}
and therefore $U_t[\bvtheta_{t:T}] = U_t[\btheta_{t:T}]$ a.s., that is both $\btheta_{t:T}$ and $\bvtheta_{t:T}$ generate the same comonotonic uniform rv.

Finally, applying this result and continuing from \eqref{eqn:\RC-intermediate}, we have
\begin{align*}
      \RC_{t,i}[\bvtheta] &= 
      \,
   c_t^{\btheta}\;\E\left[
   \theta_{t,i}  \left(\dX_{t,i} +
   \frac{X_{t+1,i}}{\btheta_{t+1}^\intercal \bX_{t+1}}
   \RM_{t+1}[\btheta_{t+1:T}]\right)
    \gamma_t \big(
    U_{t}[\btheta_{t:T}]\big)
    ~\Big|~\F_t\right] = 
    c_t^{\btheta}\;\RC_{t,i}[\btheta_{t:T}]\,,
\end{align*}  
as required.
\hfill \Halmos \endproof

Consequently, the risk contributions of the induced self-financing strategy are by definition positive homogeneous, as stated in Proposition \ref{prop:homo--RC}, and satisfy the full allocation.

\section{Dynamic Risk Budgeting Portfolios}\label{sec:dynamic-risk-budgeting}

Using the dynamic risk contributions defined in the last section, we now define a dynamic risk budgeting strategy.

\begin{definition}\label{def:risk-parity}
\fontfamily{lmss}\selectfont
Let $\{\rho_t\}_{t \in \T}$ be a DRM. A strategy $\btheta_{0:T} \in\A$, $c\ge 0$,  is called a \emph{dynamic risk budgeting strategy} with budget $B = (b_{t,i})_{t\in\T, i\in\N}$ satisfying 
$b_{t,i}>0$ and $\sum_{i\in\N}b_{t,i} = 1$, for all $\tT$, if
\begin{equation}\label{eq:risk-budgeting}
    \RC_{t,i}[\btheta_{t:T}] 
    =
    b_{t,i}\;
    \RM_t[\,\btheta_{t:T}]\,,
    \quad \forall \tT \quad \text{and} \quad i\in\N\,.
\end{equation}
\end{definition}

A dynamic risk budgeting strategy is therefore a strategy such that at each time $t \in \T$ the risk contribution of investment $i \in \N$ is equal to $b_{t,i}\, \%$ of the risk-to-go at time $t$. For example, if the risk budget is $b_{t,i} =\frac{1}{n}$ for all $i\in\N$ and $t\in\T$, then we call the risk budgeting strategy \emph{risk parity}, which means equal risk contributions, since it satisfies
\begin{equation*}
    \RC_{t,i}[\btheta_{t:T}]  = \RC_{t,j}[\btheta_{t:T}] \,,
    \quad \forall i,j \,\in \N  \text{ and } 
    \forall \, t\in \T\,.
\end{equation*}

\begin{proposition}\label{prop:risk-budgeting-iff}
Let $\{\rho_t\}_{t \in \T}$ be a coherent distortion DRM, $\btheta$ an admissible strategy and $\bvtheta$ its corresponding induced self-financing strategy. If $\btheta$ is a risk budgeting strategy with risk budget $B$, then $\bvtheta$ is a risk budgeting strategy with risk budget $B$.
\end{proposition}

\proof{Proof:}
The case when $t=0$ is trivial since $\RM_0[\btheta_{0:T}] = \RM_0[\bvtheta_{0:T}]$ and $RC_{0,i}[\btheta_{0:T}] = RC_{0,i}[\bvtheta_{0:T}]$. Next, let $t>0$ and
assume that $\btheta_{0:T}$ is a risk budgeting strategy. Then, for each $\tT/\{0\}$ and $i \in \N$, it holds that
\begin{align*}
    \RC_{t,i}[\bvtheta_{t:T}] 
    &=
    \left(\prod_{s = 0}^{t-1} w_s^\btheta\right)
      \RC_{t,i}[\btheta_{t:T}]
     =
    \left(\prod_{s = 0}^{t-1} w_s^\btheta\right)
      b_{t,i} \;\RM_{t}[\btheta_{t:T}]
    =
      b_{t,i} \left(\prod_{s = 0}^{t-1} w_s^\btheta\right)\RM_{t}[\btheta_{t:T}]
     = 
    b_{t,i}\;
    \RM_t[\,\bvtheta_{t:T}]
    \,,
\end{align*}
where we used Proposition \ref{prop:\RC-vtheta} in the first equation, then the fact that $\btheta_{0:T}$ is a risk budgeting strategy, and finally Proposition \ref{prop:RM-vtheta}.
\hfill \Halmos\endproof

The next result pertains to the characterisation of self-financing risk budgeting strategies as a unique solution of a series of strictly convex and recursive (backward in time) optimisation problems. Moreover, we show under technical conditions that if a self-financing risk budgeting strategy with initial wealth of $1$ exists, then it is given by a rescaled version of the solution to the series of convex optimisation problems. 

\begin{theorem}\label{thm:opt}
Let $\{\rho_t\}_{t\in\T}$ be a coherent distortion DRM with weight functions $\{\gamma_t\}_{\tT}$ and $B = (b_{t,i})_{t\in\T, i\in\N}$ a risk budget. For $c>0$ consider the recursive optimisation problems 
\begin{align}
      \btheta_t^* :&= 
      \argmin_{\btheta_t\in \At} \E\left[
      \RM_t[(\btheta_t, \btheta_{t+1:T}^*)]
    -
    \sum_{i\in\N} \,b_{t,i}\,
      \log \theta_{t,i} \,
      \right]\,, \qquad 
       \forall \tT\,.
    \tag{\text{$P$}}
      \label{eqn:P-prime}
\end{align}
There exists a unique solution to \eqref{eqn:P-prime}. Furthermore, if $\btheta^*\in\mathcal{A}_{\tilde{c}}$, for some $\tilde{c}>0$, then it satisfies $\RM_t[\btheta^*_{t:T}] = 1$ a.s., for all $\tT$, and moreover
\begin{enumerate}

    \item[(a)] \label{thm:b-self-financing}
    the self-financing strategy $\bvtheta^*_{0:T}$, induced by $\btheta^*_{0:T}$, is in $\Ao$ and is a self-financing risk budgeting strategy with risk budget $B$;

    \item[(b)] \label{thm:c-normalised}
    the normalised strategy $\bvtheta^{\dagger}_{0:T}:=\frac{1}{\bvtheta_0^{*\intercal}\bX_0}\bvtheta^*_{0:T}$ is in $\Ao$ and is a self-financing risk budgeting strategy with the same risk  budget $B$ and initial wealth $1$.
\end{enumerate}

\end{theorem}

\proof{Proof:} 
As the risk-to-go process is convex, recall that the one-step risk measures are convex, and the $(-\log)$ is strictly convex, the objective functional is strictly convex. Moreover, the set of admissible strategies is convex, thus a unique solution to \eqref{eqn:P-prime} exists. The proof that the risk-to-go process is a.s. equal to 1 follows from the proof of part (a).

\underline{Part (a):} Let the unique solution to \eqref{eqn:P-prime} be denoted by $\btheta^*$ and for simplicity assume that $\btheta^*\in\A$. For
$\tT$, consider the objective function
(where for ease of notation we suppress the dependence of $L_t$ on $\btheta_{t+1:T}^*$)
\begin{align}\label{eq:L-opt}
    L_t[\btheta_t]
    :=
    \E\left[\RM_t\big[(\btheta_t, \btheta_{t+1:T}^*)\big]
    -
    \sum_{i\in\N} \,b_{t,i}\,
      \log \theta_{t,i} \,\right]\, \quad \text{for} \quad \btheta_t \in \At\,. 
\end{align}
Next, take any $\btheta_t'\in\At$. Then, by Lemma \ref{lemma:admissble-combination} it holds that for all $\ep\in[0,1]$, the strategy $\btheta_t + \ep \,  \big(\btheta_t' - \btheta_t\big) = (1-\ep  )\,\btheta_t + \ep \, \btheta_t' \in \At$. By Proposition \ref{aux:lemma:interchange}, we can interchange limit and expectation, thus the one-sided \Gat derivative of $ L_t[\btheta_t]$ in direction $\dbtheta:=(\btheta_{t}' - \btheta_{t})$ is
\begin{align}
&\hspace*{-3em}
\lim_{\ep \downarrow 0}
\frac{1}{\ep} \Big(L_t[\btheta_t + \ep\, \dbtheta] - L_t[\btheta_t ]\Big)
    \nonumber
    \\
    &= 
    \sum_{i\in\N}\;\E\left[
    \D_{i}^{{\dtheta_i}}\,\RM_t[(\btheta_t, \btheta_{t+1:T}^*)]
    - 
    b_{t,i} \;\frac{{\dtheta_i}}{ \theta_{t,i}}\,\right]\,
    \nonumber
    \\
    &= 
    \sum_{i\in\N}\;{\E\left[ \left.\E\left[
    \D_{i}^{{\dtheta_i}}\,\RM_t[(\btheta_t, \btheta_{t+1:T}^*)]
    - 
    b_{t,i} \;\frac{{\dtheta_i}}{ \theta_{t,i}}~\right|~\F_t\right]\, \right]\,}
    \nonumber
    \\
    &= \;
    \sum_{i\in\N}\; \E\left[\dtheta_i\;\;
    \E\left[
    \,\left(\dX_{t,i} +
   \frac{X_{t+1,i}}{\btheta_{t+1}^{*\, \intercal} \bX_{t+1}}
   \RM_{t+1}\left[\btheta_{t+1:T}^*\right]\right)
    \gamma_t \big(U_{t}[(\btheta_t,\btheta^*_{t+1:T})]\big)
    - \frac{b_{t,i}}{ \theta_{t,i}}
    ~\Big|~\F_t\right]   
    \;\right]\,.
    \label{eqn:Gateaux-Lagrangian}
\end{align}
As $\At$ is a convex set, the unique optima $\btheta_t^*$ is attained where the one-sided \Gat derivative is non-negative for all $\btheta_t'\in\At$, see e.g., Thm. 23.2 in \cite{Rockafellar2015book}, i.e.,
\begin{equation*}
\lim_{\ep \downarrow 0}
\frac{1}{\ep} \Big(L_t[\btheta_t^* + \ep\, (\btheta'_t-\btheta_t^*)] - L_t[\btheta_t^* ]\Big) \ge 0    \,, \qquad \forall\; \btheta'_t \in\At\,,
\end{equation*}
which by Equation \eqref{eqn:Gateaux-Lagrangian} is equivalent to 
\begin{equation}\label{eqn:Gateaux-inequality}
    \sum_{i\in\N}\; \E\left[\,(\theta_{t,i}' - \theta_{t,i}^*) \; \frac{\RC_{t,i}[\btheta^*_{t:T}]-b_{t,i}}{\theta_{t,i}^*}\,\right] \ge 0    \,, \qquad \forall\; \btheta_t'\in\At\,.
\end{equation}

We next show that $\btheta^*_t$ fulfils \eqref{eqn:Gateaux-inequality} if, and only if, it satisfies 
\begin{equation}\label{pf:eq:risk-budget}
    \RC_{t,i}[\btheta^*_{t,T}]=b_{t,i}\,, \qquad \forall \; i \in \N\,.
\end{equation}
To prove this, note that if $\btheta_t^*$ satisfies \eqref{pf:eq:risk-budget}, then \eqref{eqn:Gateaux-inequality} holds. 
To show that \eqref{eqn:Gateaux-inequality} implies \eqref{pf:eq:risk-budget}, we proceed by contradiction. Suppose the equality in \eqref{pf:eq:risk-budget} does not hold a.s. for some $i=k$ and let $B^+\in\F$, $B^-\in\F$, $B^0\in\F$ denote the sets on which $(\RC_{t,k}[\btheta^*_{t,T}]-b_{t,k})$ is positive, negative, and zero respectively. By assumption, $\P(B^+)+\P(B^-)>0$. Then, define $\btheta'_t$, s.t. $\theta'_{t,i}=\theta_{t,i}^*$ for all $i\ne k$, and 
\[
\theta'_{t,k}:=2\, \theta^*_{t,k} \Id_{B^-} + \tfrac12(c+\theta^*_{t,k})\, \Id_{B^+} 
+ \theta^*_{t,k} \Id_{B^0},
\]
Note that  $\btheta'_t> c$ a.s., moreover, $\btheta_t'\le 2\,\btheta^*_t$. Hence, we have that $\E[(\btheta'^{\intercal}_t\bX_t)^2]\le
4\E[(\btheta^{*\intercal}_t\bX_t)^2]<+\infty$, and $\frac{\btheta_{t}^{'\intercal}\bX_{t+1}}{\btheta_{t+1}^{*\intercal}\bX_{t+1}}
\le 
\frac{2\btheta_{t}^{*\intercal}\bX_{t+1}}{\btheta_{t+1}^{*\intercal}\bX_{t+1}}=
2\,w^{\btheta^*}_t<+\infty$. Therefore, $\btheta'_t\in\At$.
 However, 
\begin{align*}
&\hspace*{-3em}
\lim_{\ep \downarrow 0}
\frac{1}{\ep} \Big(L_t[\btheta_t^* + \ep\, \big(\btheta_t' - \btheta_t^*\big)] - L_t[\btheta_t^* ]\Big) 
\\
&=
 \E\left[\,
 \Big(\theta_{t,k}^*\,\Id_{B^-}+\tfrac12(c-\theta_{t,k}^* ) \,\Id_{B^+}
 \Big) 
  \frac{(\RC_{t,k}[\btheta^*_{t,T}]-b_{t,k})}{\theta_{t,k}^*}\,\right]
 < 0,
\end{align*}
and therefore $\btheta_t^*$ does not satisfy \eqref{eqn:Gateaux-inequality}, and we arrive at a contradiction.

Thus, $\btheta^*_t$ must satisfy Equation \eqref{pf:eq:risk-budget} and by Corollary \ref{thm:euler-theta}, it holds 
\begin{equation*}
    \RM_t[\btheta^*_{t:T}] 
    =
    \sum_{i\in\N}     \RC_{t,i}[\btheta^*_{t:T}]
    =
    \sum_{i\in\N} b_{t,i}
    =
    1\,,
    \quad \text{which implies that}\quad
    \RC_{t,i}[\btheta^*_{t:T}] = b_{t,i}\;\RM_t[\btheta^*_{t:T}]\,,
\end{equation*}
and $\btheta_{t:T}^*$ satisfies 
the risk budgeting equation \eqref{eq:risk-budgeting} for all $\tT$.

Finally by Proposition \ref{prop:risk-budgeting-iff}, the self-financing strategy $\bvtheta^*_{0:T}$ induced by $\btheta_{0:T}^*$ is a risk budgeting strategy with budget $B$. Moreover, if $\btheta^*_{0:T} \in \A$, we have $\bvtheta^*_{0:T} \in \Ao$.

\underline{Part (b):} 
Define $\bvtheta_{0:T}^\dagger := \frac{1}{\bvtheta_0^{*\, \intercal} \bX_0} \bvtheta^*_{0:T}$, clearly $\bvtheta_{0:T}^\dagger$ has initial wealth of 1, is self-financing, and is an element of $\Ao$, as $\bvtheta_0^{*\, \intercal} \bX_0 >0$, and we claim that it is a risk budgeting strategy with budget $B$. Indeed, we have for all $\tT$ and $i\in\N$
\begin{equation*}
   \RC_{t,i} [\bvtheta_{t:T}^\dagger]
   =
    \frac{\RC_{t,i} [\bvtheta_{t:T}^*]}{\bvtheta_0^{*\, \intercal} \bX_0}\, 
    = 
    \frac{b_{t,i}}{\bvtheta_0^{*\, \intercal} \bX_0}\, \RM_t[\bvtheta_{t:T}^*]
    = 
    b_{t,i}\left( \frac{1}{\bvtheta_0^{*\, \intercal} \bX_0}\RM_t[\bvtheta_{t:T}^*]    \right)
    =
    b_{t,i} \,\RM_t[\bvtheta_{t:T}^\dagger]\,,
\end{equation*}
where we applied positive homogeneity of the risk contributions, see Proposition \ref{prop:homo--RC}, and the fact that $\bvtheta_{0:T}^*$ is a risk-budgeting strategy.
Thus, $\bvtheta_{0:T}^\dagger$ is in $\Ao$ and a self-financing risk budgeting strategy with risk budget $B$ and initial wealth of 1.
\hfill \Halmos\endproof

The above result states that the unique optimiser of \eqref{eqn:P-prime} is a risk budgeting strategy. 
In the next theorem, whose proof is delegated to Appendix \ref{app:proof-uniqueness}, we show that, under technical conditions, any self-financing risk budgeting strategy with initial wealth of $1$ is a rescaled version to the solution of the optimisation problem \eqref{eqn:P-prime}, and in particular given in Theorem \ref{thm:opt}(b).

\begin{theorem}[Uniqueness]\label{thm:uniquness}
Let $\{\rho_t\}_{t\in\T}$ be a coherent distortion DRM. 
Consider an admissible self-financing dynamic risk budgeting strategy $\bvphi_{0:T} \in \A$, for some $c>0$, that has initial wealth 1 and budget $B$. If the corresponding risk-to-go process satisfies $0<c_R\le \RM_t[\bvphi_{t:T}] \le c^R< + \infty$ for all $\tT$, then the risk budgeting strategy is the unique solution to \eqref{eqn:P-prime} with lower bound on the strategy $\frac{c}{c^R}$, and characterised by Theorem \ref{thm:opt}(b).
\end{theorem}

\section{Approximation of Risk Budgeting Strategies}\label{sec:nuermical-implemenatation}

While Theorems \ref{thm:opt} and \ref{thm:uniquness} provide a characterisation of risk budgeting strategies as solutions to a sequence of convex optimisation problems, they do not provide a methodology for finding them. Thus, we develop a deep learning approach that leverages the flexibility of neural networks (NNs) to approximate high dimensional functions (see e.g., \cite{goodfellow2016deep}), together with new techniques that have been developed for optimising convex DRMs (\cite{coache2021reinforcement} and \cite{coache2022conditionally}). The latter works solve portfolio allocation and algorithmic trading problems by  making explicit use of the dual representation of convex risk measures.
In contrast, our proposed approach relies on the analytical results developed for the class of coherent distortion DRM and in particular their risk contributions given in Theorem \ref{thm:\RC-theta}.

To optimise the performance criterion $L_t[\btheta_{t:T}]$ (given in Equation \eqref{eq:L-opt}) of \eqref{eqn:P-prime} we make use of gradient descent methods. The \Gat derivative is related to the risk contributions (see Theorem \ref{thm:\RC-theta}) and requires evaluating the risk-to-go $\RM_t[\btheta_{t:T}]$ and the rv $U_{t}[\btheta_{t:T}]$. Therefore, we develop algorithms for estimating and sampling from  $\RM_t[\btheta_{t:T}]$ and  $U_{t}[\btheta_{t:T}]$, and for estimating the \Gat derivative of $L_t[\btheta_{t:T}]$. 

The general strategy behind our actor-critic algorithm is:
\begin{enumerate}
\item[1.)] Parameterise the strategy $\btheta_{0:T}$ by a NN with parameters $\ba$.

\item[2.)] For estimating $U_t[\btheta_{t:T}]$, we approximate the conditional cdf of $g_t:=\btheta^\intercal_t\Delta\bX_t + w_t^\btheta \RM_{t+1}$ given $\F_t$, i.e. $F_{g_t|\F_t}(z):=\mathbb{P}(g_t\le z | \F_t)$, by a NN with parameters $\mff$.

\item[3.)] For estimating $\RM_t[\btheta_{t:T}]$, we approximate the risk-to-go by a NN with parameters $\rNN$.
\end{enumerate}

To implement points 2) and 3), we use the notion of elicitability -- see Section \ref{subsec:elicitable} --,  which provides a numerically efficient alternative to  nested simulations when calculating conditional risk measures and cdfs \citep{coache2022conditionally}. Specifically, we use Proposition \ref{prop:prob-score} for approximating $F_{g_t|\F_t}(\cdot)$ and Proposition \ref{prop:score} for approximating $\RM_t[\btheta_{t:T}]$.

The parameterised strategies $\btheta_{0:T}$ are the so-called ``actors'', as they are at the investor's discretion. The approximation of the conditional cdf and the risk-to-go are the ``critics'', as they evaluate the effectiveness of a strategy. 
In the following subsections, we provide details on the algorithm and the gradients  for training the NNs (Section \ref{subsec:gradient}), how conditional elicitability is leveraged in the algorithm (Section \ref{subsec:elicitable}), and the specific NN architectures used (Section \ref{subsec:NN-structure}).

\subsection{Algorithms and Gradient Formulas}\label{subsec:gradient}

Here, we explain the general structure of the algorithm. Optimisation of the criterion $L_t[\btheta_{t:T}]$ proceeds in an iterative fashion, where for each iteration we update (i) the risk-to-go $m_r$-times, then (ii) the conditional cdf $m_f$-times, and then (iii) the strategy once. For the risk-to-go we employ a main and a target network (with parameters $\rNN$ and $\rNN'$, respectively), which is known to improve stability and convergence rates of actor-critic methods  \citep{mnih2015human}. The main network of the risk-to-go is updated using stochastic gradient descent, while the target network is updated using soft-updates, i.e., via a convex combination of the current main and target network parameters, $\rNN'\leftarrow (1-\tau)\rNN'+\tau\,\rNN$, with soft-update rate $\tau \in (0,1)$ \citep{fujita2021chainerrl}. The detailed algorithm is provided in Algorithms \ref{algo:drb} and \ref{algo:sim}.

\begin{algorithm}[tbp]
    \SetKwRepeat{Do}{do}{while}
	\caption{Actor-critic algorithm for learning risk budgeting strategies}
	\label{algo:drb}
	\KwIn{
        NN parameters $\ba$, $\rNN$, $\rNN'$, and $\mff$, number of risk-to-go updates $m_r\ge 1$ per iteration, number of conditional cdf  updates $m_f\ge 1$ per iteration, learning rates, soft-update rate $\tau$
    }
    \Do{not converged}
    {
        \For{$i = 1, 2, \ldots, m_r$}{
            get simulations of $\bX_{0:T}$, $\bTheta_{0:T}$, and $(\RM_t)_{\tT}$ using Algorithm \ref{algo:sim}\;
            compute the expected score $S_{\rho}$ \eqref{eqn:score-rho} using the risk-to-go with main parameters $\rNN$\;
            use a gradient step to minimise the score and update the risk-to-go  main parameters parameters $\rNN$\;
            perform a soft-update of the target to the main NN parameters $\rNN'\leftarrow (1-\tau)\, \rNN' + \tau\,\rNN$\;
        }
        \For{$j = 1, 2, \ldots, m_f$}{
            get simulations of $\bX_{0:T}$, $\bTheta_{0:T}$, and $(\RM_t)_{\tT}$ using Algorithm \ref{algo:sim}\;
            compute the expected score $S_{cdf}$ \eqref{eqn:score-cdf} using  conditional cdf parameters $\mff$\;
            use a gradient step to minimise the score and update the cdf parameters $\mff$\;
        }
        get simulations of $\bX_{0:T}$, $\bTheta_{0:T}$, and $(\RM_t)_{\tT}$ using Algorithm \ref{algo:sim}\;
        perform a gradient step to minimise the objective \eqref{eq:L-opt} using \eqref{eqn:grad-L} and update the strategy parameters $\ba$\; 
    }
\end{algorithm}

\begin{algorithm}
    \caption{Simulate asset prices, strategy, and risk-to-go.}
    \label{algo:sim}
    \SetKwInOut{KwOut}{Output}
    \KwIn{
        NN parameters $\ba$ and $\rNN'$\;
    }
    simulate asset prices $\bX_{0:T}$\; 
    use asset prices to generate samples of the strategy $\bTheta_{0:T}$ with policy parameters $\ba$\;
    compute the risk-to-go for all simulations and time points with NN target parameters $\rNN'$\;
    \KwOut{
    asset prices: $\bX_{0:T}$, strategies: $\bTheta_{0:T}$, risk-to-go: $(\RM_t)_{\tT}$\;
    }
\end{algorithm}

Next, we discuss how the NN parameters of the strategy $\btheta_{0:T}$ are trained. Recall that the strategy's parameters are a set of vectors $(\ba_t)_{\tT}$, where for all $\tT$, we have $\ba_t\in\B$, $\B\subset \R^m$. We write, with slight abuse of notation, $\btheta_{t}^{\ba_t}=\btheta_t(X_0,\dots,X_t;\ba_t)$, where $\btheta_t\colon \R^{n \times t} \times \B \to \R$. We call $\ba_{0:T}$ a policy, as it parametrises the investor's strategy. Next, for $\tT$ we view the criterion in Equation \eqref{eq:L-opt} as a function of $\ba_t$, and aim to minimise it over these parameters. To this end, we write the time $t$ loss function as
\begin{equation*}
\begin{split}
    L_t[\ba_t;\, \ba_{t+1:T}^*]
    :=&\,    \E\Bigg[\RM_t\big[(\btheta_t^{\ba_t},\btheta_{t+1}^{\ba_{t+1}^*},\dots,\btheta_T^{\ba_T^*})\big]
    -\sum_{i\in\N} \,b_{t,i}\,
      \log \theta_{t,i}^{\ba_t} \,\Bigg]\,,
      \end{split}
\end{equation*}
where $\ba_{t+1:T}^*$ are the optimal policies from time $t+1$ onwards. 
We adopt for the methodology of deterministic policy gradient and, starting from the current estimate of the parameters, update the parameters using the gradient step rule
\[
\ba_t \;\leftarrow\; \ba_t - \eta\,\nabla_{\ba} L_t[\ba;\ba_{t+1:T}^*]\Big|_{\ba=\ba_t}\,,
\]
where $0<\eta\ll 1$ is a learning rate. Hence, we require an efficient way to estimate the gradient of $L_t$. By the chain rule, (as in the proof of Theorem \ref{thm:\RC-theta}), we have
\begin{align}
    \begin{split}
    \partial_{\beta_k} L_t[\ba; \ba_{t+1:T}^*]
    =\;&
    \E\Bigg[ 
       \partial_{\beta_k}\theta_{t,i}^{\ba} 
       \Bigg\{\left(\dX_{t,i} +
       \frac{X_{t+1,i}}{\big(\btheta_{t+1}^{\ba_{t+1}^*}\big)^\intercal \bX_{t+1}}
       \RM_{t+1}\Big[\big(\btheta_{t+1}^{\ba_{t+1}^*},\dots,\btheta_T^{\ba_T^*}\big)\Big]\right)
       \\
       & \hspace*{5em}
        \times \gamma_t \Big(U_{t}[(\btheta_t^{\ba},\btheta_t^{\ba_{t+1}^*},\dots,\btheta_T^{\ba_T^*})]\Big)
    -
    \frac{b_{t,i}}{\theta_{t,i}^{\ba}}\,\Bigg\}
      \Bigg]\,.         
    \end{split}
    \label{eqn:grad-L}
\end{align}

Given a policy $\ba$, we estimate the expectation in \eqref{eqn:grad-L} using its sample average, which however, requires samples of $\RM_{t+1}\Big[\big(\btheta_t^{\ba_{t+1}^*}\dots,\btheta_T^{\ba_T^*}\big)\Big]$ and $U_{t}[(\btheta_t^{\ba},\btheta_{t+1}^{\ba_{t+1}^*}\dots,\btheta_T^{\ba_T^*})]$. Thus, we next elaborate on how we leverage the notion of conditional elicitability to simulate from these rvs.

\subsection{Conditional Elicitability}\label{subsec:elicitable}
As the risk-to-go $     \RM_{t}[\btheta_{t:T}] 
    =
    \rho_t\left(
    \btheta_t^\intercal \bdX_t + 
    w^\btheta_t\;\RM_{t+1}[\btheta_{t+1:T}]
    \right)$ consists of one-step (conditional) risk measures, both the risk-to-go and the conditional cdf could be estimated via nested simulations, which, however is numerically expensive. To overcome this, we use the notion of elicitable functionals which circumvents the need for nested simulations. The key concept is that a functional is elicitable, e.g., mean, $\VaR_\alpha$, $\ES_\alpha$, if it is the minimiser of the expected value of a scoring function. Thus, an elicitable functional may be estimated by solving a convex optimisation problem. Furthermore, conditional elicitable functionals can be estimated by minimising the expected score  over arbitrary functions of the conditioning variables (see e.g., Proposition \ref{prop:score-argmin}). For completeness, in Appendix \ref{sec:appendix-elicitability} we collect known results on elicitability and conditional elicitability that are relevant to the exposition, as well as new results pertaining to the specific risk measures considered here.

A large class of one-step distortion risk measures are elicitable -- in particular those with piecewise constant weight function -- see \cite{Fissler2016AS} for the static and \cite{coache2022conditionally} for the dynamic setting. In the numerical examples we consider a subclass of coherent distortion DRM given by the weighted average of the ES and expected value, but our approach can be generalised to other elicitable coherent distortion DRMs. 
Specifically, we consider the family of coherent distortion DRMs $\{\rho_t\}_{\tT}$ parametrised by $p\in[0,1]$
\begin{equation}\label{eq:rho-example}
    \rho_t (Z)
    := 
    p\;\ES_\alpha(Z\, |\,\F_t)  + (1-p) \; \E[\,  Z \,|\, \F_t] \,, \quad  Z \in \Z_{t+1}
\end{equation}
which for each $\tT$ is a one-step distortion risk measure with weight function $\gamma_t(u) = p \,\frac{1}{1-\alpha} \Id_{u \ge \alpha} + (1-p)$. For each $p\in[0,1]$, $\rho_t(\cdot)$ is coherent since the distortion weight function $\gamma_t(\cdot)$ is increasing. If $p = 0$, then $\rho_t (\cdot) = \E [\cdot |\; \F_t\;]$ and if $p = 1$, then $\rho_t (\cdot) = \ES_\alpha(\cdot |\; \F_t\;)$.

For the remainder of this subsection, let $(\bX, Y)$ be a $\F$-measurable random vector with joint cdf $F_{\bX, Y}$, where $Y$ is a univariate rv with cdf $F_Y$ and $\bX$ an $n$-dimensional random vector with cdf $F_\bX$. The next result explains how the one-step risk measures in \eqref{eq:rho-example} can be estimated using conditional elicitability.

\begin{proposition}[Mean-ES risk measure]\label{prop:score-argmin}
Let $\rho_t$ be given in \eqref{eq:rho-example} with $p \in (0,1)$. Denote by $\mathcal{G}:= \{g~|~ g \colon \R^n \to \R\}$.
Then it holds
\begin{equation*}
    \big(\VaR_{\alpha}(Y|\bX) \,, \ES_\alpha(Y|\bX), \, \rho_t(Y|\bX)\, \big)
    \;=
    \argmin_{(g_1, g_2, g_3) \in \mathcal{G}\times \mathcal{G}\times \mathcal{G}} \;\E\big[ S_{\rho}(g_1(\bX), g_2(\bX), g_3(\bX), Y)
   \big]\,,
\end{equation*}
where, for $0<D<+\infty$, $S_\rho$ is the strictly consistent scoring function given by
\begin{align}
\label{eqn:score-rho}
   S_\rho(z_1, z_2, z_3, y)
   :&= \log \Big(\frac{z_2 + D}{y+D}\Big) 
    - \frac{z_2}{z_2+D}
      + 
    \frac{\big(\Id_{\{y\le z_1\}} - \alpha\big)z_1   + \Id_{\{y> z_1\}}\, y}{(z_2+D)(1-\alpha)}
    +
    \Big( \frac{z_3 - p \, z_2}{1-p} - y \Big)^2
\,.
\end{align}
     
\end{proposition}
Thus, to estimate $\rho_t(Y|\bX)$ we proceed similarly to \cite{Fissler2023EJOR} and \cite{coache2022conditionally} and parameterise each function $g_i$ with a NN $\mfg^i$ with parameters $\bnu_i$, $i  = 1,2,3$. Then, the NN parameters are estimated via
\begin{equation*}
    \big(\hat{\bnu}_1,\, \hat{\bnu}_2,\, \hat{\bnu}_3\big)
     = 
     \argmin_{(\boldsymbol{\nu_1},\, \boldsymbol{\nu_2},\, \boldsymbol{\nu_3})}
     \; \frac1N \sum_{k = 1}^N S_\rho\Big(\mfg^1_{\boldsymbol{\nu_1}}(\bx^{(k)}), \, \mfg^2_{\boldsymbol{\nu_2}}(\bx^{(k)})\, , \mfg^3_{\boldsymbol{\nu_3}}(\bx^{(k)})\,, y^{(k)}\Big)\,,
\end{equation*}
over independent simulated mini-batches $(\bx^{(k)}, y^{(k)})_{k\in\{1, \ldots, N\}}$ of $(\bX, Y)$. As $\ES_\alpha$ is always larger than $\VaR_\alpha$, we set $\mfg^2_{\bnu_2}:=\mfg^1_{\bnu_1}+\mfg^4_{\bnu_2}$, where $\mfg^4_{\bnu_2}$ is a NN with parameters $\bnu_2$ and a softplus output layer, so that $\mfg^4_{\bnu_2}(\cdot) \ge 0$.

Elicitability of cdfs was shown in \cite{Gneiting2007JASA}, here we provide its conditional version; see also Proposition \ref{prop:prob-score} in the appendix.
\begin{proposition}[Conditional Distribution Function]
Denote by $\mathcal{H}:= \{h~|~ h \colon \R^n\times\R \to [0,1],\; h(\cdot,z) \text{ increasing in $z$}\}$.
Then it holds that
\begin{equation}\label{eqn:score-cdf}
F_{Y|\bX}(\cdot) = \argmin_{F\in\mathcal{H}}
\E\big[ S_{cdf}(F(\bX,\cdot), Y) \big]\,,
\qquad \text{where}
\quad
S_{cdf}(\bx,y):=
\int_\R \left(F(\bx,z) - \Id_{z\ge y}\right)^2 \, dz\,.
\end{equation}
\end{proposition}
As above, we parameterise the functions $F\in\mathcal{H}$ by NNs $\mfF$ with parameters $\mff$ and estimate the parameters by 
\begin{equation*}
    \hat{\mff} = \argmin_{\mff}
    \frac1N\sum_{k=1}^N 
    \sum_{l=1}^L \Big(\mfF_\mff\big(\bx^{(k)},z_l\big)-\Id_{z_l\ge y^{(k)}}\Big)^2\,\Delta z,
\end{equation*}
over independent simulated mini-batches $(\bx^{(k)}, y^{(k)})_{k\in\{1, \ldots, N\}}$, of $(\bX, Y)$, and where $z_l:=\underline{z}+(l-1)\Delta z$, $l = 1, \ldots, L$ with $\Delta z := \frac1{L-1}(\overline{z}-\underline{z})$ for truncation limits $\underline{z},\overline{z}$, such that $\underline{z}<\overline{z}$. The inner summation is an approximation to the Riemann integral in \eqref{eqn:score-cdf}, while the outer summation is an empirical approximation to the expectation.

\subsection{Neural Network Approximators}\label{subsec:NN-structure}

This section focuses on {NN architectures for the strategy $\btheta_{0:T}$, the risk-to-go process $\RM_{0:T}$, and the uniform rvs $(U_{t}[\btheta_{t:T}])_{\tT}$. 
As it is not clear whether the optimal strategy or the risk-to-go process are Markovian in asset prices, we proceed using non-Markovian parameterisations. In the context of NN approximations, recurrent NNs (RNNs) can be used to accomplish this goal. Our implementation employs gated recurrent units (GRUs) to encode non-Markovian features, though long short-term memory (LSTM) networks (attention networks are viable alternatives). Below we describe the architecture for the actor critic approach in detail.

First, the actor (strategy) NN (visualised in Figure \ref{fig:tikz-theta}) consists of a five layered GRU, with each layer consisting of hidden states of dimension $n$ (recall that $n$ is the asset dimension). The input features into the GRU are time, the wealth process of the induced self-financing strategy, and asset prices. We denote them by $y_t=(t,\;\bvtheta_{t-1}^\intercal \bX_t,\;\bX_t)\in\R^{n+2}$, $\tT$, and call them the state. At each time $\tT$, the output from all hidden layers from the previous time step, denoted by $h_{t-1}$, and the state from the current time step, $y_t$, are concatenated and passed through a five layer feed forward NN (FFN) to produce an $n$-dimensional output corresponding to $\btheta_t$. The internal layers of the FFN have sigmoid linear units (SiLU) activation functions, while, to ensure the strategy is long only, the last layer has a softplus activation function. 

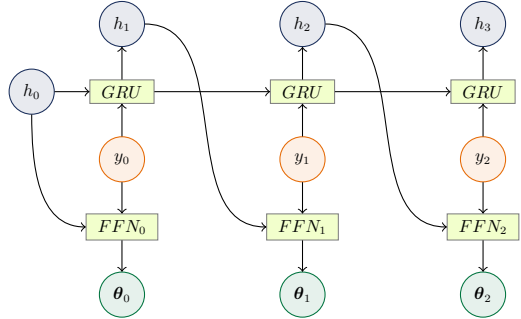
\begin{figure}[H]
    \centering
    \begin{tikzpicture}[scale=0.6,every node/.style={transform shape},minimum width=1.0cm]
    
    \node[hidden] (h0) at (-2,0) {$h_0$};
    
    \node[state] (y0) at (0,-1.5) {$y_0$};
    \node[hidden] (h1) at (0,1.5) {$h_1$};
    \node[gru] (gru0) at (0,0) {$~GRU~$};
    
    \draw [->] (h0) to (gru0);
    \draw [->] (y0) to (gru0);
    \draw [->] (gru0) to (h1);
    
    \node[state] (y1) at (4,-1.5) {$y_1$};
    \node[hidden] (h2) at (4,1.5) {$h_2$};
    \node[gru] (gru1) at (4,0) {$~GRU~$};
    
    \draw [->] (gru0) to (gru1);
    \draw [->] (y1) to (gru1);
    \draw [->] (gru1) to (h2);    
    
    \node[state] (y2) at (8,-1.5) {$y_2$};
    \node[hidden] (h3) at (8,1.5) {$h_3$};
    \node[gru] (gru2) at (8,0) {$~GRU~$};
    
    \draw [->] (gru1) to (gru2);
    \draw [->] (y2) to (gru2);
    \draw [->] (gru2) to (h3);        
    
    \node[FFN] (ANN0) at (0,-3) {$~FFN_0~$};
    \node[theta] (theta0) at (0,-4.5) {$\btheta_0$};
    
    \draw [->] (y0) to (ANN0);
    \draw [->] (ANN0) to (theta0);
    \draw [->] (h0) to [out=-90,in=180] (ANN0.west);
    
    \node[FFN] (ANN1) at (4,-3) {$~FFN_1~$};
    \node[theta] (theta1) at (4,-4.5) {$\btheta_1$};
    
    \draw [->] (h1) to [out=0,in=180]  (ANN1.west);
    \draw [->] (y1) to (ANN1);
    \draw [->] (ANN1) to (theta1);
    
    \node[FFN] (ANN2) at (8,-3) {$~FFN_2~$};
    \node[theta] (theta2) at (8,-4.5) {$\btheta_2$};
    
    \draw [->] (h2) to [out=0,in=180]  (ANN2.west);
    \draw [->] (y2) to (ANN2);
    \draw [->] (ANN2) to (theta2);    
    
    
    \end{tikzpicture}
    
    \caption{Directed graph representation for encodings and parameterisation of $\btheta_{0:T}$ functions.}
    \label{fig:tikz-theta}
    
\end{figure}

Next, the critic (risk-to-go) NN (visualised in Figure \ref{fig:tikz-V}) has the same GRU and FFN structure as the strategy network, however, the final output of the FFN is three dimensional corresponding to the conditional VaR ($\VaR_t$ in Figure \ref{fig:tikz-V} and $\mfg^1$ in Proposition \ref{prop:score-argmin}), the difference between the conditional ES and the conditional VaR ($\ES_t$ in Figure \ref{fig:tikz-V} and $\mfg^4$ in Proposition \ref{prop:score-argmin}), and the conditional risk measure ($\RM_t$ in Figure \ref{fig:tikz-V} and $\mfg^3$ in Proposition \ref{prop:score-argmin}). There is no activation function in the final layer for VaR and the risk measure, while we have a softplus activation for the difference of ES and VaR to ensure ES is always larger or equal to VaR.

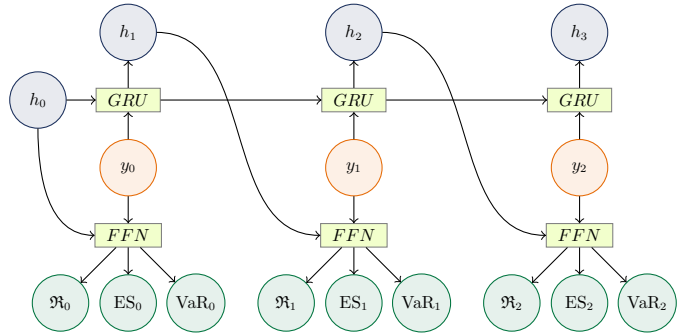
\begin{figure}[H]
    \centering
    \begin{tikzpicture}[scale=0.6,every node/.style={transform shape},minimum width=1.25cm]
    
    \node[hidden] (h0) at (-2,0) {$h_0$};
    
    \node[state] (y0) at (0,-1.5) {$y_0$};
    \node[hidden] (h1) at (0,1.5) {$h_1$};
    \node[gru] (gru0) at (0,0) {$~GRU~$};
    
    \draw [->] (h0) to (gru0);
    \draw [->] (y0) to (gru0);
    \draw [->] (gru0) to (h1);
    
    \node[state] (y1) at (5,-1.5) {$y_1$};
    \node[hidden] (h2) at (5,1.5) {$h_2$};
    \node[gru] (gru1) at (5,0) {$~GRU~$};
    
    \draw [->] (gru0) to (gru1);
    \draw [->] (y1) to (gru1);
    \draw [->] (gru1) to (h2);    
    
    \node[state] (y2) at (10,-1.5) {$y_2$};
    \node[hidden] (h3) at (10,1.5) {$h_3$};
    \node[gru] (gru2) at (10,0) {$~GRU~$};
    
    \draw [->] (gru1) to (gru2);
    \draw [->] (y2) to (gru2);
    \draw [->] (gru2) to (h3);        
    
    \node[FFN] (ANN0) at (0,-3) {$~FFN~$};
    \node[theta] (theta0a) at (-1.5,-4.5) {$\RM_0$};
    \node[theta] (theta0b) at (0,-4.5) {$\ES_0$};
    \node[theta] (theta0c) at (1.5,-4.5) {$\VaR_0$};

    \draw [->] (h0) to [out=-90,in=180] (ANN0.west);
    \draw [->] (y0) to (ANN0);
    \draw [->] (ANN0) to (theta0a);
    \draw [->] (ANN0) to (theta0b);
    \draw [->] (ANN0) to (theta0c);
    
    \node[FFN] (ANN1) at (5,-3) {$~FFN~$};
    \node[theta] (theta1a) at (3.5,-4.5) {$\RM_1$};
    \node[theta] (theta1b) at (5,-4.5) {$\ES_1$};
    \node[theta] (theta1c) at (6.5,-4.5) {$\VaR_1$};    

    \draw [->] (h1) to [out=0,in=180]  (ANN1.west);
    \draw [->] (y1) to (ANN1);
    \draw [->] (ANN1) to (theta1a);
    \draw [->] (ANN1) to (theta1b);
    \draw [->] (ANN1) to (theta1c);
    
    \node[FFN] (ANN2) at (10,-3) {$~FFN~$};
    \node[theta] (theta2a) at (8.5,-4.5) {$\RM_2$};
    \node[theta] (theta2b) at (10,-4.5) {$\ES_2$};
    \node[theta] (theta2c) at (11.5,-4.5) {$\VaR_2$};

    \draw [->] (h2) to [out=0,in=180]  (ANN2.west);
    \draw [->] (y2) to (ANN2);
    \draw [->] (ANN2) to (theta2a);
    \draw [->] (ANN2) to (theta2b);
    \draw [->] (ANN2) to (theta2c);
    
    \end{tikzpicture}
        \caption{Directed graph representation for encodings and parameterisation of the risk-to-go $\RM_t$, conditional  $\ES$, and conditional $\VaR$.}
        \label{fig:tikz-V}
 \end{figure}

%

\begin{figure}[H]  

    \centering
    \begin{tikzpicture}[scale=0.6,every node/.style={transform shape},minimum width=1.0cm]
    
    \node[hidden] (h0) at (-2,0) {$h_0$};
    
    \node[state] (y0) at (0,-1.5) {$y_0$};
    \node[hidden] (h1) at (0,1.5) {$h_1$};
    \node[gru] (gru0) at (0,0) {$~GRU~$};
    
    \draw [->] (h0) to (gru0);
    \draw [->] (y0) to (gru0);
    \draw [->] (gru0) to (h1);

    \node[state] (y1) at (4,-1.5) {$y_1$};
    \node[hidden] (h2) at (4,1.5) {$h_2$};
    \node[gru] (gru1) at (4,0) {$~GRU~$};
    
    \draw [->] (gru0) to (gru1);
    \draw [->] (y1) to (gru1);
    \draw [->] (gru1) to (h2);    
    
    \node[state] (y2) at (8,-1.5) {$y_2$};
    \node[hidden] (h3) at (8,1.5) {$h_3$};
    \node[gru] (gru2) at (8,0) {$~GRU~$};
    
    \draw [->] (gru1) to (gru2);
    \draw [->] (y2) to (gru2);
    \draw [->] (gru2) to (h3);        
    
    \node[FFN] (ANN0) at (0,-3) {$~FFN~$};
    \node[state] (z0) at (-2,-4) {$z_0$};
    \node[theta] (theta0) at (0,-5) {$F_0$};

    \draw [->] (h0) to [out=-90,in=180] (ANN0.west);
    \draw[->] (z0) to (ANN0);
    \draw [->] (y0) to (ANN0);
    \draw [->] (ANN0) to (theta0);
    
    \node[FFN] (ANN1) at (4,-3) {$~FFN~$};
    \node[state] (t1) at (2,-4) {$z_1$};    
    \node[theta] (theta1) at (4,-5) {$F_1$};

    \draw [->] (h1) to [out=0,in=180]  (ANN1.west);
    \draw[->] (t1) to (ANN1);
    \draw [->] (y1) to (ANN1);
    \draw [->] (ANN1) to (theta1);
    
    \node[FFN] (ANN2) at (8,-3) {$~FFN~$};
    \node[state] (t2) at (6,-4) {$z_2$};        
    \node[theta] (theta2) at (8,-5) {$F_2$};

    \draw [->] (h2) to [out=0,in=180]  (ANN2.west);
    \draw[->] (t2) to (ANN2);
    \draw [->] (y2) to (ANN2);
    \draw [->] (ANN2) to (theta2);    
    
    \end{tikzpicture}
    \caption{Directed graph representation for encodings and parameterisation of $F_t(\cdot):=F_{g_t|\F_t}(\cdot)$.}
    \label{fig:tikz-F}
    
\end{figure}
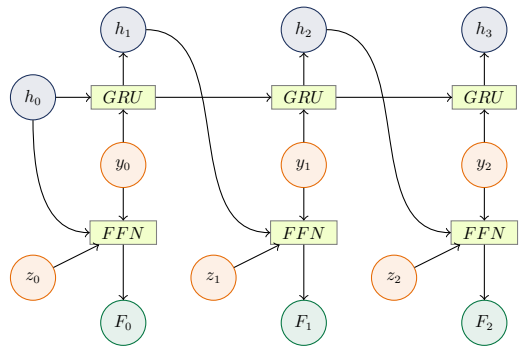

\vspace{-1em}
The second critic's NN architecture (for the conditional cdf $F_{g_t|\F_t}(\cdot)$) is provided in Figure \ref{fig:tikz-F} and is similar to that of the risk-to-go. Two important differences are (i) we concatenate not only the  hidden layers from the previous time step and the state, but also the value $z$ corresponding to $F_{g_t|\F_t}(z)$, and (ii) the output activation function is a sigmoid to ensure that $F_{g_t|\F_t}(z)\in(0,1)$. We also add the additional penalty $\int \Id_{\{\partial_z F(\bx, z)<0\}} (\partial_z F(\bx, z))^2\,dz$ to the score in Proposition \ref{prop:prob-score} to ensure that $F_{g_t|\F_t}(z)$ is increasing in $z$.

\section{Numerical Illustrations}\label{sec:num-examples}

In this section, we explore a stochastic volatility market model and an investor searching for a risk parity allocation, i.e., $b_{t,i}=\frac1n$ for all $t$ and $i$, and a risk budgeting strategy with budget $\boldsymbol{b}_{t} = (\frac{1}{15},\frac{2}{15},\frac{3}{15},\frac{4}{15},\frac{5}{15})$, under the coherent distortion DRM given in \eqref{eq:rho-example}. We consider $n = 5$ assets and a time horizon of $T + 1 = 12$, which corresponds to one year and monthly time steps. First, we describe the market model and then the optimal risk budgeting strategy.

\subsection{Market Model and NN Hyperparameters}

We use a discrete time version of a Heston inspired market model over a time horizon of $T +1 = 12$ months, where asset returns have a student-t copula dependence. The investor updates their strategy monthly, corresponding to time index $\tT$ in our optimisation problem. For simulating the market model we use the real time variable $t_k:=k\,\Delta t$ (with $\Delta t=\frac1{48}\,year$) such that $\tT$ corresponds to a real time point of $t_{4t}$ -- i.e., we take four discretisation steps between decision points. Specifically, we use a model  inspired by the Milstein discretisation of the Heston model and assume
\begin{subequations}
\begin{align*}
    \log \left( \frac{X_{t_{k+1},i}}{X_{t_{k},i}}\right) &= 
    (\mu_i-0.5(v_{t_{k},i})_+^2)\,\Delta t 
    + \sqrt{(v_{t_{k},i})_+} \;\Delta W^X_{t_{k},i}\,,
    \\
    v_{t_{k+1},i} &= \theta_i + \big((v_{t_{k},i})_+-\theta_i\big)\,e^{-\kappa_i\Delta t}
    + \eta_i\, \sqrt{(v_{t_{k},i})_+}\;\Delta W^v_{t_{k},i} + \tfrac14\eta^2_i\big((\Delta W^v_{t_{k},i})^2
    -\Delta t\big)\,.
\end{align*}%
\end{subequations}
Here, $(\cdot)_+:=\max(\cdot,0)$, $(\Delta W_{t_{k},i}^X,\Delta W_{t_{k},i}^v)_{i\in\N}$ are independent across $k$  but not $i$ rvs. They are marginally normal with mean zero and variance $\Delta t$. For $i\ne j$ and $\tT$, we have that (a) $\Delta W^X_{t_{k},i}$ and $ \Delta W^v_{t_{k},j}$ are independent, and (b) $\Delta W^v_{t_{k},i}$ and $\Delta W^v_{t_{k},j}$ are independent. Moreover, $(\Delta W^X_{t_{k},i}, \Delta W^X_{t_{k},j})_{i,j \in \N}$ have a student-t copula with $4$ degrees of freedom and $(\Delta W^X_{t_{k},i}, \Delta W^v_{t_{k},i})_{i \in \N}$ follow a Gaussian copula. The corresponding correlation matrix for the dependence structure and the additional market model parameters are provided in Appendix \ref{sec:parameters}.

Figure \ref{fig:heston-terminal} shows the distribution of the terminal log return while Table \ref{tab:asset-stats} provides basic statistics of the asset's total return. As can be seen in Figure \ref{fig:heston-terminal}, the distributions are all left skewed and volatility and expected return increases with the asset index label $i$.
\vspace{1em}

\begin{figure}
    \centering
        \includegraphics[width=0.5\textwidth]{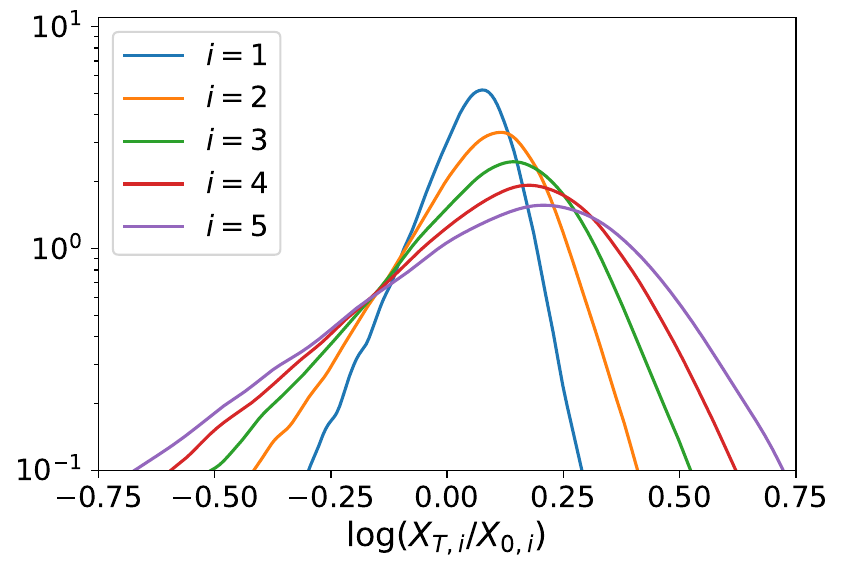}
            \caption{Distribution of log returns of various assets.}
            \label{fig:heston-terminal}
\end{figure}

\begin{table}[h]         \label{tab:asset-stats}
    \centering
      \begin{tabular}{lrrrrrrrr}
      \toprule \toprule
               &       &    std.    & \multicolumn{1}{l}{Sharpe}       & \multicolumn{5}{c}{correlation} \\
        Asset & \multicolumn{1}{l}{mean} & \multicolumn{1}{l}{dev.} & \multicolumn{1}{l}{ratio} & \multicolumn{1}{l}{$i=1$} & \multicolumn{1}{l}{$i=2$} & \multicolumn{1}{l}{$i=3$} & \multicolumn{1}{l}{$i=4$} & \multicolumn{1}{l}{$i=5$} \\
        \midrule
                $i=1$ &          0.05  & 0.10  & 0.49  & 1     &       &       &       &  \\
        $i=2$ &          0.08  & 0.16  & 0.48  &          0.17  & 1     &       &       &  \\
        $i=3$ &          0.11  & 0.22  & 0.47  &          0.16  &          0.15  & 1     &       &  \\
        $i=4$ &          0.13  & 0.29  & 0.46  &          0.16  &          0.15  &          0.14  & 1     &  \\
        $i=5$ &          0.16  & 0.35  & 0.46  &          0.16  &          0.15  &          0.15  &          0.15  & 1 \\
        \bottomrule\bottomrule
     \end{tabular}
    \caption{Statistics of the asset's total return $\frac{X_{T,i}}{X_{0,i}}-1$.}
\end{table}


When training the NNs we use a learning rate of $0.001$, a soft-update parameter of $\tau=0.001$, the ADAMW method for computing gradient updates of the NN parameters, and a scheduler that decreases the learning rate by a multiplicative factor of $0.99$ every 20 outer iterations of Algorithm \ref{algo:drb}. As well, use $m_r=20$ iterations for updating the risk-to-go and $m_f=5$ iterations for updating the conditional cdfs. The specific NN architectures we employ are as follows: (i) the GRUs have five layers with each layer having five hidden states, and (ii) the feed forward layers all have five layers with thirty two hidden nodes in each layer. We refer to Appendix \ref{sec:computational-times} for computational time metrics. 

\subsection{Risk budgeting strategy}

In Figure \ref{fig:RC-R-to-go-iterate}, we provide convergence results for the case $p=0.5$ and $\alpha=0.75$. The x-axis in the figures are iterations and, for each $\tT$, we plot the risk contributions for all assets, the sum of risk contributions across assets, and the risk-to-go. The left column contains the results for case when $\boldsymbol{b}_{t}=(\frac{1}{5},\frac{1}{5},\frac{1}{5},\frac{1}{5},\frac{1}{5})$ and the right panel when $\boldsymbol{b}_{t}=(\frac{1}{15},\frac{2}{15},\frac{3}{15},\frac{4}{15},\frac{5}{15})$.
The shaded region shows a measure of the confidence in the estimator, that is the standard deviation of the last 200 estimates (resulting from the learnt NN approximation of the risk-to-go and the simulated values of the risk contributions), while the solid lines show the moving average using the last 200 estimates. As the figure shows, the risk-to-go all converge to the value of $1$, a result of Theorem \ref{thm:opt}, and the risk contributions all converge to $b_{t,i}$ a result of Equation \eqref{pf:eq:risk-budget}. As the risk-to-go all converge to the theoretical value of $1$ and the risk contributions converge to their targeted values, this brings confidence that the numerical scheme has converged to a good approximation to the true solution of the problem. Interestingly, the risk contributions converge faster to their target values than the risk-to-go.

\begin{figure}
    \centering
\begin{tabular}{c|c}
    \centering
    equal $b_{t,i}$
    &
    unequal $b_{t,i}$
    \\
    \includegraphics[width=0.48\textwidth]{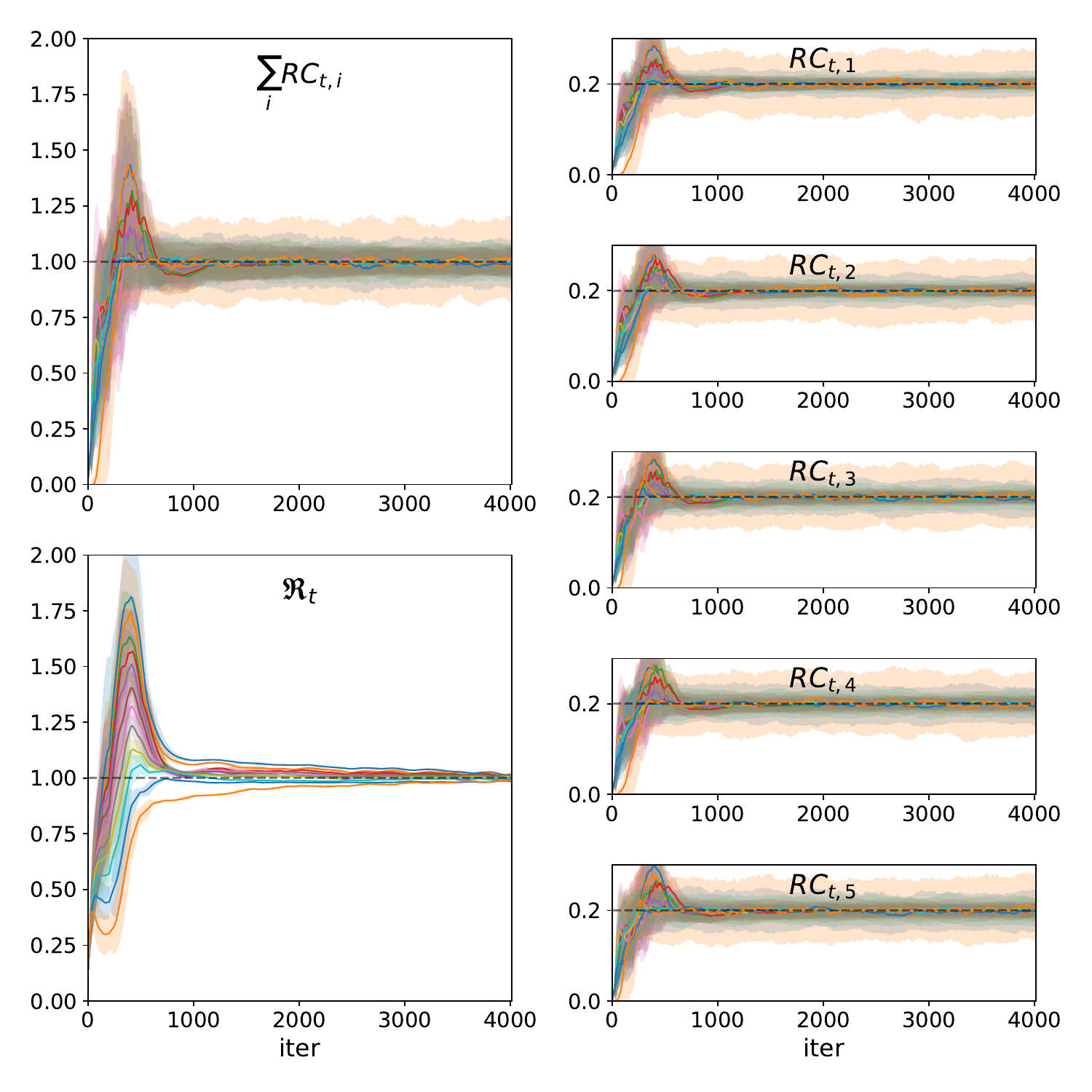} 
    & 
    \includegraphics[width=0.48\textwidth]
    {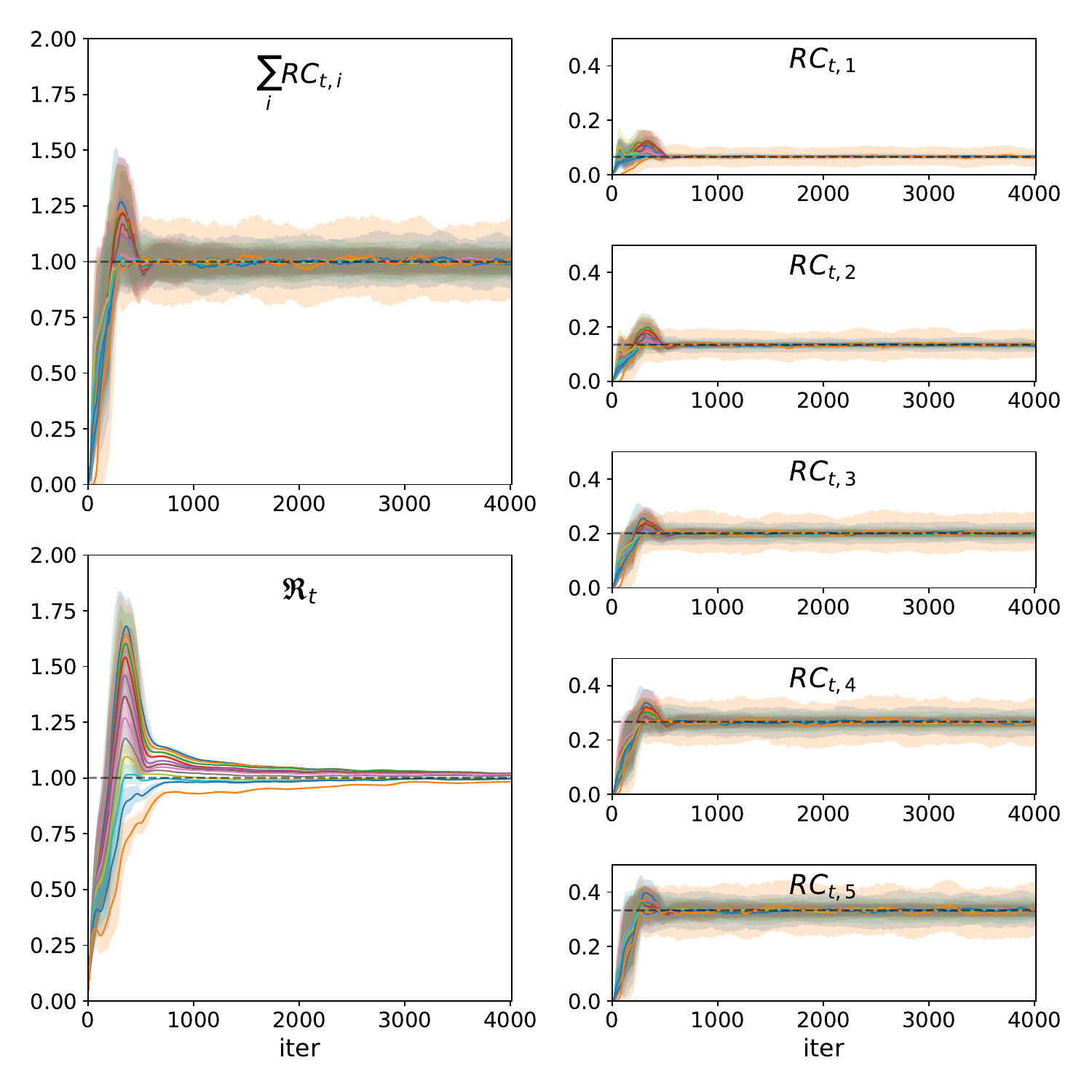} \\
  \end{tabular}
    \caption{Risk contributions and risk-to-go versus iterations for the optimal strategy $\btheta^*_{0:11}$ when $p=50\%$ and $\alpha=75\%$. Left: equal risk contributions, right: $\boldsymbol{b}_{t} = (\frac{1}{15},\frac{2}{15},\frac{3}{15},\frac{4}{15},\frac{5}{15})$.  $RC_{t,i}[\btheta_{t:T}]$ are estimated using $500$ simulations at each iteration. The bands and lines correspond to the moving standard deviation  and moving average (100 lags) of the corresponding quantities.}
    \label{fig:RC-R-to-go-iterate}
\end{figure}

To gain a deeper understanding of the learnt risk budgeting strategy, we present histograms in Figure \ref{fig:beta-hist} showing the percentage held in each asset across the twelve time steps. Each column in the figure represents a fixed choice of $p$ and choice of the risk budget $b_{t,i}$, while the rows correspond to different assets. 

\begin{figure}[H]
    \centering

    \centering
    \begin{tabular}{c|c}
    \begin{tabular}{cc}
         \multicolumn{2}{c}{equal $b_{t,i}$}  
         \\
         $p=50\%$ & $p=90\%$
         \\
        \includegraphics[width=0.2\textwidth]{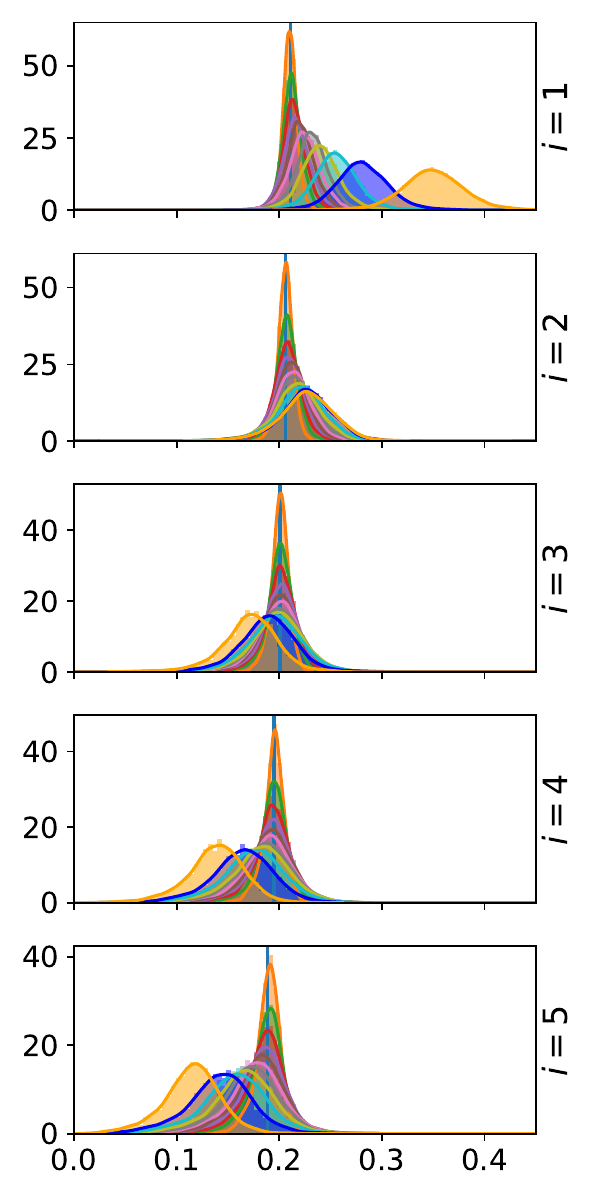}     
        &
        \includegraphics[width=0.2\textwidth]{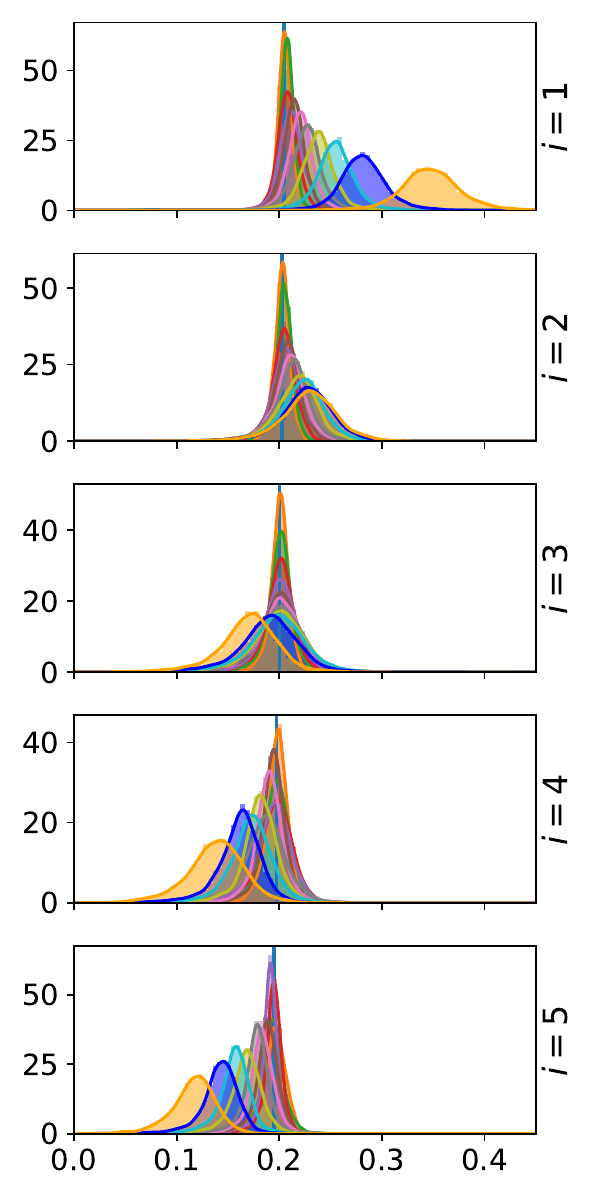}             
    \end{tabular}
    &
    \begin{tabular}{cc}
         \multicolumn{2}{c}{unequal $b_{t,i}$}  
         \\
         $p=50\%$ & $p=90\%$
         \\
        \includegraphics[width=0.2\textwidth]{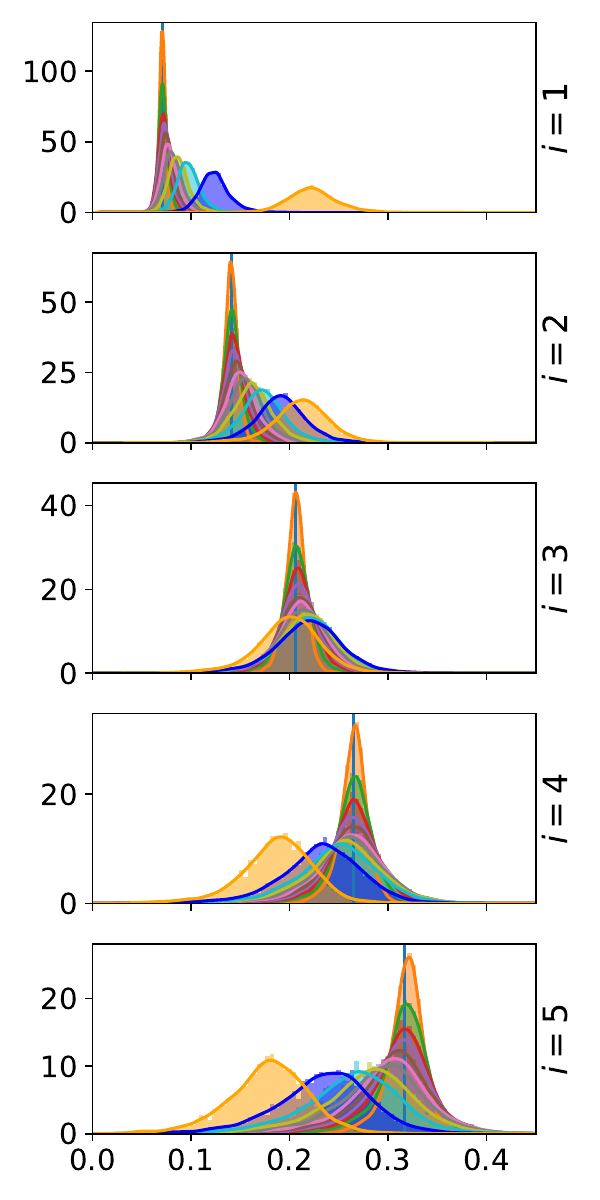}     
        &
        \includegraphics[width=0.2\textwidth]{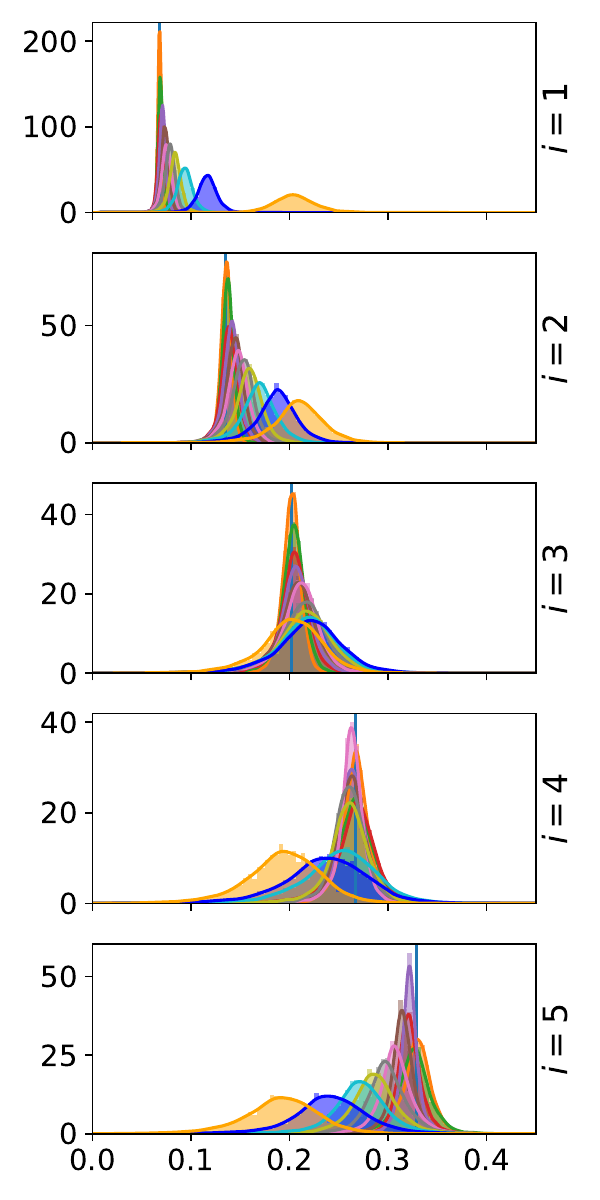}             
    \end{tabular}
    \end{tabular}  
    \caption{Percentage of wealth invested in each asset for each point in time for $\alpha = 0.75$, $p=50\%$ and $90\%$, and  $\boldsymbol{b}_{t}=(\frac{1}{5},\frac{1}{5},\frac{1}{5},\frac{1}{5},\frac{1}{5})$ and $\boldsymbol{b}_{t} = (\frac{1}{15},\frac{2}{15},\frac{3}{15},\frac{4}{15},\frac{5}{15})$.}
    \label{fig:beta-hist}
\end{figure}
In the first column of Figure \ref{fig:beta-hist}, which corresponds to the case $p=0.5$ and $b_{t,i}=\frac15$, we observe a general trend, whereby the investment in asset-$i$ decreases as $i$ increases. This trend is consistent with the fact that assets become increasingly volatile as the index-$i$ increases, making it reasonable to allocate less capital to the riskier assets to generate an equal risk budgeting portfolio. For the less risky assets $i = 1,2,3$, as time increases, investments become more disperse. Contrastingly, for the more risky assets $i=4,5$, as time increases, investments become more concentrated. This is sensible, as the investor aims to have a risk parity portfolio at all point in times and hence needs to deleverage the more risky assets. It is more challenging to provide a full description of how the distributions vary with $p$, as there are a number of competing factors that are difficult to disentangle. If we fix e.g., the fifth row, $i = 5$, and compare the $p=50\%$ case to the $p=90\%$ case, where the investor puts more weight on the $\ES$, the investment becomes less variable and more left skewed meaning that they invest less in the most risky asset. If we fix, e.g., the fourth row, $i = 4$, we observe that as $p$ increases, the distribution of the percentage of wealth at time $2$ becomes more variable, but shifts to the left; once again indicating a deleverage. Focusing on the unequal $b_{t,i}$ cases in Figure \ref{fig:beta-hist}, we see that  the percentage of wealth invested in asset-$i$ increases as $i$ increases, which is inline with the choices of $b_{t,i}$; they are now willing to take on more risk in the assets with a higher index.

\begin{figure}
    \centering
    \begin{tabular}{cc}
         \multicolumn{2}{c}{equal $b_{t,i}$}  
         \\
         $p=50\%$ & $p=90\%$
         \\
        \includegraphics[width=0.3\textwidth]{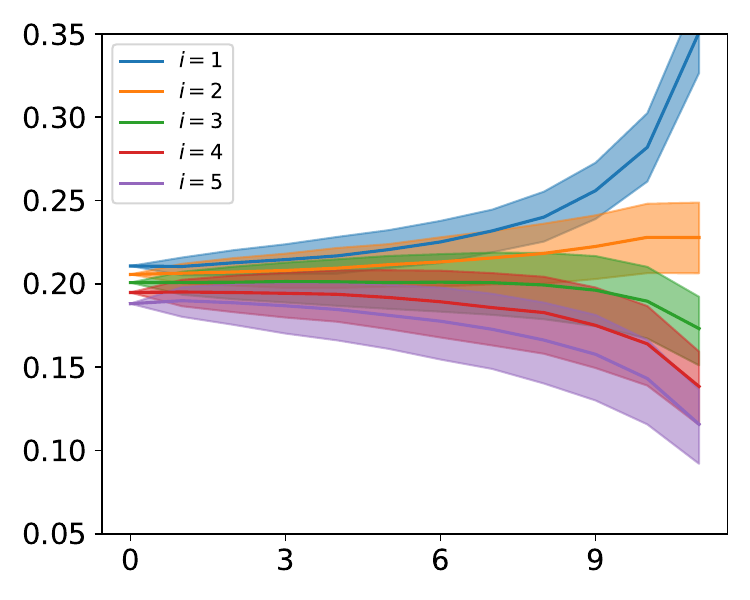}     
        &
        \includegraphics[width=0.3\textwidth]{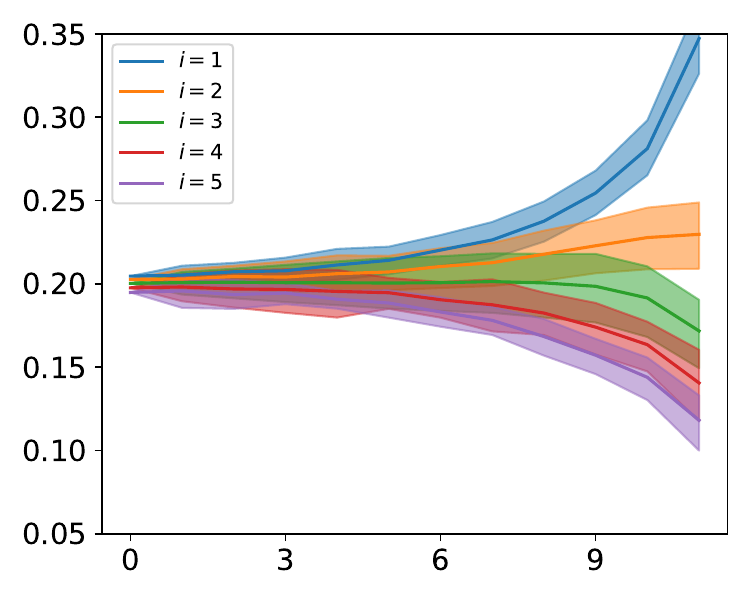}             
    \\
    \midrule
    \\
         \multicolumn{2}{c}{unequal $b_{t,i}$}  
         \\
         $p=50\%$ & $p=90\%$
         \\
        \includegraphics[width=0.3\textwidth]{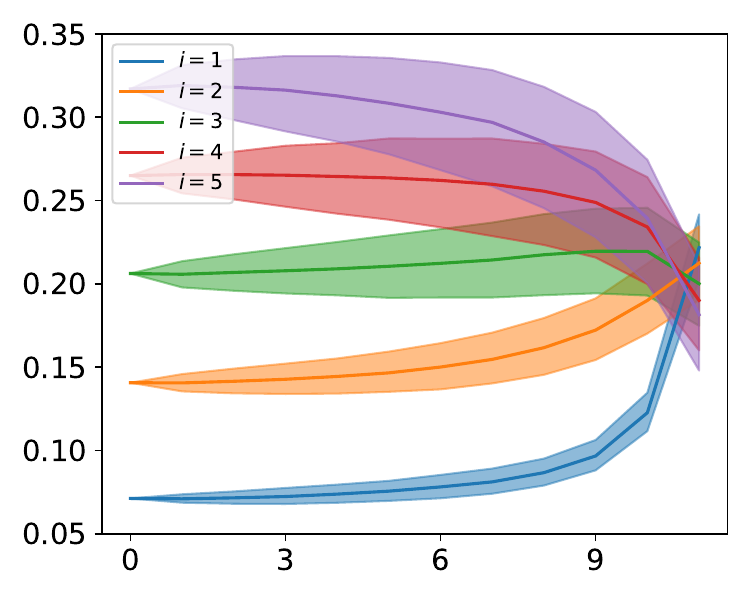}     
        &
        \includegraphics[width=0.3\textwidth]{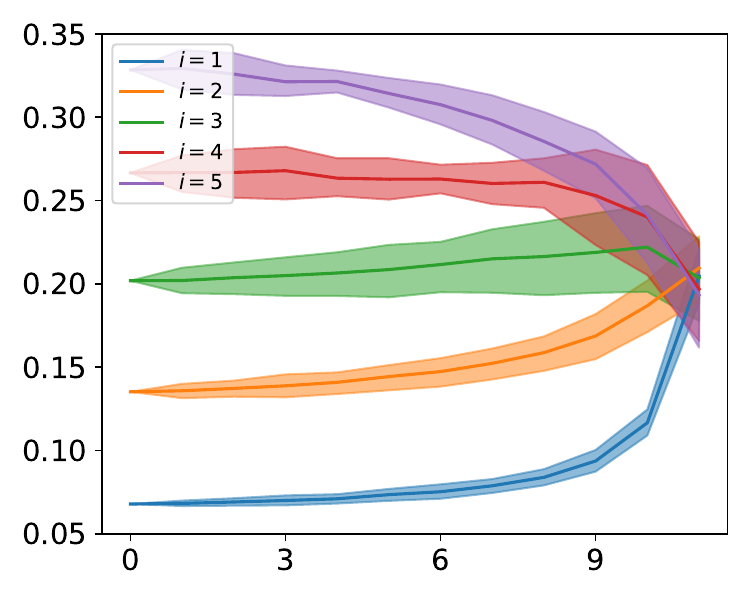}             
    \end{tabular}

    \caption{Percentage of wealth invested in each asset for each point in time for $\alpha = 0.75$, $p=50\%$ and $90\%$, and  $\boldsymbol{b}_{t}=(\frac{1}{5},\frac{1}{5},\frac{1}{5},\frac{1}{5},\frac{1}{5})$ and $(\frac{1}{15},\frac{2}{15},\frac{3}{15},\frac{4}{15},\frac{5}{15})$. The lines show the median and the bands show the 20\% and 80\% quantiles.}
    \label{fig:beta-bands}
\end{figure}
Figure \ref{fig:beta-bands} provides an alternative view of the evolution of the weights invested in each asset. It shows the medianm percentage of wealth invested for all assets as a function of time, together with the 20\% to 80\% quantile bands for the same cases shown in Figure \ref{fig:beta-hist}. 
Figure \ref{fig:beta-bands} shows that increasing $p$ generally induces less variability in the percentage of wealth invested. Moreover, when the risk contributions are equal, the amounts invested start almost equal weighted, but then spread out, while when the risk contributions are unequal, the amounts invested start unequal but move towards an equally weighted portfolio.

\section{Conclusion}

In this work, we show how an investor can allocate investments in risky assets to attain a predefined risk budget in a dynamic setting. To do so, we first propose a notion of risk contributions for coherent distortion DRM and demonstrate that they satisfy the full allocation property. For the class of coherent distortion DRM, we derive explicit formulae for risk contributions and prove that strategies that attain a particular risk budget are specified by the solution to a collection of convex optimisation problems. Leveraging elicitability of coherent distortion DRM, we further provide a deep learning approach for solving those optimisation problems. Finally, we demonstrate the stability of the numerical scheme through several examples using a stochastic volatility market model.

%
%
%

\begin{APPENDIX}{}

\section{Auxiliary Definition and Results}\label{app:aux-res}

The following definition of one-step distortion risk measures is in terms of the Choquet integral.
\begin{definition}[One-step Distortion Risk Measures]\label{def:distortion-Choquet}
\fontfamily{lmss}\selectfont
 For each $\tT$, let $g_t: [0,1] \times \Omega \to [0,1]$  be a (state dependent) distortion function such that for all $\omega\in\Omega$, the function $g_t(\cdot, \omega)$ is non-decreasing and satisfies $g_t(0, \omega) = 0$ and $g_t(1, \omega) = 1$. Further, we assume that the rv $g_t(x,\cdot) : \Omega \to [0,1]$ is $\F_{t}$-measurable for every $x\in[0,1]$ and for all $\tT$. Then, the one-step (conditional) distortion risk measure with distortion functions $\{g_t\}_{\tT}$ is the family $\{\rho_t\}_{\tT}$, where for each $\tT$ and $Z\in \Z_{t+1} $, $\rho_t$ is defined as
\begin{equation*}
    \rho_t(Z)
    :=
    -\int_{-\infty}^0\Big[ 1- g_t\big(1-F_{Z|F_t}(x)\big) \Big]\, dx  + \int_0^{+\infty} g_t\big(1-F_{Z|F_t}(x)\big)\, dx\,.
\end{equation*} 
If $g_t(\cdot, \omega)$ is absolutely continuous for all $\omega \in \Omega$, then the one-step risk measure admits representation \eqref{eq:distortion-rm},
where the distortion weight function $\gamma_t: (0,1) \times \Omega \to \R_+$ is given by $\gamma_t(u, \omega):= \frac{\partial^-}{\partial x} g(x, \omega)|_{x = 1-u}$, where $\frac{\partial^-}{\partial x}$ is the left derivative with respect to $x$, for all $u \in (0,1)$ and $\omega \in \Omega$, see e.g., \cite{Dhaene2012EAJ} for a proof in the static setting. 
\end{definition}

\begin{lemma}
\label{lemma:rho-finite}
Under Assumption \ref{assumption:gamma}, any coherent one-step distortion risk measure $\rho_t$ is a mapping  $\rho_t \colon \Z_{t+1} \to \Z_t$, $\tT/\{0\}$ and $\rho_0 \colon \Z_1 \to \R$.
\end{lemma}
\proof{Proof:}
Take $Z\in\Z_{t+1}$, then we have by \eqref{eq:distortion-rm} that $\rho_t(Z)$ is $\F_t$-measurable, and
\begin{align*}
\E[|\rho_t(Z)|] &= \E\left[\left|\E\left[Z\,\gamma_t(U_{Z|\F_t}) | \F_t\right]\right|\right] 
\\
&\le 
\E\left[
\left(\E\left[Z^2 | \F_t\right]\right)^\frac12
\left(\E\left[(\gamma_t(U_{Z|\F_t})^2 | \F_t\right]\right)^\frac12
\right]
\le 
C^\frac12 \left(\E\left[
\E\left[Z^2 | \F_t\right]\right]\right)^\frac12 < +\infty,
\end{align*}
where the first inequality follows from Cauchy-Schwartz, the second from Assumption \ref{assumption:gamma}, and third from $Z\in\Z_{t+1}$.
\hfill \Halmos\endproof

\begin{lemma}
\label{lemma:admissble-combination}
For any $c\ge0$, the set of admissible strategies $\A$ is a convex set.
\end{lemma}

\proof{Proof:}
Let $\btheta^{(0)}$ and $\btheta^{(1)}$ be two admissible strategies and define $\btheta^{(a)} := a \, \btheta^{(0)} + (1-a)\,\btheta^{(1)} $ for $a \in [0,1]$. First note that $\btheta^{(a)} \in \bmZ_{0:T}$.

\underline{Part 1:} $\btheta_{0:T}^{(a)}$ is lower bounded by $c$. Indeed for $\tT$, $\btheta_{t}^{(a)}=  a \, \btheta^{(0)}_t + (1-a)\,\btheta^{(1)}_t
    > a c + (1-a) c= c$.

\underline{Part 2:} 
it holds that $w^{\btheta^{(a)}} \in \L_{t+1}^\infty$. 
For this note that
\begin{align*}
    w^{\btheta^{(a)}}_t = 
    \frac{\btheta^{(a)\intercal}_{t} \bX_{t+1}}{\btheta^{(a)\intercal}_{t+1} \bX_{t+1}}
    &= a\, \frac{\btheta^{(0)\intercal}_{t} \bX_{t+1}}{\left(a\,\btheta^{(0)}_{t+1}+(1-a)\,\btheta^{(1)}_{t+1}\right)^\intercal \bX_{t+1}}
    + (1-a) \frac{\btheta^{(1)\intercal}_{t} \bX_{t+1}}{\left(a\,\btheta^{(0)}_{t+1}+(1-a)\,\btheta^{(1)}_{t+1}\right)^\intercal \bX_{t+1}}
    \\
    &\le 
    a\,\frac{\btheta^{(0)\intercal}_{t} \bX_{t+1}}{a\,\btheta^{(0)\intercal}_{t+1}\bX_{t+1}}
    + (1-a)\,\frac{\btheta^{(1)\intercal}_{t} \bX_{t+1}}{(1-a)\, \btheta^{(1)\intercal}_{t+1} \bX_{t+1}}
    = w_t^{\btheta^{(0)}} + w_t^{\btheta^{(1)}} < +\infty\,,
\end{align*}%
where the first inequality follows as $\btheta^{(i)\intercal}_t \bX_t\ge0$ and the last by admissibility of  $\btheta^{(0)}$ and $\btheta^{(1)}$.

\underline{Part 3:}
For every $\tT$, it holds that
\begin{align*}
\left(\E\left[\big(\btheta_t^{(a)\intercal} \Delta \bX_t\big)^2\, \right]\right)^\frac12
    &=
    \left(\E\left[\Big(\big(a \, \btheta^{(0)\, }_t + (1-a)\,\btheta^{(1)}_t \big)^\intercal \Delta \bX_t\Big)^2\, \right]\right)^\frac12
    \\
    &\le 
    \left(\E\bigg[\Big(a \, \btheta^{(0)\, \intercal}_t \Delta \bX_t \Big)^2 \, \bigg]\right)^\frac12
    +
    \left(\E\bigg[\Big((1-a)\,\btheta^{(1)\, \intercal}_t\Delta \bX_t \Big)^2\, \bigg]\right)^\frac12
    \\[0.5em]
    &< + \infty\,,
\end{align*}
where the inequality is the triangle inequality and the strict inequality by admissibility of $\btheta^{(0)}$ and $\btheta^{(1)}$. Hence, we have that
$\sum_{\tT} \E\left[\big(\btheta_t^{(a)\intercal} \Delta \bX_t\big)^2\,\right] <+\infty$.

Parts 1--3 imply that $\theta^{(a)}$ is admissible.
\hfill \Halmos\endproof

\begin{lemma}
\label{lemma:risk-to-go-finite}
For any $c\ge0$, if the risk measure is a coherent distortion DRM  satisfying Assumption \ref{assumption:gamma}, then any admissible strategy $\btheta \in \A$ satisfies $\RM_t[\btheta_{t:T}] \in \Z_t$, $\tT/\{0\}$, and $\RM[\btheta_{0:T} ]<+\infty$. 
\end{lemma}

\proof{Proof:}
Recall that for all $\tT / \{0\}$, the one-step distortion risk measures are mappings $\rho_t \colon \Z_{t+1} \to \Z_t$. Next, we proceed by induction backwards in time. At time $T$, the risk is
\begin{equation*}
    \RM_T[\btheta_{T}]
    =
    \rho_T\big(\btheta_T^\intercal \Delta \bX_T \big)\,.
 \end{equation*}
By admissibility of $\btheta$, we have $\btheta_T^\intercal \Delta \bX_T \in \Z_{T+1}$ thus $\RM_T[\btheta_{T}]\in \Z_T$. Next assume that $\RM_{t+1}[\btheta_{t+1}] \in \Z_{t+1}$, then the risk-to-go at time $t$ is
\begin{equation*}
    \RM_t[\btheta_{t:T}]
    =
    \rho_t\big(\btheta_t^\intercal \Delta \bX_t + w_t^\btheta \, \RM_{t+1}[\btheta_{t+1:T}]\big)\,.
 \end{equation*}
By admissibility of $\btheta$, we have $\btheta_t^\intercal \Delta \bX_t \in \Z_{t+1}$, $w^\btheta_t \in \L^\infty_{t+1}$, and thus $w^\btheta_t  \RM_{t+1}[\btheta_{t+1:T}]\in \Z_{t+1}\,$, which implies that $\RM_t[\btheta_{t:T}]\in \Z_t$. A similar argument yields that $\RM[\btheta_{0:T}]< +\infty$.
\hfill \Halmos\endproof

\begin{lemma}\label{lemma:RC-in-L1} For any $c\ge0$ and $\btheta_{0:T}\in\A$, we have that $\E[|\RC_{t,i}[\btheta_{t:T}]|]<+\infty$.
\end{lemma}

\proof{Proof:}
Note that 
\begin{align*}
\E\left[\left| \RC_{t,i}[\btheta_{t:T}]
    \right|\right]
    &\le
    \E\left[\;
    \left|\theta_{t,i} \dX_{t,i}\gamma_t \big(U_t[\btheta_{t:T}]\big)\right| 
    +
    \left|    \frac{\theta_{t,i}\,X_{t+1,i}}{\btheta_{t+1}^\intercal \bX_{t+1}}\;\RM_{t+1}[\btheta_{t+1:T}]\gamma_t \big(U_t[\btheta_{t:T}]\big) \right|\;
    \right]
    \\
    &\le 
    \left(\E\left[
    \left(
    \theta_{t,i} \dX_{t,i}\right)^2
    \right]
    \E\left[\left(\gamma_t \big(U_t[\btheta_{t:T}]\big)\right)^2\right]
    \right)^{\frac12}
    \\
    &\quad+
    \left(\E\left[
    \left(
    \frac{\theta_{t,i}\,X_{t+1,i}}{\btheta_{t+1}^\intercal \bX_{t+1}}\;\RM_{t+1}[\btheta_{t+1:T}] \right)^2\;
    \right]\;
    \E\left[
    \left(\gamma_t \big(U_t[\btheta_{t:T}]\big) \right)^2
    \right]
    \right)^{\frac12}
    \\
    &\le C^{\frac12}\left(\E\left[
    \sum_{\tT}
    \left(
    \btheta_{t}^\intercal \bdX_{t}\right)^2\,
    \right]\right)^{\frac12}
    +C^{\frac12}
    \left(\E\left[
    \left(
    w^\btheta_t\;\RM_{t+1}[\btheta_{t+1:T}] \right)^2\,
    \right]\right)^{\frac12}<+\infty,
\end{align*}
where the last inequality follows as (i) $w_t^\btheta$ are in $\L^\infty$, (ii) $\btheta$ is admissible, and (iii) $\RM_{t+1}[\btheta_{t+1:T}]\in \Z_{t+1}$ from Lemma \ref{lemma:admissble-combination}.
\hfill \Halmos\endproof

\begin{proposition}[Lipschitz continuity of One-step Coherent Distortion Risk Measures]\label{aux:prop-lipschitz}
Let $\rho_t$ be a one-step coherent distortion risk measure with distortion weight function $\gamma_t$ satisfying Assumption \ref{assumption:gamma}. Then, $\rho_t$ is Lipschitz continuous w.r.t the conditional $L^2$ norm, i.e. for any $X,Y \in \Z_{t+1}$, it holds that
\begin{equation*}
    \big| \rho_t(X) - \rho_t(Y)\big|
    \le 
    C^\frac12\;  \|X-Y\|_{t} \quad a.s.\,,
\end{equation*}
where the conditional norm $\|\cdot\|_t$ is defined for all $Z\in \Z_{t+1}$ by $\|Z\|_t := \big(\E\big[Z^2\big| \F_t\big]\big)^\frac12$, $\forall\;\tT/\{0\}$. Furthermore, $\rho_0 \colon \Z_1 \to \R$.
\end{proposition}

\proof{Proof}
This is an adaption of Lemma 2.1 in \cite{Inoue2003JMAA} to one-step distortion risk measures. By sub-additivity of the conditional distortion risk measure we have a.s.
\begin{align*}
    \rho_t(X) - \rho_t(Y)  
    &\le 
    \rho_t\big(X-Y\big)
    =
    \E\Big[(X-Y) \gamma_t(U_{X-Y|\F_t})\Big| \F_t\Big] 
    \\
    &\le 
    \big(\E\big[(X-Y)^2 \big| \F_t\big]\big)^\frac12\; 
     \left(\E\Big[\big(\gamma_t(U_{X-Y|\F_t})\big)^2 \Big| \F_t\Big]\right)^\frac12
     \le
     C^\frac12\;  ||X-Y||_t\,.
\end{align*}
where in the second inequality we use the conditional Cauchy-Schwarz inequality, and the last inequality follows by assumption on $\gamma_t$. Interchanging $X$ and $Y$ concludes the proof. 

The fact that $\rho_0 \colon Z_1 \to \R$ follows by e.g., Def. 6.37 in \cite{Shapiro2021book}.
\hfill \Halmos\endproof

\begin{proposition}[Interchanging Expectation and Limits]\label{aux:lemma:interchange}
Suppose that $\btheta=(\btheta_t,\btheta_{t+1:T}^*)$ and $\btheta'=(\btheta_t',\btheta_{t+1:T}^*)$  are admissible strategies, and let $\dbtheta:=\btheta'_{t}-\btheta_{t}$, then we have that
\[
\lim_{\ep\downarrow0} \frac1\ep\left(L_t[\btheta_t+\ep \dbtheta]-L_t[\btheta_t]\right) = 
   \sum_{i\in\N} \; \E\left[\;
\D_{i}^{{\dtheta_i}}\,\RM_t[(\btheta_t, \btheta_{t+1:T}^*)]
    - 
    b_{t,i} \;\frac{{\dtheta_i}}{ \theta_{t,i}}\;\right].
\]
\end{proposition}

\proof{Proof:}
For any $\ep\in[0,1]$, let  $\btheta^\ep:=\btheta+\ep \,\dbtheta = (1-\ep)\,\btheta+\ep\,\btheta'$. By  Lemma \ref{lemma:admissble-combination}, $\btheta^\ep$ is admissible. Then, by definition, we have that
\begin{align}
    & \hspace*{-3em}
    \lim_{\ep\downarrow0} \frac1\ep\left(L_t[\btheta_t+\ep \dbtheta]-L_t[\btheta_t]\right)
    \nonumber
    \\
    &= \lim_{\ep\downarrow0}\frac1\ep
    \E\Big[
    \RM_t[(\btheta_{t}^\ep\,, \btheta_{t+1:T}^*)]
    -
    \RM_t[(\btheta_t\,, \btheta_{t+1:T}^*)]
    - \sum_{i\in\N}b_{t,i} \;\left(\log\theta_{t,i}^\ep-\log \theta_{t,i}\right)\;\Big]
    \nonumber
    \\
    &
    =\lim_{\ep\downarrow0}\frac1\ep
    \E\Big[
    \underbrace{\rho_t\big(\btheta_{t}^{\ep\,\intercal} \bdX_t + w_t^{\btheta_\ep}\,\RM_{t+1}[\btheta_{t+1:T}^*]\big)
    -\rho_t\big(\btheta_t^\intercal \bdX_t + w_t^{\btheta}\,\RM_{t+1}[\btheta_{t+1:T}^*]\big)}_{=:A_t}
    \nonumber
    \\
    &\hspace{5em}
     \underbrace{-\sum_{i\in\N}b_{t,i} \;\left(\log\theta_{t,i}^\ep-\log \theta_{t,i}\right)}_{=:B_t}\;\Big]\,.
     \label{eqn:lim-outside}
\end{align}
We next show that $\frac1\ep|A_t|\le\mfA_t$  s.t. $\mfA_t$ is independent of $\ep$ and $\E\big[\mfA_t\,\big]<+\infty$, and similarly for $\frac1\ep|B_t|\le\mfB_t$ s.t.
$\mfB_t$ is independent of $\ep$ and $\E\big[\mfB_t\,\big]<+\infty$.

To this end, by Proposition \ref{aux:prop-lipschitz}, we have that
\begin{align*}
\frac1\ep
\big|A_t
    \big|\;
&\le \frac1\ep \,C^{\frac12}\,
\Big\|\ep\dbtheta^\intercal \bdX_{t} + (w_t^{\btheta^\ep}-w_t^\btheta) \RM_{t+1}[\btheta_{t+1:T}^*] \Big\|_t
\\
&= 
C^{\frac12} \,\Big\|\dbtheta^\intercal \bdX_{t} + \frac{\dbtheta^\intercal \bX_{t+1}}{\btheta_{t+1}^{*\intercal} \bX_{t+1}} \,\RM_{t+1}[\btheta_{t+1:T}^*] \Big\|_t   
\\
\text{\tiny (by $\triangle$ inequality) }
&\le C^{\frac12}\left(
    \,\big\|\dbtheta^\intercal \bdX_{t} \big\|_t+ \Big\|
    \frac{\dbtheta^\intercal \bX_{t+1}}{\btheta_{t+1}^{*\intercal} \bX_{t+1}}   \, \RM_{t+1}[\btheta_{t+1:T}^*] \Big\|_t\right) =: \mfA_t    
\qquad \text{a.s.}
\end{align*}
Note $\mfA_t$ is independent of $\ep$, furthermore,
\begin{align*}
C^{-\frac12}\; \E[ \mfA_t \,]
&\le
\,\left(\E\left[\E\left[\left(\dbtheta^\intercal \bdX_{t} \right)^2\,|\,\F_t\right] \, \right] \right)^\frac12
\qquad \text{\tiny (by def. $\|\cdot\|_t$ and Jensen) }
\\
&\qquad
+ \left(\E\left[\left.\E\left[\,\left(\frac{\dbtheta^\intercal \bX_{t+1} }{\btheta_{t+1}^{*\intercal}  \bX_{t+1}}   \RM_{t+1}[\btheta_{t+1:T}^*]\right)^2 \,\right|\,\F_t\right] \,\right] \right)^\frac12 
\\
&=
\left\|\dbtheta^\intercal \bdX_{t}  \right\|_0
+ \left\|\frac{\dbtheta^\intercal \bX_{t+1} }{\btheta_{t+1}^{*\intercal} \bX_{t+1}}   \RM_{t+1}[\btheta_{t+1:T}^*] 
\right\|_0
\\
\text{\tiny (by $\triangle$ inequality) }&
\le 
\left\|\btheta_{t}^\intercal\bdX_{t} \right\|_0
+
\left\|\btheta_{t}'^\intercal\bdX_{t}\right\|_0
\\
&\qquad
+\left\|\frac{\btheta_{t}^\intercal \bX_{t+1}}{\btheta_{t+1}^{*\intercal} \bX_{t+1}}   \RM_{t+1}[\btheta_{t+1:T}^*] 
\right\|_0
+\left\|\frac{\btheta_{t}'^\intercal \bX_{t+1}}{\btheta_{t+1}^{*\intercal} \bX_{t+1}}   \RM_{t+1}[\btheta_{t+1:T}^*] 
\right\|_0
\\
&
= 
\left\|\btheta_{t}^\intercal\bdX_{t} \right\|_0
+
\left\|\btheta_{t}'^\intercal\bdX_{t}\right\|_0
+\left\|w^\btheta \,\RM_{t+1}[\btheta^*_{t+1:T}] 
\right\|_0
+\left\|w^{\btheta^1} \, \RM_{t+1}[\btheta^*_{t+1:T}] 
\right\|_0
\\
&
\le
\left(\E\,\left[\textstyle\sum_{\tT}(\btheta_{t}^\intercal\bdX_{t})^2\,\right]\right)^\frac12
+
\left(\E\,\left[\textstyle\sum_{\tT}(\btheta_{t}^{1,\intercal}\bdX_{t})^2\,\right]\right)^\frac12
\\
&\qquad
+\big\|w^\btheta \,\RM_{t+1}[\btheta^*_{t+1:T}] 
\big\|_0
+\big\|w^{\btheta^1} \, \RM_{t+1}[\btheta^*_{t+1:T}] 
\big\|_0
\\
&< +\infty\,,
\end{align*}
where the last inequality follows from the admissibility of $\btheta$ and $\btheta^1$ and from Lemma \ref{lemma:risk-to-go-finite}.

Next, as $|\log(x)-\log(y)|\le \frac1c|x-y|$ for all $x,y\ge c>0$, we have that
\begin{align*}
    |\tfrac1\ep B_t| \le \sum_{i\in\N}\tfrac c\ep|\theta_{t,i}^\ep-\theta_{t,i}| = c\sum_{i\in\N}|\delta \theta_{t,i}|
    \le \sum_{i\in\N}c(|\theta_{t,i}'|+|\theta_{t,i}|) =: \mfB_t
\end{align*}
where $c$ is the lower bound for admissible strategies and $\mfB_t$  is independent of $\ep$. As $\btheta$ and $\btheta'$ are in $\bmZ$, we have that $\E[\mfB_t]<+\infty$.

Putting these bounds together, we have by Lebesgue dominated convergence that the $\lim_{\ep\downarrow0}$ in \eqref{eqn:lim-outside} may be moved under the expectation, and the claim follows.
\hfill \Halmos\endproof

\begin{proposition}\label{aux:prop-assumption}
Let $\rho_t$ be a one-step coherent distortion risk measures with distortion weight function $\gamma_t$ satisfying Assumption \ref{assumption:gamma}. Let $\tT$ and $Y, W \in \Z_{t+1}$, where we assume that $(Y, W)$ has a joint density, though the proof can be generalised to include point masses. If $\ep \to F^{-1}_{Y + \ep W}(u)$ is differentiable in a neighbourhood around $\ep=0$ with bounded derivative, for all $u \in (0,1)$, then it holds that
\begin{equation}
    \lim_{\ep \to 0}\frac{\rho_t\left(Y + \ep\, W\right) -\rho_t(Y)}{\ep}
    =
    \E\left[\,W \, \gamma_t \left(U_{Y|\F_t}\right)| \F_t\,\right]\,,
\end{equation}
\end{proposition}

\proof{Proof:}
First we define the conditional cdfs $F(y) := \P(Y \le y\;|\;\F_t)$ and $F(y, \ep) := \P(Y  + \ep \,W\le y\;|\;\F_t)$ and their corresponding densities by $f(y)$ and $f(y, \ep)$. We further write $F^{-1}(u)$ and $F^{-1}(u, \ep)$ for the quantile functions of $F(\cdot)$ and $F(\cdot, \ep)$, respectively. 

Next using $\rho_t\left(Y + \ep\, W\right) = \E\left[F^{-1}(U, \ep)\, \gamma_t(U)\;|\;\F_t\right]$, for a uniform rv $U \in \F_t$, the integrability assumption on $\gamma_t$, and the differentiability assumption on $F^{-1}(u,\ep)$, the mean value theorem together with Lebesgue dominated convergence allows us to interchange expectation and limit to obtain 
\begin{equation}\label{eq:derivative-rm}
    \lim_{\ep \to 0}\frac{\rho_t\left(Y + \ep\, W\right) -\rho_t(Y)}{\ep}
    =
    \E\left[\partial_\ep F^{-1}(U, \ep)\, \gamma_t(U)\;|\;\F_t\right]
    \Big|_{\ep = 0}\,.
\end{equation}
By taking a derivative with respect to $\ep$ of the equation $F(F^{-1}(u, \ep), \ep) = u$, we obtain for all $u \in (0,1)$,
\begin{equation}\label{eq:derivative-inverse}
    \partial_\ep F^{-1}(u, \ep)
    = 
    - \frac{\partial_\ep F(y, \ep) }{f(y, \ep)}\Big|_{y = F^{-1}(u, \ep)}
\end{equation}
Next, we calculate the derivative $\partial_\ep F(y, \ep)$. For this note that 
\begin{align}
    \partial_\ep F(y, \ep)
    &= 
    \partial_\ep \E[\Id_{Y + \ep W \le y}\;|\;\F_t]
    = 
    \lim_{\ep \to 0} \frac1\ep\;
    \E\left[\Id_{Y + \ep W \le y} - \Id_{Y \le y}\;|\;\F_t\right]
    \nonumber
    \\
    &= 
    \lim_{\ep \to  0} \frac1\ep\;
    \E\left[\E\left[\Id_{Y \in( y - \ep W , y]} \; |\: W\right]\;|\;\F_t\right]
    = 
    \lim_{\ep \to 0} \frac1\ep\; 
    \E\left[\left.\int_{y - \ep\, W}^{y}dF_{Y|W}(y')\;\right|\;\F_t\right]
    \nonumber
    \\
    &= 
    -\,\E\big[W\, f_{Y|W}(y)\;|\;\F_t\big]\,,
    \label{eq:derivative-cdf}
\end{align}
where, $F_{Y|W}$ and $f_{Y|W}$ are the distribution and density, respectively, of $Y$ conditional on $W$.
Plugging \eqref{eq:derivative-inverse} and \eqref{eq:derivative-cdf} into Equation \eqref{eq:derivative-rm}, we obtain
\begin{align*}
    \lim_{\ep \to 0}\frac{\rho_t\left(Y + \ep\, W\right) -\rho_t(Y)}{\ep}
    &=
    \E\left[\left.
    \frac{\E\left[W\, f_{Y|W}(y)\right]}{f(y)}\Big|_{y = F^{-1}(U)}\;
    \gamma_t(U)\;\right|\;\F_t\right]
    \\
    &=
    \E\left[\left.
    \E\left[W\;|\; Y = y \right]\Big|_{y = F^{-1}(U)}\;
    \gamma_t(U)\;\right|\;\F_t\right]
    \\
    &=
    \E\left[\left.
    \E\left[W\;|\; F(Y) = U \right]\;
    \gamma_t(U)\;\right|\;\F_t\right]
    \\
    &=
    \E\left[
    W\; \gamma_t(U_{Y|\F_t})\;|\;\F_t\right]\,.
\end{align*}
\hfill \Halmos 
\endproof

\section{Additional Proofs}\label{app:sec:proof}
\subsection{Proof of Proposition \ref{prop:impact-of-decision}}\label{app:impact-of-decision}

First we generalise the \Gat derivative as follows.
For a functional $F_t:\bmZ_{t:T}\to \Z_t$, $t\in \T$ and $s \ge t$, we denote by $\D_{s,i}^{\dZ} \,F_t$, its \Gat derivative of the $i^\text{th}$ component at time $s$ in direction $\dZ \in\Z_s$. That is, for $s,\tT$, $s \ge t$, and $\bZ_{t:T}\in\bmZ$
\begin{equation*}
    \D_{s,i}^{\dZ}\, F_t[\bZ_{t:T}] := 
    \lim\limits_{\ep\to 0} \frac{1}{\ep}\Big(F_t[\bZ_{t:T} + \ep\,\1_{s,i} \dZ] - F_t[\bZ_{t:T}]\Big)\,.
\end{equation*}

Next, note that 
\begin{align}
    RC_{t,i}[\btheta_{t:T}]
    &= 
    \D_{i}^{\theta_{t,i}}\,    \rho_t\left(
    \btheta_t^\intercal \bdX_t + 
    \;\RM_{t+1}[w^\btheta_t\, \btheta_{t+1:T}]
    \right)
    \nonumber
    \\
    &= 
    \lim_{\ep \to 0}\frac{1}{\ep}
    \Big\{
    \rho_t\Big(
    \ep \theta_{t,i} \dX_{t,i} 
    + 
   \ep  \D_{t,i}^{\theta_{t,i}}\RM_{t+1}[w^\btheta_t\, \btheta_{t+1:T}]
   \nonumber
   \\
   &
   \qquad \qquad \qquad
    +    
    \btheta_t^\intercal \bdX_t + 
    \;\RM_{t+1}[w^\btheta_t\, \btheta_{t+1:T}]
    \Big)
    -
    \RM_t[\btheta_{t:T}]
    \Big\}
    \nonumber
    \\
    &\stackrel{\text{by \eqref{eq:proof-derivative-rm}}}{=} 
    \E\left[
    \left(\theta_{t,i} \dX_{t,i} 
    + 
   \D_{t,i}^{\theta_{t,i}}\RM_{t+1}[w^\btheta_t\, \btheta_{t+1:T}]
   \right)\, 
   \Gamma_t^\btheta
   ~ \Big| ~ \F_t
    \right]\,.
    \label{eqn:RC-in-terms-of-one-step-Gateaux}
\end{align}

Next, we show that for all $s \in\{ t, \ldots, T\}$, that 
\begin{equation}\label{eq:proof-gateaux-Rs}   \D_{t,i}^{\theta_{t,i}}\;\RM_{s}\left[w_t^\btheta\, \btheta_{s:T}\right]
    =
    \sum_{r = s}^T\,
    \E\left[
        \frac{\theta_{t,i}\, X_{t+1,i}}{\btheta_{t+1}^\intercal \bX_{t+1}} \; (\btheta_r^\intercal \bdX_r) \;\Gamma^\btheta_s \cdots\Gamma^\btheta_r 
    ~\Big|~\F_s\right]
\end{equation}
While we require the above equation for the case $s=t+1$ only, to make the proof easier to understand, we introduce an additional variable $s$ and use mathematical induction over $s$. First we show that Equation \eqref{eq:proof-gateaux-Rs} holds for $s = T$. To see this we calculate
\begin{align*}
    \D_{t,i}^{\theta_{t,i}}\;\RM_{T}\left[w_t^\btheta\, \btheta_{T}\right]
    &=
    \lim_{\ep \to 0}\frac{1}{\ep}
    \left\{
    \rho_T\left(w_t^\btheta (\btheta_T^\intercal \bdX_T)  + \ep \,
    \frac{\theta_{t,i}\, X_{t+1, i}}{\btheta_{t+1}^\intercal \bX_{t+1}} (\btheta_T^\intercal \bdX_T)\right)
    -
    \rho_T\left(w_t^\btheta (\btheta_T^\intercal \bdX_T)  \right)
    \right\}
    \\
    &=
    \E\left[\frac{\theta_{t,i} X_{t+1, i}}{\btheta_{t+1}^\intercal \bX_{t+1}} \; (\btheta_T^\intercal \bdX_T) \; \Gamma_T^\btheta ~\Big|~ \F_T
    \right]\,,
\end{align*}
where we applied Equation \eqref{eq:proof-derivative-rm} to obtain the last equality. Next, assuming Equation \eqref{eq:proof-gateaux-Rs} holds for all $r \in\{s+1, \ldots, T\}$, we show that it also hols for $s$. Indeed, using the definition of a \Gat derivative, i.e., that $F_s[\bZ_{s:T} + \ep\,\1_{t,i} \dZ] = \ep\, \D_{t,i}^{\dZ}\, F_s[\bZ_{s:T}] 
    + F_s[\bZ_{s:T}] + o(\ep)$, we obtain
\begin{align*}
    \D_{t,i}^{\theta_{t,i}}\;\RM_{s}\left[w_t^\btheta\, \btheta_{s:T}\right]
    &\stackrel{\text{by \eqref{eq:risk-to-go-theta-t}} }{=}
    \D_{t,i}^{\theta_{t,i}}\;
    \rho_s\left(w_t^\btheta \, \btheta_s^\intercal \bdX_s + \RM_{s+1} 
    \left[w_t^\btheta \, \btheta_{s+1:T}\right]\right)
    \\
    &=
    \lim_{\ep \to 0}\frac{1}{\ep}
    \bigg\{
    \rho_s\bigg(
    \ep \,
    \frac{\theta_{t,i}^\intercal X_{t+1, i}}{\btheta_{t+1}^\intercal \bX_{t+1}} \; \btheta_s^\intercal \bdX_s
    +\, 
    w_t^\btheta \, \btheta_s^\intercal \bdX_s 
    \\
    & \qquad \qquad \qquad
    + \ep\; \D_{t,i}^{\theta_{t,i}}\;\RM_{s+1}\left[w_t^\btheta\, \btheta_{s+1:T}\right]
    + \RM_{s+1} 
    \left[w_t^\btheta \, \btheta_{s+1:T}\right]
    \bigg)
    -
    \RM_{s}\left[w_t^\btheta\, \btheta_{s:T}\right]
    \bigg\}
    \\
    & \stackrel{\text{by \eqref{eq:proof-derivative-rm}}}{=} 
    \E\left[
    \left(\frac{\theta_{t,i}^\intercal X_{t+1, i}}{\btheta_{t+1}^\intercal \bX_{t+1}} \; \btheta_s^\intercal \bdX_s
    +  \D_{t,i}^{\theta_{t,i}}\;\RM_{s+1}\left[w_t^\btheta\, \btheta_{s+1:T}\right]
    \right)
    \Gamma_s^\btheta 
    ~\Big|~ \F_s    
    \right]
    \\
    &= \E\left[
    \left(\frac{\theta_{t,i}^\intercal X_{t+1, i}}{\btheta_{t+1}^\intercal \bX_{t+1}} \; \btheta_s^\intercal \bdX_s\right.\right.
    \\
    & \qquad
    \left.\left.
    +  \sum_{r = s+1}^T\,
    \E\left[
        \frac{\theta_{t,i} X_{t+1,i}}{\btheta_{t+1}^\intercal \bX_{t+1}} \; \btheta_r^\intercal \bdX_r \;\Gamma^\btheta_{s+1} \cdots\Gamma^\btheta_r 
    ~\Big|~\F_{s+1}\right]
    \right)
    \Gamma_s^\btheta 
    ~\Big|~ \F_s    
    \right]
    \\
    &= 
    \sum_{r = s}^T\,
    \E\left[
    \E\left[
        \frac{\theta_{t,i} X_{t+1,i}}{\btheta_{t+1}^\intercal \bX_{t+1}} \; \btheta_r^\intercal \bdX_r \;\Gamma^\btheta_{s+1} \cdots\Gamma^\btheta_r 
    ~\Big|~\F_{s+1}\right]
    \Gamma_s^\btheta 
    ~\Big|~ \F_s    
    \right]
    \\
        &= 
    \sum_{r = s}^T\,
    \E\left[
        \frac{\theta_{t,i} X_{t+1,i}}{\btheta_{t+1}^\intercal \bX_{t+1}} \; \btheta_r^\intercal \bdX_r \;\Gamma^\btheta_s \cdots\Gamma^\btheta_r 
    ~\Big|~ \F_s    
    \right]\,,
\end{align*}
where we use the induction argument in fourth equation and that $\Gamma_s^\btheta \in \F_{s}$ in the last equality. This concludes the proof of Equation \eqref{eq:proof-gateaux-Rs}.

Finally combining  \eqref{eq:proof-gateaux-Rs} with  \eqref{eqn:RC-in-terms-of-one-step-Gateaux}, noticing that $\Gamma_t^\btheta \in \F_t$, and using the law of iterated expectations concludes the proof.
\hfill \Halmos

\subsection{Proof of Theorem \ref{thm:uniquness}}\label{app:proof-uniqueness}
\proof{Proof: }
Let $\bvphi_{0:T} \in \A$ denote a self-financing risk budgeting strategy with budget $B$ and initial wealth of $1$, whose risk-to-go satisfies $c_R \le\RM_t[\bvphi_{t:T}]\le c^R$ a.s. for all $\tT$. We show that  $\bvphi_{0:T} = \frac{1}{\bvtheta_0^{*\, \intercal}\bX_0}\bvtheta_{0:T}^*$, where $\bvtheta_{0:T}^*$ is the induced self-financing strategy of the unique solution to optimisation problem \eqref{eqn:P-prime}, i.e., given by Theorem \ref{thm:opt}(b) with lower bound $c' := \frac{c}{c^R}$. For this we proceed by contradiction. Assume  $\bvphi_{0:T} \neq \frac{1}{\bvtheta_0^{*\, \intercal}\bX_0}\bvtheta_{0:T}^*$ and define the (not necessarily self-financing) strategy $\bpsi_{0:T}$ via
\begin{equation}
\label{eqn:theta-vartheta}
    \bpsi_t
    := 
    \frac{1}{\RM_t[\bvphi_{t:T}]}
    \,\bvphi_t\,,
    \qquad 
    \forall \tT\,.
\end{equation}
It holds that $\bpsi \in \mathcal{A}_{c'}$. This follows as the bounds on $\RM_t[\bpsi_{t:T}]$ implies that $w_t^\bpsi \in \L_t^\infty$, and $\RM_t[\bpsi_{t:T}]\le c^R$ guarantees that $\bpsi_t  = \frac{1}{\RM_t[\bvphi_{t:T}]}
    \,\bvphi_t \ge c' $, for all $\tT$. 
Next, we show that the risk-to-go process of $\bpsi_{0:T}$ satisfies
\begin{equation}\label{eq:proof:uniquness-RM}
    \RM_t[\bpsi_{t:T}]
    = 
    1\,,
    \qquad 
    \forall \tT\,.
\end{equation}
We proceed by induction. At time $T$, by positive homogeneity of $\RM_T$, and since $c_R\le \RM_T[\bvphi_T] \le c^R$:
\begin{equation*}
    \RM_T[\bpsi_T] 
    =
    \frac{1}{\RM_T[\bvphi_{T}]}\,
    \RM_T[\bvphi_T] 
    =
    1\,.
\end{equation*}
Assume Equation \eqref{eq:proof:uniquness-RM} holds for $t+1$, then, again using $c_R\le \RM_t[\bvphi_{t:T}] \le c^R$,
\begin{align*}
    \RM_t[\bpsi_{t:T}]
    &=
    \rho_t\left(
    \bpsi_t^\intercal \bdX_t +
    w_t^\bpsi\; \RM_{t+1}[\bpsi_{t+1:T}]
    \right)
    \\
    &=
    \rho_t\left(
    \bpsi_t^\intercal \bdX_t +
    \frac{\bpsi_t^\intercal \bX_{t+1}}{\bpsi_{t+1}^\intercal \bX_{t+1}}\, 
    \right)
    \\
    &=
    \rho_t\left(
    \frac{1}{\RM_{t}[\bvphi_{t:T}]}\bvphi_t^\intercal \bdX_t 
    +
    \frac{\RM_{t+1}[\bvphi_{t+1:T}]}{\RM_{t}[\bvphi_{t:T}]}\,
    \frac{\bvphi_t^\intercal \bX_{t+1}}{\bvphi_{t+1}^{\intercal} \bX_{t+1}}
    \right)
    \\
    \text{\tiny(by positive homogeneity of $\rho_t(\cdot)$ and $\varphi$ is self-financing)}&=
    \frac{1}{\RM_{t}[\bvphi_{t:T}]}    \rho_t\left(
    \bvphi_t^{\intercal} \bdX_t 
    +
    \RM_{t+1}[\bvphi_{t+1:T}]
    \right)
    \\
    &=1\,,
\end{align*}
Thus, Equation \eqref{eq:proof:uniquness-RM} holds for all $\tT$.

Next, we show that $\bpsi_{0:T}$ is a risk budgeting strategy with budget $B$. By positive homogeneity of  risk contributions (Proposition \ref{prop:homo--RC}) we obtain
\begin{equation*}
    RC_{t,i}[\bpsi_{t:T}]
    =
    \frac{1}{\RM_t[\bvphi_{t:T}]}
    RC_{t,i}[\bvphi_{t:T}]
    =
    \frac{1}{\RM_t[\bvphi_{t:T}]}
    b_{t,i} \, \RM_t[\bvphi_{t:T}]
    =
    b_{t,i} 
    =
    b_{t,i}\, \RM_t[\bpsi_{t:T}]\,.
\end{equation*}
Thus, $\bpsi_{0:T}$ is not only a risk budgeting strategy but also a solution to optimisation problem \eqref{eqn:P-prime} with lower bound $c'$, that is for all $\tT$ it satisfies Equations \eqref{pf:eq:risk-budget}. As $\bpsi\in \mathcal{A}_{c'}$ is a solution to optimisation problem \eqref{eqn:P-prime} with $\mathcal{A}_{c'}$, it induces a self-financing, risk budgeting strategy, with initial wealth of $1$ as given in Theorem \ref{thm:opt}(b), and denoted here by $\bvtheta_{0:T}$. Specifically, we have 
\begin{equation*}
\bvtheta_0 
    : =
\frac{1}{\bpsi^{ \intercal}_0 \bX_0} \bpsi_0\,,
\quad \text{and} \quad 
\bvtheta_t 
    : = 
\frac{1}{\bpsi^{ \intercal}_0 \bX_0}  \left( \prod_{s = 0}^{t-1} w_s^{\bpsi} \right)\, \bpsi_t\,,
\quad
\forall \tT/\{0\}\,,
\end{equation*}

Finally, we show that $\bvtheta_{0:T} = \bvphi_{0:T}$. For this recall that $\bvphi_{0:T}$ is a self-financing strategy with initial wealth of 1, thus
\begin{equation*}
    \bvtheta_0 
    =
    \frac{1}{\bpsi^{ \intercal}_0 \bX_0} \bpsi_0
        =
    \frac{\RM_0[\bvphi_{0:T}]}{\bvphi^{ \intercal}_0 \bX_0} \; \frac{\bvphi_0}{\RM_0[\bvphi_{0:T}]}
    =
    \bvphi_0\,.
\end{equation*}
For $t \in \T/\{0\}$, we have
\begin{align*}
    \bvtheta_t 
    &= 
    \frac{1}{\bpsi^{ \intercal}_0 \bX_0}   \left(\prod_{s = 0}^{t-1} \frac{\bpsi_s^\intercal \bX_{s+1}}{\bpsi_{s+1}^{\intercal}\bX_{t+1}} \right)\, \bpsi_t
    \\
    &=
    \frac{\RM_0[\bvphi_{0:T}]}{\bvphi^{ \intercal}_0 \bX_0} \;
    \left(\prod_{s = 0}^{t-1} \frac{\RM_{s+1}[\bvphi_{s+1:T}]}{\RM_{s}[\bvphi_{s:T}]}
    \; \frac{\bvphi_s^{ \intercal} \bX_{s+1}}{\bvphi_{s+1}^{\intercal}\bX_{t+1}}\right)
    \frac{\bvphi_t}{\RM_t[\bvphi_{t:T}]}
   \\
   \text{\tiny(as $\bvphi_{0:T}$ is self-financing)} &=
    \RM_0[\bvphi_{0:T}] \;
    \left(\prod_{s = 0}^{t-1} \frac{\RM_{s+1}[\bvphi_{s+1:T}]}{\RM_{s}[\bvphi_{s:T}]}
    \; \right)
    \frac{\bvphi_t}{\RM_t[\bvphi_{t:T}]}
    \\
    &=
    \bvphi_t\,.
\end{align*}
Thus, $\bvphi_{0:T} = \bvtheta_{0:T}$ and as $\bvtheta_{0:T}$ is given by Theorem \ref{thm:opt}(b), we arrive at a contradiction.  
Uniqueness follows from strict convexity of the optimisation problem \eqref{eqn:P-prime}. 
\hfill \Halmos\endproof

\section{Elicitability}
\label{sec:appendix-elicitability}

We first recall the classical definition of elicitability and scoring functions, see e.g., \cite{Fissler2016AS}. For this first define the set of cdfs $\M := \{F_{Y} ~|~ Y \in \F_{T+1}\}$.
\begin{definition}
[Elicitability]
\label{def:elicitability}
\fontfamily{lmss}\selectfont
A functional $\mfT \colon \M \to A$, $A \subset \R^k$, is called $k$-elicitable on $\widetilde{\M} \subseteq \M$, if there exists a measurable function $S \colon A \times \R \to [0,\infty]$ -- called a strictly consistent scoring function -- if for all $F\in \widetilde{\M}$ and for all $z\in A$ 
\begin{equation}\label{eq:elicitable}
    \int  S\big(\mfT(F), y\big) \,dF(y)
    \le 
    \int  S(z, y)\, dF(y)\,, 
\end{equation}
and equality in \eqref{eq:elicitable} holds only if $z = \mfT(F)$. 
\end{definition}

In the numerical examples, we consider the family of coherent distortion DRM $\{\rho_t\}_{\tT}$, a convex combination of the mean and ES, given in \eqref{eq:rho-example}. We first illustrate that its static version
\begin{equation}\label{eq:rho-example-static}
    \rho(Z) := p \,\ES_\alpha(Z) + (1-p)\, \E[X]\,, \qquad Z \in \Z_{T+1}\,, 
\end{equation}
is elicitable.
While the mean is well-known to be 1-elicitable, the $\ES_\alpha$ is only jointly elicitable together with $\VaR_\alpha$ at the same $\alpha$-level. We recall these well-known results.

\begin{proposition}[Mean -- \cite{Gneiting2011JASA}]\label{prop:score-mean}
Let $\phi \colon \R\to \R$ be strictly convex with subgradient $\phi'$ and denote by $\M^\dagger \subset \M$ be the class of cdfs with finite mean such that $\int |\phi(y)|\,\mathrm{d}F(y) <+\infty$ for all $F\in\M^\dagger$. Then 
\begin{equation*}
    S_{\E}(z,y) :=  \phi'(z)(z-y) - \phi(z)+ \phi(y)
    \,,\qquad z,y\in\R\,,
\end{equation*}
is strictly $\M^\dagger$-consistent for the mean. 
\end{proposition}

\begin{proposition}[$(\VaR, \ES)$ -- \cite{AcerbiSzekely2014, Fissler2016AS}]\label{prop:score-VaR-ES}
Let $\alpha \in (0,1)
$, $A := \{(z_1, z_2) \in \R^2 \,:\, z_1 \ge z_2 \}$ and define the scoring functions $S_{\VaR, \ES}: A \times \R \to \R$ by
\begin{align*}
S_{\VaR, \ES}(z_1,z_2,y)
&=  
\big(\Id_{\{y\le z_1\}} - \alpha\big)\big(g(z_1) -g(y)\big)  \\ 
& \quad 
+ \Phi'(z_2)\Big(z_2 - \tfrac{1}{1 - \alpha} S^+_\alpha(z_1,y)\Big)- \Phi(z_2) + \Phi(y)\,,
\end{align*}
where $S^+_\alpha(z_1,y) = (\Id_{\{y\le z_1\}} - \alpha)z_1 - \Id_{\{y\le z_1\}}y + y$,
$\Phi\colon\R\to\R$ is strictly convex with subgradient $\Phi'$ and $g\colon\R\to\R$ is such that for all $z_2\in\R$
\begin{align*}
    &z_1\mapsto g(z_1) - z_1\Phi'(z_2)/(1 - \alpha)
\end{align*}
is strictly increasing. 

Let $\M^{\ddagger} \subset \M$ be the cdfs with unique $\alpha$-quantile, finite mean, and such that $\int |g(y)|\,dF(y)<+\infty$ and $\int |\Phi(y)|\,dF(y)<+\infty$ for all $F\in\M^{\ddagger}$. Then $S_{\VaR, \ES}$ is strictly $\M^{\ddagger}$-consistent for the couple $(\VaR_\alpha, \ES_\alpha)$. 
\end{proposition}

Next, we show that for $p\in(0,1)$, the risk measure $\rho$ given in \eqref{eq:rho-example-static} is 3-elicitable, that is jointly elicitable together with $\VaR_\alpha$ and $\ES_\alpha$. The following proposition is different to Corollary 5.4 in \cite{Fissler2016AS}, which states that $\rho$ is 4-elicitable, specifically they show that $(\VaR_\alpha, \ES_\alpha, \E, \rho)$ is jointly elicitable.

\begin{proposition}[Mean-ES risk measure]\label{prop:score}
For $p \in (0,1)$, let $\rho$ be given in \eqref{eq:rho-example-static} and let the assumptions of Propositions \ref{prop:score-mean} and \ref{prop:score-VaR-ES} be enforced. Then the function $S_\rho \colon A \times \R^2 \to [0, \infty]$ given by 
\begin{align}\label{eqn:S-rho}
    S_{\rho}(z_1, z_2, z_3, y)
    :&=
   S_{\VaR, \ES}(z_1,z_2,y)
   +
   S_{\E}\left( \frac{z_3 - p \, z_2}{1-p} ,\; y\right)
   \,,
\end{align}
is a strictly $\M^{\ddagger}$-consistent scoring function for the triplet $(\VaR_\alpha, \ES_\alpha, \rho)$. 
\end{proposition}

\proof{Proof:}
From Propositions \ref{prop:score-mean} and \ref{prop:score-VaR-ES}, the functionals $(\VaR_\alpha, \ES_\alpha)$ and $\E$ are elicitable. Using Lemma 2.6 in \cite{Fissler2016AS}, we obtain that $(\VaR_\alpha, \ES_\alpha, \E)$ is elicitable with consistent scoring function given by 
\begin{equation*}
S_{\VaR, \ES, \E}(z_1,z_2,z_3, y) 
=
S_{\VaR, \ES}(z_1,z_2,y) + S_{\E}( z_3,y)    \,.
\end{equation*}
Next, we apply Osband's \textit{revelation principle}, see e.g., Theorem 4 in \cite{Gneiting2011JASA}. First, define the bijective function $g \colon \R^3 \to \R^3$ by $g(z_1, z_2, z_3) = \big(z_1, \,z_2, \,p z_2 + (1-p) z_3 \big)^\intercal$ with  inverse $g^{-1}(a_1, a_2, a_3) = \left(a_1, a_2, \frac{a_3 - p\, a_2}{1-p}\right)^\intercal$. The revelation principle states that $g (\VaR, \ES,\E) = (\VaR ,\ES, \rho)^\intercal$ is elicitable with scoring function 
\begin{equation*}
    S_{\VaR, \ES, \rho}(z_1,z_2,z_3, y) 
    =
    S_{\VaR, \ES, \E}\left(g^{-1}(z_1,z_2,z_3)^\intercal, y\right) 
    =
    S_{\VaR, \ES}(z_1,z_2,y) + S_{\E}\left( \frac{z_3 -p\,z_2}{1-p},\,y\right) \,.
\end{equation*}
Moreover, if the scoring functions $S_{\VaR, \ES}(z_1,z_2, y) $ and $S_{\E}(z_3, y) $ are strictly consistent for $(\VaR, \ES)$ and $\E$, respectively, then $S_{\VaR, \ES, \rho}(z_1,z_2,z_3, y) $ is strictly consistent for $(\VaR, \ES, \rho)$.
\Halmos
\endproof
In our implementations, we make the specific choices of $\phi(z)=z^2$, $g(z)=C$ and $\Phi(z)=-\log(z+C)$, where $C>0$.

Finally, we need to elicit the conditional cdf $F_{Y|\bX}\colon \R \to [0,1]$, defined by $F_{Y|\bX}(y): = \P(Y \le y \:|\: \bX = \bx)$, for any $Y \in \Z_{t+1}$, $\bX \in \bmZ_t$,  and $\bx \in \R^n$. 
Cdfs are known to be elicitable with the \emph{continuous ranked probability score}, see e.g. Equation (20) in \cite{Gneiting2007JASA}. Here we recall the key result.

\begin{proposition}[Distribution Function -- \cite{Gneiting2007JASA}]\label{prop:prob-score}
Let $\M^\dagger$ be the set of cdfs with finite mean. Then the scoring function $S_{cdf}\colon \M^\dagger \times \R\to [0, \infty]$ given by
\begin{equation}\label{eqn:S-cdf}
S_{cdf}(F, y)
:=
\int_\R \left(F(z) - \Id_{z\ge y}\right)^2 \, dz\,,
\end{equation}
is a strictly consistent scoring function for $F_{Y}$. In particular, it holds that 
\begin{equation*}
    \argmin_{F\in \M^\dagger} \E[\;S_{cdf}(F,Y)\; ]\,,
\end{equation*}
is attained by the cdf $F_{Y}\colon \R\to [0,1]$.
\end{proposition}

Next, we consider the concept of conditional elicitability.
Let $(\bX, Y)$ be a random vector with joint cdf $F_{\bX, Y}$, where $Y$ is a univariate rv with cdf $F_Y$ and $\bX$ an $n$-dimensional random vector with cdf $F_\bX$. Further, let $S\colon A \times \R \to [0, \infty]$ be a strictly consistent scoring function  that elicits the functional $\mfT:\M\to A$. Next define the expected score $\mfS \colon \M_{n+1} \to [0,\infty]$ by
\begin{equation*}
    \mfS(F_{\bX,Y},g) := \int S\big(g(\bx),y\big)\,F_{\bX,Y}(d\bx,dy) 
    = \int\left(\int S\big(g(\bx),y\big)\, F_{Y|\bX=\bx}(dy)\right) F_\bX(d\bx)\,,
\end{equation*}
where $\M_{n+1}$ is the space of cdfs of $(n+1)$--dimensional random vectors, and $\mathcal{G} := \{g ~|~ g \colon \R^n \to \R\,, \text{ with } \mfS(F,g) < + \infty\}$. Then the expected score satisfies
\begin{align*}
    \mfS(F_{\bX,Y},g) 
    &= \int\left(\int S\big(g(\bx),y\big)\, F_{Y|\bX=\bx}(dy)\right) F_\bX(d \bx)
    \\
    \text{\tiny(by def. \ref{def:elicitability}) }&\ge
    \int\left(\int S\big(\mfT(F_{Y|\bX=\bx}\big)\,,\,y) \, F_{Y|\bX=\bx}(dy)\right) F_\bX(d\bx)
    \\
    &= \int S\big(\mfT(F_{Y|\bX=\bx})\,,\,y\big)\;
    F_{\bX,Y}(d\bx,dy)
    = \mfS\big(F_{\bX,Y}\,,\,\mfT(F_{Y|\bX})\big)\,.
\end{align*}
Therefore, $\mfT(F_{Y|\bX})$ minimises the expected  score $\mfS(F,g)$ over all functions $g$. Note, we may also write $\mfS(F,g):=\E[S(g(\bX),Y)]$ with the expectation  taken over a probability measure where $(\bX,Y)$ has cdf $F_{\bX,Y}$. We can approximate the minimiser of this expected score by seeking over a rich parameterised class of functions (such as NNs) and estimate the expectation using the empirical mean from simulations. This is what we use when eliciting approximations of $\RM_t[\btheta_{t:T}]$ and $U_t[\btheta_{t:T}]$.

\section{Additional Information on Numerical Implementation}\label{app:numerical}
\subsection{Parameters used in Market Model Simulation.}
\label{sec:parameters}

This section contains further details on the simulated market model simulation. In particular, Table \ref{tab:Market-para} specifies the market model parameters and Table \ref{tab:corr} gives the correlation matrix of the dependence structure. 

\begin{table}[H]
  \centering
  \caption{Market model parameters.}
    \begin{tabular}{l 
    S[table-format=3.2]
     S[table-format=3.2]
      S[table-format=3.2]
       S[table-format=3.2]
        S[table-format=3.2]}
    \toprule\toprule
          & \multicolumn{1}{l}{$i=1$} & \multicolumn{1}{l}{$i=2$} & \multicolumn{1}{l}{$i=3$} & \multicolumn{1}{l}{$i=4$} & \multicolumn{1}{l}{$i=5$} \\
    \midrule
    $\kappa_i$ & 4     & 4.5   & 5     & 5.5   & 6 \\
    $\theta_i$ & 0.01  & 0.0225 & 0.04  & 0.0625 & 0.09 \\
    $\eta_i$ & 0.5   & 0.875 & 1.25  & 1.625 & 2 \\
    $\mu_i$ & 0.05  & 0.075 & 0.10 & 0.125 & 0.15 \\
    \bottomrule\bottomrule
    \end{tabular}%
  \label{tab:Market-para}%
\end{table}%

\begin{table}[H]
  \centering
  \caption{Correlation matrix of dependence structure for the market model. Only non-zero entries are shown.}
  \setlength{\tabcolsep}{5pt}
    \begin{tabular}{lrrrrrrrrrr}
    \toprule\toprule
          & {$X_{1}$} & {$X_{2}$} & {$X_{3}$} & {$X_{4}$} & {$X_{5}$} & {$v_{1}$} & {$v_{2}$} & {$v_{3}$} & {$v_{4}$} & {$v_{5}$} \\
    \midrule
    $X_{1}$ & 1.0   & 0.3   & 0.3   & 0.3   & 0.3   & -0.5  &    &    &    &  \\
    $X_{2}$ & 0.3   & 1.0   & 0.3   & 0.3   & 0.3   &    & -0.5  &    &    &  \\
    $X_{3}$ & 0.3   & 0.3   & 1.0   & 0.3   & 0.3   &    &    & -0.5  &    &  \\
    $X_{4}$ & 0.3   & 0.3   & 0.3   & 1.0   & 0.3   &    &    &    & -0.5  &  \\
    $X_{5}$ & 0.3   & 0.3   & 0.3   & 0.3   & 1.0   &    &    &    &    & -0.5 \\
    $v_{1}$ & -0.5  &    &    &    &    & 1.0   &    &    &    &  \\
    $v_{2}$ &    & -0.5  &    &    &    &    & 1.0   &    &    &  \\
    $v_{3}$ &    &    & -0.5  &    &    &    &    & 1.0   &    &  \\
    $v_{4}$ &    &    &    & -0.5  &    &    &    &    & 1.0   &  \\
    $v_{5}$ &    &    &    &    & -0.5  &    &    &    &    & 1.0 \\
    \bottomrule\bottomrule
    \end{tabular}%
  \label{tab:corr}%
\end{table}%

\subsection{Computation Times}\label{sec:computational-times}

Table \ref{tab:exec-times} shows the average time the algorithm takes to execute one full outer iteration, which includes $m_r=20$ iterations for updating the risk-to-go, $m_f=5$ iterations for updating the conditional cdf, and one iteration for updating the strategy. Models were trained on a Intel(R) Xeon(R) CPU E5-2630 v4@2.20GHz (from 2016) with 64GB of RAM equipped with an NVDIA TITAN RTX GPU (from 2018). As the results show, the timing scales approximately linearly with time steps, but increases only marginally with number of assets. The largest limitation is the size of memory on the GPUs for computing gradients.

\begin{table}[H]
  \centering
  
  \caption{Average execution time (in seconds) per outer iteration with $m_r=20$ and $m_f=5$.\label{tab:exec-times} }
  \setlength{\tabcolsep}{5pt}
    \begin{tabular}{rrrrrr}
    \toprule\toprule
          &       & \multicolumn{4}{c}{T+1} \\
           \cmidrule{3-6}
    \multicolumn{1}{l}{d} &       & \multicolumn{1}{c}{2}     & \multicolumn{1}{c}{4}     & \multicolumn{1}{c}{8}     & \multicolumn{1}{c}{16} \\
    \cmidrule{1-1}\cmidrule{3-6}
    2     &       & 0.71  & 1.44  & 3.26  & 8.05 \\
    4     &       & 0.73  & 1.46  & 3.28  & 8.83 \\
    8     &       & 0.75  & 1.54  & 3.39  & 10.10 \\
    \bottomrule\bottomrule
    \end{tabular}%
\end{table}%
\end{APPENDIX}

\ACKNOWLEDGMENT{SJ and SP acknowledge support from the Natural Sciences and Engineering Research Council of Canada (grants RGPIN-2018-05705, RGPAS-2018-522715, and DGECR-2020-00333, RGPIN-2020-04289). RT acknowledges the support from CNPq (200293/2022-2) and FAPERJ (E-26/201.350, E-26/211.426, E-26/211.578). YS acknowledges the support from CNPq (306695/2021-9) and FAPERJ (E-26/201.375/2022 272760) }
\vspace*{2em}


\bibliographystyle{informs2014}
\bibliography{main.bib}

\end{document}